\documentclass[fleqn,usenatbib]{mnras}

\usepackage{newtxtext,newtxmath}

\usepackage[T1]{fontenc}

\DeclareRobustCommand{\VAN}[3]{#2}
\let\VANthebibliography\thebibliography
\def\thebibliography{\DeclareRobustCommand{\VAN}[3]{##3}\VANthebibliography}

\usepackage{graphicx}	
\usepackage{amsmath}	
\usepackage{xcolor}
\usepackage{hyperref}
\graphicspath{{./}{figures/}}
\usepackage[normalem]{ulem} 

\usepackage{tikz}
\usetikzlibrary{svg.path}
\definecolor{orcidlogocol}{HTML}{A6CE39}
\tikzset{orcidlogo/.pic={
 \fill[orcidlogocol] svg{M256,128c0,70.7-57.3,128-128,128C57.3,256,0,198.7,0,128C0,57.3,57.3,0,128,0C198.7,0,256,57.3,256,128z};
 \fill[white] svg{M86.3,186.2H70.9V79.1h15.4v48.4V186.2z}
 svg{M108.9,79.1h41.6c39.6,0,57,28.3,57,53.6c0,27.5-21.5,53.6-56.8,53.6h-41.8V79.1z M124.3,172.4h24.5c34.9,0,42.9-26.5,42.9-39.7c0-21.5-13.7-39.7-43.7-39.7h-23.7V172.4z}
 svg{M88.7,56.8c0,5.5-4.5,10.1-10.1,10.1c-5.6,0-10.1-4.6-10.1-10.1c0-5.6,4.5-10.1,10.1-10.1C84.2,46.7,88.7,51.3,88.7,56.8z};
}}
\newcommand\orcidicon[1]{\href{https://orcid.org/#1}{
\begin{tikzpicture}[yscale=-0.04,xscale=0.04,transform shape]
\pic{orcidlogo};
\end{tikzpicture}
}}

\DeclareMathOperator{\sech}{sech}

\newcommand{\msun}{M$_{\odot}$}

\title[Deprojection and modelling of BP/X bars]{Deprojection and stellar dynamical modelling of  boxy/peanut bars in edge-on discs}

\author[S. Dattathri et al.]{
Shashank Dattathri~\orcidicon{0000-0002-7941-1149}\!,\!$^{1,2}$\thanks{E-mail: shashank.dattathri@yale.edu}
Monica Valluri~\orcidicon{0000-0002-6257-2341}\!,\!$^{1}$\thanks{E-mail: mvalluri@umich.edu}
Eugene Vasiliev~\orcidicon{0000-0002-5038-9267}\!,\!$^{3}$
Vance Wheeler~\orcidicon{0000-0003-4679-4435}\!$^{1,4}$
{and Peter Erwin~\orcidicon{0000-0003-4588-9555}\!$^{5}$}
\\
$^{1}$Department of Astronomy, University of Michigan, 1085 S. University Avenue, Ann Arbor, MI, 48109, USA\\
$^{2}$Department of Astronomy, Yale University, PO. Box 208101, New Haven, CT 06520, USA\\
$^{3}$Institute of Astronomy, University of Cambridge, Madingley Road, CB3 0HA, UK\\
$^{4}$Department of Physics, University of Chicago, Chicago, Illinois 60637, USA\\
$^{5}$Max-Planck-Insitut f\"{u}r extraterrestrische Physik, Giessenbachstrasse, D-85748 Garching, Germany
}

\date{Accepted XXX. Received YYY; in original form ZZZ}

\pubyear{2023}

\begin{document}
\label{firstpage}
\pagerange{\pageref{firstpage}--\pageref{lastpage}}
\maketitle

\begin{abstract}
We present a new method to infer the 3D dimensional luminosity distributions of edge-on barred galaxies with boxy-peanut/X (BP/X) shaped structures from their 2D surface brightness distributions. Our method relies on forward modeling of newly introduced parametric 3D density distributions for the BP/X bar, disc and other components using an existing image fitting software package (IMFIT). We validate our method using an N-body simulation of a barred disc galaxy with a moderately strong BP/X shape. For fixed orientation angles the derived 3D BP/X shaped density distribution is shown to yield a gravitational potential that is accurate to at least 5\% and forces that are accurate to at least 15\%, with average errors being $\sim 1.5 \%$ for both. When additional quantities of interest, such as the orientation of the bar to the line-of-sight, its pattern speed, and the stellar mass-to-light ratio are unknown they can be recovered to high accuracy by providing the parametric density distribution  to the Schwarzschild modelling code FORSTAND. We also explore the ability of our models to recover the mass of the central supermassive black hole. This method is the first to be able to accurately recover both the orientation of the bar to the line-of-sight and its pattern speed even when the disc is perfectly edge-on. 
\end{abstract}




\section{Introduction}

Stellar bars are found in nearly $\sim$ 50-65\% of nearby disc galaxies \citep[e.g.][]{Knapen99,eskridge_etal_02,marinova_jogee_07, Irina2007, Sheth08,Barazza2008,Aguerri2009, Erwin2018}. Bars are relatively easier to identify in low inclination galaxies by the presence of non-axisymmetric isophotes. For highly inclined (near edge-on) galaxies identification of bars is more difficult but can be done e.g. using stellar kinematic data. In particular, the distribution of the skewness parameter (corresponding to $h_3$ coefficient of the Gauss-Hermite expansion) of the line-of-sight velocity distribution (LOSVD) is correlated with the rotation velocity in the bar region but anti-correlated in axisymemtric galaxies  \citep{Bureau1999,Bureau2005,Palicio2018}. Another prominent (though not universal) signature of a bar in an edge-on galaxy is the presence of a boxy-peanut/X (hereafter BP/X) shaped bulge in the central region \citep{Kuijken1995,Bureau1999_2,Fragkoudi2017}. Bars are considered important drivers of secular evolution in disc galaxies \citep{Sellwood1993,Kormendy2004}  therefore understanding their structure -- especially their three-dimensional mass distribution -- is crucial to understanding their dynamics and evolution.

The deprojection of the surface brightness distribution of a galaxy into its 3D luminosity distribution poses several challenges, as it is inherently an ill-posed inverse problem. For non-spherical distributions there are no unique solutions other than for special inclinations. Since most galaxies are not spherical their 3D luminosity distributions cannot be reconstructed from simple one-dimensional inversion of the surface brightness profile. \citet{Rybicki1987} show that for axisymmetric galaxies of inclination $i$, the Fourier slice theorem leads to a ``cone of ignorance'' of half-opening angle $90^\circ-i$, inside of which the observed surface brightness yields no information. Inside this cone of ignorance, there exist an infinite number of densities (called ``konus densities``) that project to zero surface brightness \citep{Gerhard1996,Kochanek1996}. However, the total mass of a konus density is zero, and \citet{vdb1997} show that konus densities can contribute at most a few percent to the mass profile in the central regions of elliptical galaxies. In addition, \citet{Magorrain1999} showed that although discy konus density components are invisible photometrically, they have strong kinematic signatures, making it possible to constrain their properties by the observed line-of-sight kinematics.

If multiple assumptions are made about the density profile (e.g. with respect to its symmetry properties,  representing the density as a series expansion, etc.), then it is often possible deproject a given surface brightness into a 3D density.  \citet{palmer1994} show that if the density of an axisymmetric galaxy can be represented as a finite sum of spherical harmonics, then it is possible to uniquely deproject the surface brightness, assuming that the true inclination is known. A major development in this direction came with the development of Multi-Gaussian Expansion (MGE) \citep{Bendinelli1991, Monnet1992, Emsellem1994}. The basic algorithm of the MGE method is that if we assume the 3D density profile can be stratified onto concentric ellipsoids, then the surface brightness distribution can be represented as a sum of 2D Gaussians assuming known projection angles. Each of the 2D Gaussian components of the surface brightness can then be deprojected to a 3D Gaussian density distribution. It is important to note that the MGE method does not remove the inherent degeneracy problem in deprojection and gives just one possible solution. Each component of an MGE can be deprojected for a certain range of viewing angles, and thus the entire surface brightness can be deprojected for any assumed orientation in the intersection of these ranges. For a given set of viewing angles, the 3D density can be uniquely recovered from the 2D Gaussians, and the resulting density is smooth and positive. The MGE method has been applied to study various galaxy properties (e.g. \citealt{Emsellem1994_2, vdb1998, Cappellari2008, Miller2021}). 

However, when applied to a disc galaxy  with  a bar at intermediate inclination, MGE produces a density profile that is significantly different from the true density \citep[see fig 2 in][]{VV20_conference}. Since bars may not always be ellipsoidal,  their projected surface brightness distributions cannot always be mapped onto concentric ellipsoids, especially when viewed at non face-on inclinations. To describe such profiles by a superposition of ellipsoidal components, even with varying axis ratios, one would need to make the weights of some components negative, in which case it is hard to ensure that the total density stays positive everywhere. When MGE is used to model barred galaxies, both the bar and disc are transformed into flattened ellipsoids. A few other works \citep{Lablanche2012,Tahmasebzadeh2021} use MGE to describe the photometry of barred galaxies from N-body simulations after masking the disc and modelling it separately.

BP/X shaped bulges are nearly ubiquitous features associated with bars in galaxies with stellar mass $\geq 2.5\times10^{10}$\msun~ \citep{Erwin_Debattista2017}. They are particularly prominent when the disc is viewed edge-on and the bar major axis lies between $\sim 30^\circ -\ 90^\circ$ to the line-of-sight. That our own Milky Way's bar has a BP/X bulge been known ever since the COBE/DIRBE 2.4 micron images were analyzed \citep{Blitz_spergel_1991}, even though it is viewed close to end-on. In recent years made-to-measure models of the  Galactic bar/bulge region using the 3D spatial distribution and line-of-sight kinematics of red clump stars  \citep{Wegg2013,Wegg2015} has set much tighter constraints on mass distribution and even orbital structure of boxy/peanut bulges \citep{Portail2015,Portail2017,Abbott17}.  It is now clear that the density profile of the central region of the Milky Way consists of a prominent boxy-peanut bulge which is part of a longer bar structure \citep{Wegg2013, Ness2016}. Such BP/X structures have long been observed in external disc galaxies \citep[e.g.][]{Laurikainen_etal_2011,Erwin2013,Erwin_Debattista_2016, Yoshino2015}. In addition, N-body simulations have shown that when disc galaxies form bars they can also form BP/X bulges, often following a buckling event in a  bar \citep[e.g.][]{Combes1990, Pfenniger_Friedli_1991,Raha_91}.

Due to their non-axisymmetric nature, deprojection of boxy/peanut bars is a particularly challenging task that has not been attempted for any galaxy other than the Milky Way. The analysis of near face-on barred galaxies has seen some success with deprojection \citep[e.g.][]{Gadotti2007,Li2011}.  Most notably, \citet{Tahmasebzadeh2021} recently presented a method to reconstruct the 3D density of a N-body barred galaxy by decomposing the galaxy into a bulge+bar component and a disc. They then perform MGE on each component separately. Their method yields a deprojected density that is in fair agreement with the true density. They also obtain similar orbits in their model potential and the N-body simulation. However, the model of \citet{Tahmasebzadeh2021} does not attempt to reproduce the BP/X shape. 

The formation of BP/X bulges is a widely researched topic and it is therefore of interest to more accurately model the BP/X structures of external galaxies to better understand their formation and evolution. While the dominant view is that BP/X structures form following a buckling event in the bar \citep[e.g.][]{Combes1990, Martinez-Valpuesta06, Collier2020} there is growing evidence that orbital resonances, in particular the trapping of stars by the vertical Lindblad resonance may play a prominent role in the formation of these structures \citep[e.g.][]{Quillen02, Quillen14,Sellwood_Gerhard_2020}.  Recent work has also shown that the evolution of the BP/X bulge is enhanced by the presence or early growth of a central supermassive black hole (SMBH)  and the strength of the BP/X structure itself is correlated with the bar strength \citep{Wheeler2023}. 

We emphasise that in edge-on disc galaxies the observation of a BP/X bulge is one of the primary ways of identifying the presence of a bar.
BP/X bulges may contain a significant fraction of the mass of the bar with 40-50\% of the orbits in a bar being resonant and non-resonant orbits associated with the BP/X structure \citep{Portail2015,Abbott17}, and hence it would appear reasonable that this structure must be taken into account while dynamically modelling barred galaxies. \citet{Fragkoudi2015} show that modelling a BP/X with a ``flat'' bar can introduce errors in the gravitational force up to $\sim 40 \%$ in some regions.  From a dynamical modelling perspective, correctly modeling the shape of the BP/X bulge could provide important information about both the underlying density distribution in the bar region, and insights into the formation and evolution of bars and BP/X bulges. Recently, \citet{Smirnov} introduced a method to characterize the X shape of external galaxies by introducing a Fourier distortion to the S\'ersic profile \citep{Sersic}. Although these authors study both real and simulated galaxies, they do not attempt to deproject  images of BP/X bulges nor do they compare the 3D densities between their model and snapshot. 

The two popular methods for constructing dynamical models of non-axisymmetric galaxies are the Schwarzschild method \citep{Schwarzschild1979, vdb2008,Zhu2018,VV20,Quenneville_2021,Neureiter2021} and the made-to-measure method \citep{Syer1996,deLorenzi2007,Long2010,Long2013,Portail2015b,Portail2017}. In this work we focus on the former. The Schwarzschild method has been implemented in many different codes over the years (see \citealt{VV20} for a review). While Schwarzschild codes have included the presence of a bar while modelling the Milky Way \citep{Zhao1996, Hafner2000, Wang2012, Wang2013}, applications of such codes to external barred galaxies has so far been limited. Newer codes such as SMILE \citep{Vasiliev2015} and its successor FORSTAND \citep{VV20} are capable of modelling barred galaxies, but so far have been tested only with true 3D density of N-body snapshots, rather than deprojected profiles. Recently a version of the DYNAMITE code \citep{Jethwa2020, Thater2022} was adapted for barred galaxies by \citet{Tahmasebzadeh2022}, using an MGE deprojection method to approximate the bar density. 

The underlying goal of the Schwarzschild method is to construct a dynamically self-consistent orbit-superposition model that satisfies 3D density constraints derived from the surface brightness profile and the observed kinematic constraints, which are usually represented by Gauss-Hermite (GH) coefficients of the line-of-sight velocity distribution (LOSVD) of the galaxy. The 3D luminosity distribution of stars, multiplied by some assumed mass-to-light (M/L) ratio, is used to determine the stellar gravitational potential, which together with additional unseen potential components, such as a central supermassive black hole (SMBH) and dark matter halo is then used to construct a library of orbits.  The contribution of each orbit in the library to both the 3D mass distribution and the kinematic distribution is recorded and a weighted sum of the orbits is sought that reproduces both the 2D and 3D density distribution of the stars, as well as the observed stellar kinematics. 

An important quantity that governs the secular evolution of a barred galaxy and its orbital structure is the bar pattern speed $\Omega$. While there are different definitions of pattern speed (see \citealt{Pfenniger2023}), a common definition used in N-body simulations is the angular speed of rotation of the $m=2$ Fourier mode of the galaxy \citep{Sellwood1986,Debattista2017}. In Schwarszchild modelling of barred galaxies, the orbits are usually integrated in a frame of reference corotating with the bar, in order to maintain a time-independent gravitational potential. Thus, $\Omega$ is a crucial free parameter in the fitting process. 

Observationally, the measurement of $\Omega$ is more difficult and requires certain assumptions about the galaxy model; for example subtracting a model rotation curve from the observed gas velocity and locating the points of co-rotation \citep{Font2011,Ferrer2014}. The only model-independent method that has been widely used is the \citet{Tremaine_Weinberg_1984} method. This method is based on the continuity equation and requires a measurement of both the surface brightness and the velocity field in the plane of the galaxy. As it involves integrals over radius that vanish for a plane-symmetric image, it is limited to galaxies of intermediate inclination and bar orientation \citep[e.g.,][]{Zou2019,Borodina2023}; in particular, it cannot be used with edge-on galaxies. Despite this, the Tremaine--Weinberg method and its generalizations have been successfully applied both to the Milky Way \citep{Debattista2002,Sanders2019} and surveys of external galaxies \citep{Aguerri2015,Guo2019,Oehmichen2020,Oehmichen2022}. 

Since the Schwarzschild method relies on an accurate representation of the 3D potential of the stars in order to accurately integrate the orbits, deprojecting the surface brightness to obtain the 3D density is a crucial step towards constructing realistic dynamical models of external barred galaxies. The majority of direct dynamical black hole mass measurements in external galaxies are estimated using Schwarzschild modelling \citep[e.g.][]{vdb2010,Walsh2012,Thomas2014,Thater2019,Pilawa2022, Merrell2023}. Assuming axisymmetry when modelling barred galaxies (as is commonly done) introduces biases in the measurement of black hole masses \citep{Brown2013, Onken2014}. 

In this paper we present a method to reconstruct the 3D density of edge-on N-body barred galaxies, focusing on the central BP/X shape. We then show that we can use the derived 3D BP/X shaped density distribution in the Schwarzschild modelling code FORSTAND in order to estimate quantities of interest, such as the projection angles, pattern speed of the bar, stellar mass-to-light ratio, and SMBH mass. This paper is organized as follows. In section \ref{sec:methods_sims} we describe our method for deriving the 3D distribution of BP/X barred galaxy from a 2D image. We present the results of applying this method to mock data from an N-body simulation in section \ref{sec:results_deproj}. We compare the surface brightness distribution, 3D density, and gravitational potential and forces between the input N-body galaxy and deprojected model. Section \ref{sec:results_forstand} discusses the results of dynamical modelling  with FORSTAND using the deprojected 3D density distribution and projected stellar kinematics. The recovery of the bar pattern speed, the stellar mass-to-light ratio, and the central SMBH mass is presented. We discuss the implications of our results in section \ref{sec:discussion} and conclude in section \ref{sec:conclusion}.

\section{Methods and simulations}
\label{sec:methods_sims}

Deprojection methods like  MGE for axisymmetric and triaxial distributions start by fitting the 2D surface brightness distribution with components whose 3D distributions that can be inferred from the parameters of the 2D fit once the projection angles have been assumed. This approach is robust if the 2D surface brightness profile uniquely corresponds to a 3D density that is positive everywhere, as in the case of 3D Gaussians. However, it is difficult to generalize to arbitrary shapes: in practice, only ellipsoidally stratified profiles can be uniquely deprojected, although one can use several components with different projected axis ratios to create a non-ellipsoidal total density profile. In this work we do not attempt to {\em deproject} the image of a BP/X structure to its 3D counterpart. Rather we use forward modeling and {\em assume a 3D parametric form for the density distribution of the BP/X structure} which we then project to 2D using functionality provided by the IMFIT \citep{erwin2015} image-fitting program. IMFIT's ability to project any parameteric 3D distribution through a variety of orientation angles and fit the projection to a given 2D image  to recover the best-fit parameters ensures that the 3D density distribution is always positive and finite (although it does not guarantee uniqueness). Although we refer to this procedure as ``deprojection'' we emphasise that in fact we are not attempting to solve the inverse problem, but are carrying out forward modeling. We describe the method used in IMFIT in section~\ref{sec:imfit}. We describe the components of the 3D BP/X bulge/bar and disc that we added to IMFIT in section~\ref{sec:components}. We discuss how we select initial guesses for the parameters and constraints on their values in section \ref{sec:initial_guess}. We describe our tests on mock data generated from N-body simulations of a bar with a BP/X structure in section~\ref{sec:nbody}.

\subsection{3D BP/X bulge and bar model construction with IMFIT \label{sec:imfit}}
IMFIT \citep{erwin2015} is an image-fitting program specifically designed for galaxies. Although the primary function of IMFIT is to fit 2D images of galaxies with multi-component 2D parametric models, here we use IMFIT for deriving the 3D density distribution of BP/X bulges. IMFIT is chosen for two main reasons: 
\begin{enumerate} 
    \item It also includes families of parametric 3D density profiles, which can be integrated along a specified line-of-sight. IMFIT searches the multidimensional parameter space using a maximum-likelihood method to find the projected image that provides the best fit to the input image in order to produce the best-fit model. IMFIT can either accept a fixed orientation for the 3D density profiles or search for the best fit orientation angles.
    \item The object-oriented code is easily extensible, allowing us to easily write and add additional user-defined parametric density components.
\end{enumerate}
We use the default maximum-likelihood approach of IMFIT to construct the best-fit model, which implements a $\chi^2$-minimization method using the Levenberg--Marquardt gradient search algorithm. The $\chi^2$ statistic is calculated as 
\begin{equation}
    \chi^2 = \sum_{i=1}^{N} w_i (I_{d,i}-I_{m,i} )^2 
    \label{eq:chi2}
\end{equation}
where $I_{d,i}$ and $I_{m,i}$ refer to the data and model pixel intensities respectively, and $w_i$'s are the pixel weights. The weights are given by 
\begin{equation}
w_i=1/\sigma_i^2
\end{equation}
where $\sigma_i$ is the error in each pixel. Under the Gaussian approximation of Poisson statistics, the pixel errors are related to their intensity as $\sigma_i^2 = I_{d,i}$.

In order to construct the 3D density model, we need to transform from the 2D coordinate system on the sky $(X,Y)$ to the 3D coordinate system of the galaxy $(x,y,z)$. Since the barred galaxy is non-axisymmetric, in general we need three rotation angles in order to specify the orientation of the galaxy, and it is customary to use Euler angles defined as follows. Denote the intersection of the image plane with the equatorial plane ($x$--$y$) of the model as the line of nodes. The position angle $\theta$ is the angle between the $X$ axis of the image and the line of nodes. The angle between the two planes is the inclination angle $i$, zero when the model is projected face-on and $90^\circ$ when it is projected edge-on. Finally, the bar angle $\psi$ is the angle between the major axis of the model $x$ and the line of nodes. 

\begin{eqnarray}
    \begin{pmatrix}
    x \\ y \\ z
    \end{pmatrix} &=& 
    \left[ \begin{pmatrix}
    1 & 0\\0 & \cos i \\ 0 &\sin i 
    \end{pmatrix}
    \left[ 
    \begin{pmatrix}
    \cos \theta & \sin \theta \\ -\sin \theta & \cos \theta 
    \end{pmatrix}
    \begin{pmatrix}
    X-X_0 \\ Y - Y_0
    \end{pmatrix}
    \right] 
    + s 
    \begin{pmatrix}
    0 \\ \sin i \\ -\cos i
    \end{pmatrix} \right] \times \nonumber\\
    &&\begin{pmatrix}
    \cos \psi & \sin \psi & 0 \\
    -\sin \psi & \cos \psi & 0 \\
     0 & 0 & 1
    \end{pmatrix}
\end{eqnarray}
where $(X_0,Y_0)$ are coordinates of the galaxy center, and $s$ is the distance along the line-of-sight such that $s=0$ is in the sky plane and contains the galaxy center (for more details we refer the reader to \citealt{erwin2015}). The projected image of the model is produced by integrating along $s$ to $\sim \pm 5$ times the disc scale length. 

Without loss of generality, the image of a galaxy can be rotated such that position-angle between the line-of-nodes and the image +$X$ axis is zero.
In this paper we restrict ourselves to modeling edge-on disc galaxies ($i=90$), deferring other inclination angles to a future paper. However, we allow the bar angle $\psi$ (angle between bar major axis and line-of-nodes) to be a free-parameter which is to be inferred from the modeling.

\subsection{Components of the parametric model} \label{sec:components}
Here, we describe the three components (bar, disc, bulge) of the parametric density distribution that we fit the input image to. The final density distribution is the sum of the densities of these three components. In principle additional components could be added but we use the minimum number necessary to achieve a good fit.

\subsubsection{Bar \label{sec:bar}}
We use the results of \citet{Picaud2004} and \citet{Robin2012}, who use star counts from the DENIS (Deep Near Infrared Survey of Southern Sky) survey to fit various parametric density profiles to the Milky Way bulge/bar. They find that the outer bulge/bar regions are best described by a sech$^2$ profile in scaled radius $R_s$:
\begin{equation}
\label{eqt:bar_density}
    \rho = \rho_0 \sech^2(-R_s)
\end{equation}
where, 
\begin{equation}
\label{eqt:bar_radius}
    R_s = \left( \left[  \left(\frac{x}{X_{\rm bar}} \right)^{c_\perp} +\left(\frac{y}{Y_{\rm bar}} \right)^{c_\perp} \right]^{c_\parallel/c_\perp} + \left( \frac{z}{Z_{\rm bar}}\right)^{c_\parallel} \right)^{1/c_{\parallel}}
\end{equation}
where the $(x,y,z)$ coordinates are centered at the galaxy center, $X_{\rm bar}$, $Y_{\rm bar}$, and $Z_{\rm bar}$ are the semi-axis lengths of the bar along the major ($x$), intermediate ($y$), and minor ($z$) axes respectively, where the $z$-axis is perpendicular to the disc plane. The axes lengths are related as $Y_{\rm bar}=q X_{\rm bar}$, $Z_{\rm bar}=q_z X_{\rm bar}$, where $q$ and $q_z$ are the intermediate/major and minor/major axis ratios respectively. 

\begin{figure*}
\centering
\includegraphics[width=\textwidth]{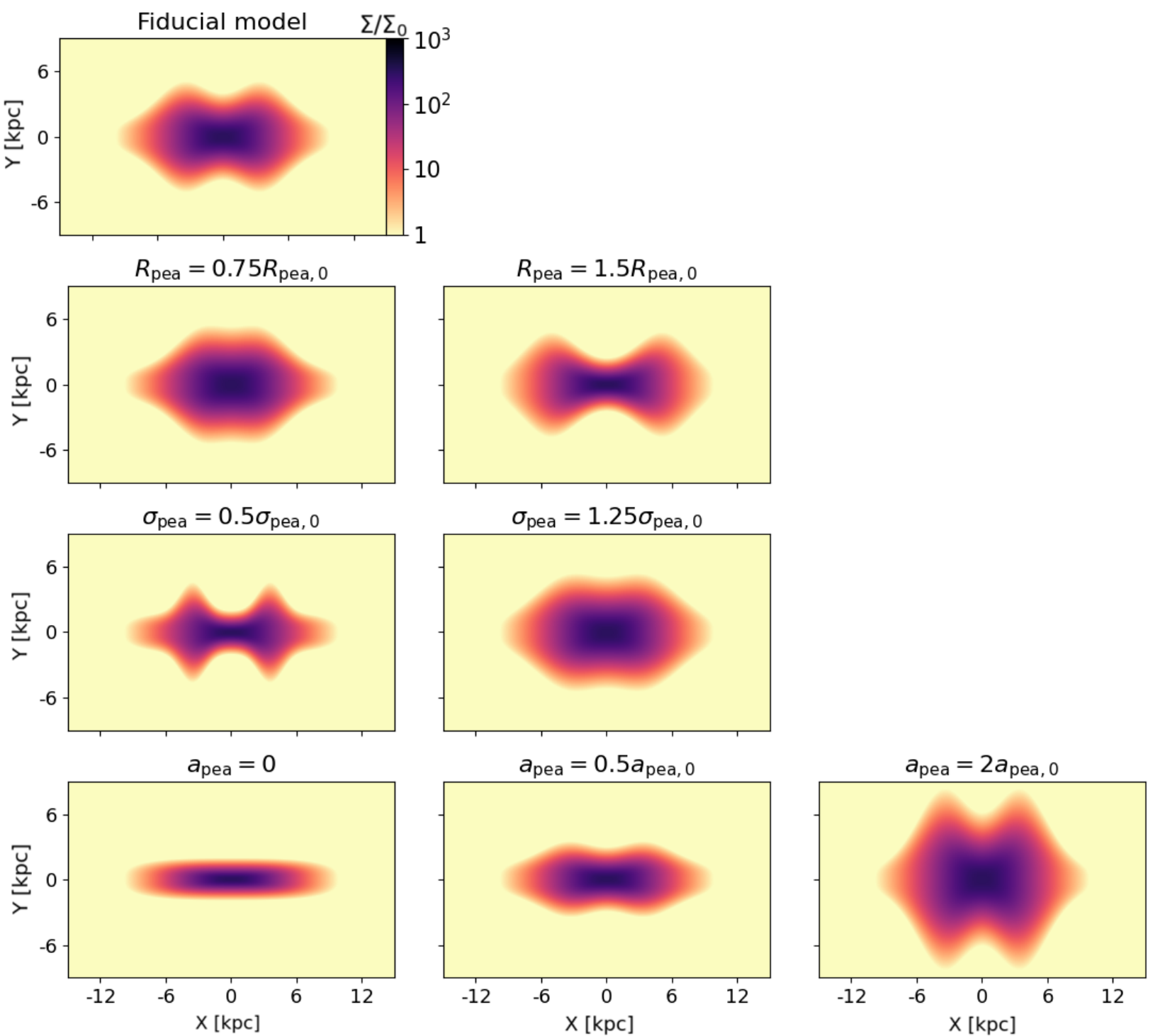}
\caption[]{Projections of different bar models that can be constructed using equations \ref{eqt:bar_density}-\ref{eqt:peanut}. The top panel shows the fiducial model, with parameters $R_{\rm pea,0}$, $\sigma_{\rm pea,0}$, and $A_{\rm pea,0}$ which are close to the best-fit parameters of the model in figure \ref{fig:2D_fits}. In each subsequent row, one of $R_{\rm pea}$, $\sigma_{\rm pea}$, and $A_{\rm pea}$ are varied. Each model is projected side-on, with surface brightness normalized to the background sky brightness.}
\label{fig:bar_models}
\end{figure*}

The parameters $c_\parallel$ and $c_\perp$ control the diskiness/boxiness of the bar (the 3D analog of \citealt{Athanassoula1990}; see \citealt{Picaud2004}) and offer a great amount of flexibility in modelling the bar shape. A pure ellipsoidal bar has values $c_\parallel=c_\perp=2$. A value $c_\parallel<2$ results in a discy side-on projection of the bar, whereas $c_\parallel>2$ results in a boxy side-on projection. Likewise, the value of $c_\perp$ controls the diskiness/boxiness of the face-on projection. The values of $c_\parallel$ and $c_\perp$ are theoretically unbounded; however, we find that IMFIT may output unreasonably large values if they are unconstrained. Therefore, we restrict the values of $c_{\parallel}$ and $c_{\perp}$ between 1.5 and 5, in line with observational studies \citep{Gadotti2009,Robin2012}. 

The above model, however, cannot model a boxy-peanut/X shape in the bar if $Z_{\rm bar}$ is held constant. Both N-body simulations \citep{Athanassoula2002} and fits to observations of the Milky Way \citep{Wegg2015} have shown that the vertical scale height of the bar ($Z_{\rm bar}$) varies along the major axis of the bar. This position-dependent scale height is what gives rise to the BP/X shape. We therefore use a double Gaussian centered at the galactic center to parametrize $Z_{\rm bar}$:
\begin{equation}
\label{eqt:peanut}
\begin{split}
    Z_{\rm bar}(x,y)&=A_{\rm pea} \exp{ \left( - \frac{(x-R_{\rm{pea}} )^2}{2 \sigma_{\rm pea}^2}  - \frac{y^2}{2 \sigma_{\rm pea}^2} \right) } + \\
    &A_{\rm pea} \exp{ \left( - \frac{(x + R_{\rm{pea}} )^2}{2 \sigma_{\rm pea}^2}  - \frac{y^2}{2 \sigma_{\rm pea}^2} \right) } + z_0
\end{split}
\end{equation}
This expression is very similar to the ``peanut height function'' described in \citet{Fragkoudi2015}, except that we constrain  the two halves of the peanut to be symmetric about the galactic center, and require that the peanuts are aligned along the major axis of the bar. 

The shape of the bar is controlled primarily by three parameters: $R_{\rm pea}$ (distance of the peanuts from the galactic center), $A_{\rm pea}$ (vertical height of the peanut from the disc plane), and $\sigma_{\rm pea}$ (width of each peanut). In figure \ref{fig:bar_models}, we show how the resulting shape of the bar changes when we vary these parameters, illustrating how the three parameters offer the versatility to model a large variety of bar shapes.

\subsubsection{Disc}
The disc is modelled as an axisymmetric density profile which follows the vertical $\sech^{2/n}$ profile of \citet{Kruit1988}. However, we find that a simple exponential fit in the radial direction is not able to accurately fit the disc density in our mock data, and  additional parameters are necessary. Therefore, we model the disc using the expression:
\begin{equation}
    \rho(R,z) = \rho_0 \ \exp{\left({ -(R/R_{\rm disc} )^{k} - R_{\rm hole}/R} \right)} \sech{(z/\alpha z_{\rm disc})}^{\alpha} 
    \label{eq:disk_dens}
\end{equation}
where $R=\sqrt{x^2+y^2}$, i.e. the radial distance in the plane of the disc and $z$ is the vertical distance perpendicular to the disc. Here $\rho_0$ (an overall normalization factor), $R_{\rm disc}$ (the radial scale length), $z_{\rm disc}$ (the vertical scale length), $k$, $R_{\rm hole}$, and $\alpha$ are all free parameters. 

This expression was used by \citet{Sormani2022} to model the stellar mass distribution of the MW. It differs from the standard exponential disc by the introduction of the radial index $k$, which controls how sharply the density falls off in the radial direction, and the parameter $R_{\rm hole}$, which models the decrease in the disc density in the inner disc regions along the axis perpendicular to the bar and within the co-rotation radius (the regions surrounding the $L_4$ and $L_5$ Lagrange points of the bar). While this hole may not be a useful model of disks in isolation, here it is used together with a separate bar component largely filling up the hole, but rearranging the stars into a non-axisymmetric structure. The central hole region may show up as a local minimum in surface density when plotted along the minor axis \citep{Freeman1970}, as illustrated in figure~\ref{fig:projections}.  In practice, we find $k \gtrsim 2$, indicating that the surface brightness decreases faster than a simple exponential or Gaussian.

We also note that we tried the built-in \textit{BrokenExponentialDisk3D} function in IMFIT, which consists of two exponential radial zones with different scale lengths joined together. However, this resulted in a worse fit compared to the disk profile in equation \ref{eq:disk_dens}. 

\subsubsection{Bulge}
The central bulge is modelled using the triaxial generalization of the \citet{Einasto} profile: 
\begin{equation}
\label{eqt:bulge}
    \rho(r) = \rho_0 \exp \left( -b_n \left( \frac{r}{R_{\rm bulge}}\right)^{1/n} -1 \right)
\end{equation}
where $r=\sqrt{x^2+(y/q)^2+(z/q_z)^2}$ is the 3D ellipsoidal radius, and $n$ is the index controlling the shape of the density profile.

For our mock data from N-body simulations, we find that a bar+disc+bulge model provides a reasonable fit to the galaxy (section \ref{sec:results_deproj}). We note that when this method is applied to real galaxies, one may need to add additional components to obtain accurate models, such as the presence of strong spiral arms, dust rings, and/or a secondary bar (e.g. \citealt{Athanassoula1990, Gadotti2007}). We also emphasise that  the  simulated barred galaxy that we use to create the mock images and kinematics used to validate our method was generated from initial conditions that did not have a classical bulge component and was initially a pure axisymmetric disc which formed a bar and boxy/peanut bulge through secular evolution. However, we find it necessary to include a spheroidal bulge component to fit both the image and the kinematics regardless of whether the bulge is strictly a classical bulge or not.

\subsection{Initial guesses and constraints for the parameters \label{sec:initial_guess}}
The Levenberg--Marquard algorithm implemented in IMFIT requires reasonable initial guesses for all the parameters that we are trying to fit. Since our disc+bar+bulge model has a total of 30 free parameters, it is difficult to provide good guesses for the parameters that describe all three components at once. We therefore use the following strategy for initializing the values:
\begin{enumerate}
    \item We completely mask out central 6 kpc of the image, which corresponds to the bar+bulge region. The resulting image is fitted to the modified exponential disc (equation \ref{eq:disk_dens}) to obtain estimates for the parameters of the disc. Since the disc-only fit contains only 6 free parameters, we can provide arbitrary initial values without the risk of getting trapped in local minima.  
    \item We then mask out the outer parts of the disc in order to focus on the central regions. We use the best-fit parameters of the disc-only fit as initial values for the disc parameters. 
\end{enumerate}
This method of strategically masking out different components in order to go from a simple model to a complex multi-component model is fairly popular (e.g. \citealt{Smirnov}), and reduces the risk of the solver getting trapped in local minima and/or producing unphysical values for the parameters.

We also assume that we have {\it apriori} knowledge of some important quantities. Since we are working with an edge-on disc galaxy, we first can rotate the image so that the disc (and bar) are aligned with the $X$-axis, thereby setting the position angle parameter ($\theta=0^\circ$). Then we assume that the true inclination is $i=90^\circ$, which is reasonable since the disc is being viewed edge-on. In the general case, the inclination $i$ of a galaxy can be estimated by various methods, for example from the shape and orientation of the disc (as well as 2D kinematic velocity field if available) \citep{Barnes2003,Cappellari2008,Tahmasebzadeh2021}, or from the distribution of HII regions \citep{GarciaGomez2002}. We then start with an initial guess for the bar angle to the line-of-sight (e.g. $\psi=45^\circ$). As we show in section \ref{sec:results_deproj}, deprojection using photometric data alone results in significant degeneracy in the measurement of $\psi$. However, since the 2D kinematics of the galaxy is sensitive to the value of $\psi$, this degeneracy can be resolved with Schwarzschild modelling. For our initial illustration of the method, we keep the bar angle $\psi$ at its true value of $\psi=45^\circ$. 

In addition, most edge-on projections of the galaxy contain no information about the bar axis ratio between the intermediate and major axes ($q=Y_{\rm bar}/X_{\rm bar}$ in equation \ref{eqt:bar_radius}). Our experiments with the value of $q$ as a free parameter showed that IMFIT alone cannot constrain this quantity. We therefore use a fixed value of $q=0.4$ in our fits. The Milky Way's bar has an axis ratio of $\sim 0.35 - 0.4$ as measured using red clump giant (RCG) stars \citep{Rattenbury2007}.  From analysing the face-on projection of our N-body mock data, we find that $q=0.4$ provides a reasonable fit to the data, and is representative of the average axis ratio of real galactic bars \citep{Sellwood1993,Gadotti2009}.

\subsection{Mock data from N-body simulation} \label{sec:nbody}

In this paper, we construct mock IFU photometric/kinematic data with the simulated disc galaxies viewed edge-on and with the disc lying along the $x$-axis of the image. This corresponds to $\theta=0^\circ$ and $i=90^\circ$. We discuss the simulated model and construction of mock photometric data here, and the construction of kinematic data is discussed in section \ref{sec:results_forstand}.

To construct mock data, we use the final snapshot of a barred disc galaxy (Model $BB_1$) from a suite of N-body simulations generated and analyzed by \citet{Wheeler2023}. These authors used the grid-based N-body simulation package \emph{GALAXY}
\citep{GALAXY} to simulate the growth of SMBHs (represented as smoothed Plummer potentials) at various stages in the formation and evolution of the bar. The initial conditions for the bar-unstable disc were generated using {GalactICS} \citep{Widrow_etal_2008,Widrow_Dubinski_2005,Kuijken_Dubinski_1995} and were previously described in detail in \citep{Debattista_etal_2017, Debattista_etal_2020,Anderson_etal_2022,Wheeler2023}. The initial conditions began as an axisymmetric exponential disc within a  Navarro--Frenk--White (NFW) \citep{Navarro_etal_1996} live dark matter halo modified to have a cut off at large radius ($r>100$ kpc). 
The disc had a total mass of $\simeq 5.37 \times 10^{10}~ \mathrm{M}_\odot$ represented by $6 \times 10^6$ equal mass particles. The dark matter halo had a total mass of $6.8\times 10^{11}$\msun\ represented by $4\times 10^6$ particles. \citet{Wheeler2023} grew a Plummer potential representing a central supermassive black hole (SMBH) with a final mass of $7.5\times 10^7$\msun~ and softening length of $\sim 33$pc at various times before, during and after the formation of the bar. SMBH were grown over a period of $378\ \mathrm{Myr}$ starting from an initial mass that was 2\% of its final mass. In the model used in this work, the SMBH was introduced 0.575~Gyr after the start of the simulation, while the bar was still growing and before it first buckled.

The edge-on projection of the galaxy can have a bar angle $\psi$ varying from $0^\circ$ (side-on projection, major axis perpendicular to the line-of-sight) to $90^\circ$ (end-on projection, major axis along the line-of-sight). The BP/X shape is distinctly visible when $\psi \lesssim 60^\circ$. For our fiducial mock image we fix the bar angle  $\psi=45^\circ$, which corresponds to an intermediate projection between side-on and end-on. The total horizontal and vertical extent of the projected image is $\pm 30$ kpc in the horizontal direction and $\pm 12$ kpc in the vertical direction. We bin the particles into square pixels, with 1000 pixels along the horizontal axis and 400 pixels along the vertical axis. This corresponds to a pixel resolution of 60~pc$\times$60~pc. 

Galactic bars may undergo buckling and become asymmetrically bent out of the galactic plane, which has been observed in N-body simulations  \citep[e.g.][]{Lokas2019} as well as observations \citep{Erwin_Debattista_2016,Xiang2021, Cuomo_etal_2023}. While it may be short-lived in some cases, the bending may be present at later times as well, resulting in a persistent bending of the bar and disc plane \citep{Wheeler2023}. This bend may result in poor fits in the bar region, since our analytic model (equations \ref{eqt:bar_radius}, \ref{eqt:peanut}) only models ``straight'' ($z$-symmetric) bars. Moreover, our dynamical modelling code FORSTAND (described in section \ref{sec:results_forstand}) is limited to models with reflection symmetry about the 3 principal axes. Therefore, we ensure that the galaxy is symmetric about the disc plane by taking $z>0$ particles and reflecting them about the z-axis. This results in considerably better photometric fits and realistic best-fit parameters. We emphasize that we only use this symmetric snapshot to produce the input photometric image for IMFIT, and later dynamical modelling steps use the unsymmetrized snapshot for both photometry and kinematics. In principle, our photometric bar model can be extended to include bent bars, but since that would add additional parameters to our model, we do not study this here.

\subsection{Comparison of 3D parametric model  with N-body snapshot}
\label{subsec:3dfit}
One may wonder whether any discrepancies between the constructed model and the N-body snapshot are due to the choice of the parametric density profile that we use, rather than the deprojection process. In order to test this, we construct a 3D density distribution using our bar+disc+bulge multi-parameter density distribution described in section
\ref{sec:components}, by directly fitting this model to the N-body snapshot instead of the projected image. 

We bin the particles of the N-body snapshot in cylindrical bins ($R$, $\phi$, $z$). The $\phi$ grid is equally spaced with 12 bins, whereas the $R$ and $z$ grids are constructed with 40 bins each with a gradually increasing spacing. This gives us sufficient resolution to study the inner bar/bulge region in detail, while also allowing us to cover the entire galaxy without an excessive number of bins. This gives us $N_{\rm snap,i}$, the number of particles in each bin $i$. We then construct a model using fiducial parameters and obtain $N_{\rm model,i}$, the expected number of particles in each bin in our model. By varying the parameters of the model, we aim to obtain the best-fit model density that matches the snapshot density. By Poisson statistics, the objective function is given by 
\begin{equation*}
    \log \mathcal{L} = \sum_i \big[- N_{\rm model,i} + N_{\rm snap,i} \times \ln(N_{\rm model,i}) \big].
\end{equation*}
We use the Nelder-Mead algorithm to minimize this objective function in order to constrain the parameters of the density profile. 

We will refer to this as the "fit-3D-snap model". By contrast, the density model recovered by IMFIT is referred to as the "deprojected model". In subsequent sections, we present the results from the fit-3D-snap model alongside the deprojected model in order to demonstrate the flexibility of our parametric density profile and the inherent limitations of deprojection.  This is also done to illustrate the value of our multi-component parametric BP/X bar model for other purposes, such as quantitative characterization of the BP/X shapes from different simulations \citep[for example see][]{Wheeler2023}.

In addition to the two parametric models using the density profiles described in Section~\ref{sec:components}, we also use the ``ground truth'' density and potential of the original snapshot, represented by the \texttt{CylSpline} potential model implemented in the AGAMA library \citep{Vasiliev2019}, which serves as a backend for the Schwarzschild modelling code FORSTAND. \texttt{CylSpline} utilizes azimuthal Fourier expansion with coefficients spline-interpolated on a 2D grid in the $\{R,z\}$ plane, and can be constructed either from an analytic density model (in particular, our parametric models described above) or directly from an $N$-body snapshot. This allows us to evaluate how well the density and potential of the deprojected/fit-3D-snap models match that of the snapshot.

\section{Results: Deprojection}
\label{sec:results_deproj}
In this section, we present our results from deprojecting the mock data obtained from the N-body simulations described in section \ref{sec:nbody}. 

\subsection{Recovered 2D and 3D structures assuming true orientation} 
\label{sec:results_trueorient}
\subsubsection{2D image fit }
\label{sec:results_2dfit}
\begin{figure*}
\centering  \includegraphics[width=\textwidth]{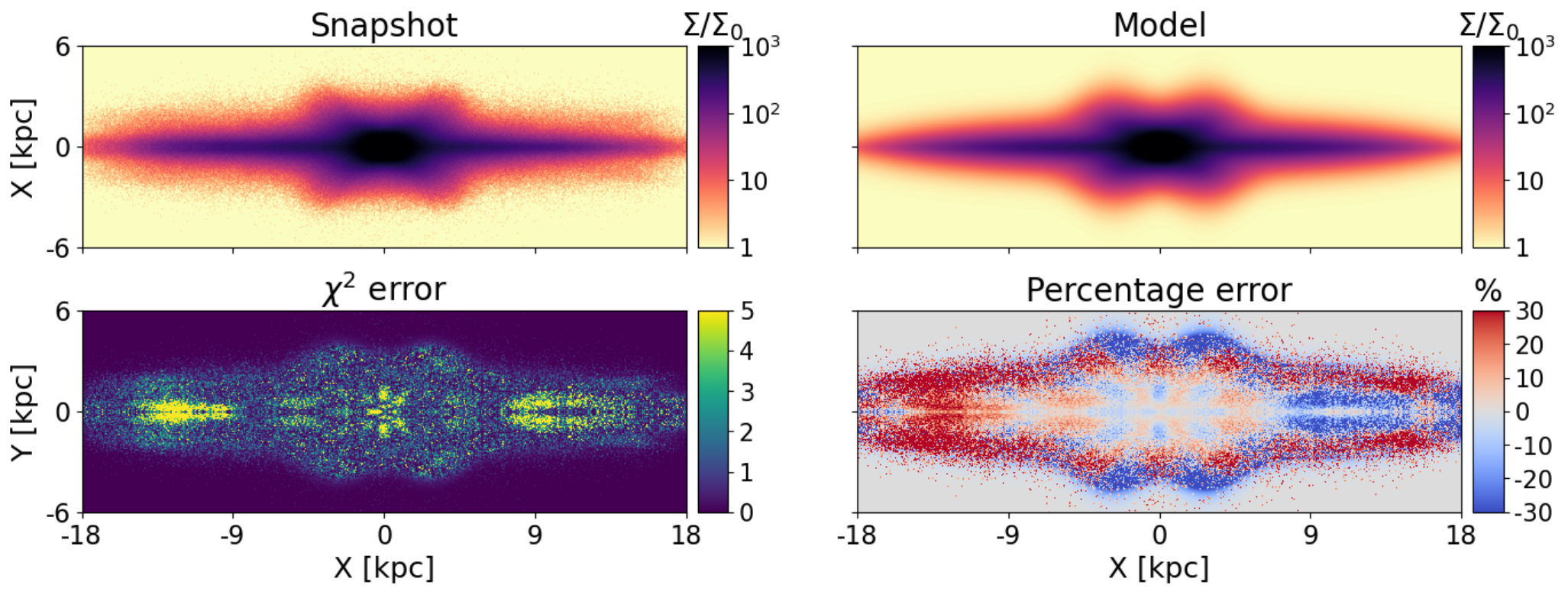}
\caption[]{Top left: projection of N-body snapshot, in units normalized by $\Sigma_0$, the background sky brightness. The snapshot is projected edge-on $i =90^\circ$ and with the bar oriented at an angle $\psi=45^\circ$ along with line of sight. Top right: the best-fit model as calculated by IMFIT, viewed along the same projection angles. Bottom left: $\chi^2$ map between the projected images of the snapshot and model, calculated as $(I_{\rm snap}-I_{\rm model})^2/I_{\rm snap}$ for each pixel. Bottom right: percentage error map, calculated as $100 \times (I_{\rm snap}-I_{\rm model})/I_{\rm snap}$. Note that the $\chi^2$ and percentage error maps trace different regions of the galaxy: the $\chi^2$ map is dominated by the high density regions while the percentage error is high in the outer low-density regions. }
\label{fig:2D_fits}
\end{figure*}

The top left panel of figure \ref{fig:2D_fits} shows the input image: an edge-on disc with the bar oriented at an angle $\psi=45^\circ$. The simulation used is Model $BB_1$ from \citet{Wheeler2023}. The peanut/X-shape of the bar is clearly visible. 

The top right panel of figure \ref{fig:2D_fits} shows the $i=90, \psi=45^\circ$ projection of the deprojected model, which is the best-fit disc+bar+bulge model derived by IMFIT from the input image in the left panel (where the values of $i$ and $\psi$ are fixed at their true values). We can see that the peanut shape is well reproduced by the model. In addition, the overall shape and surface brightness of the disc are reasonably matched, except for some low-density outer disk regions.

The bottom left and bottom right panels of figure \ref{fig:2D_fits} show the pixel-wise $\chi^2$ error map as defined in equation~\ref{eq:chi2} and the pixel-wise percentage error map ( $= 100 \times (I_{\rm snap}-I_{\rm model})/I_{\rm snap}$). It is important to note that IMFIT fits the model based on minimization of the total $\chi^2$ and not the percentage error or the pixel-wise error. We note the following features in the error maps:
\begin{itemize}
    \item The outer parts of the disc show a large percentage error, because the pixel intensity in these regions is low and is dominated by Poisson noise due to small numbers of N-body particles per pixel. These regions contain very little mass compared to the inner regions of the galaxy, and therefore are not important from a dynamical modelling point of view, unless one is interested in constraining the parameters of the dark halo (which we do not attempt here). This is reflected in the $\chi^2$ error map, as the outer regions have very little contribution to the overall $\chi^2$.
    \item Since our density profile does not model spiral arms, there are regions in the disc plane where both the percentage and $\chi^2$ errors are significant. 
\end{itemize} 

The inner region of the bar is well reproduced as determined by the small residuals in both the percentage error map and the $\chi^2$ map. It is these regions which contain the most mass and therefore contribute the most to the gravitational potential and forces.  

The reduced $\chi^2$ of the best-fit image is 0.66, and the average percentage error is $-2.04 \%$. However, the interpretation of $\chi^2$ is somewhat ambiguous, since our input image is constructed from an N-body snapshot and therefore contains no added noise apart from the Poisson noise. The primary use of $\chi^2$ in our study is in the Levenberg--Marquardt in IMFIT to recover the best-fit parameters, not to characterize the model with an absolute goodness of fit. 

Once the surface density fitting is complete we can use the best-fit parameters given by IMFIT for the 3D model (parameterized as described in section~\ref{sec:components}) to construct the density distribution. In the next section we compare various projections of the deprojected model directly with the corresponding projections of the N-body snapshot.
 
\subsubsection{Goodness of the IMFIT-recovered 3D model}

\begin{figure*}
    \centering
    \includegraphics[width=\textwidth]{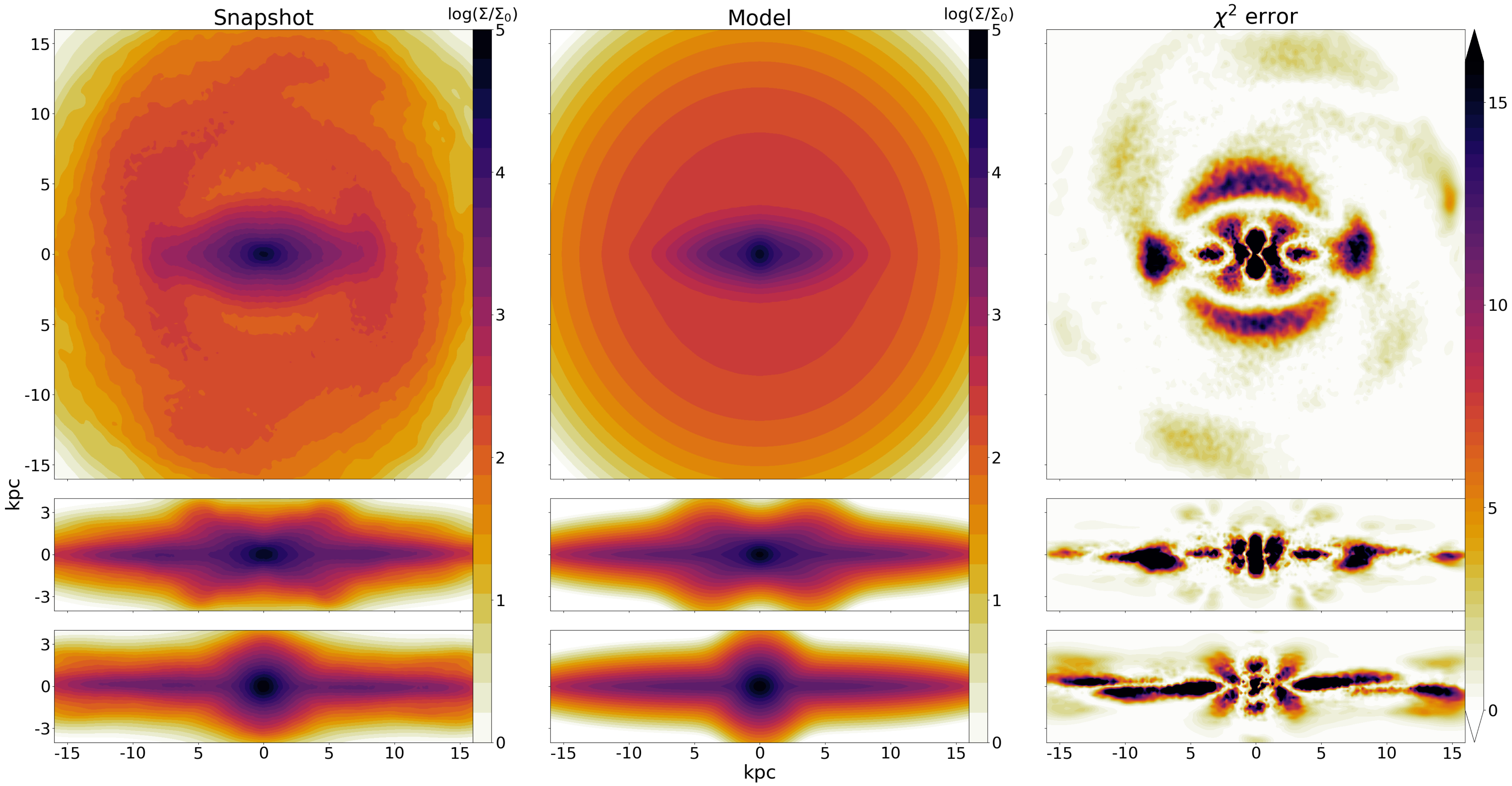}
    \caption{Left: the face-on (top), side-on (middle), and end-on (bottom) projections of the N-body snapshot. Middle: the same three projections of the deprojected model. Right: the same three projections of the $\chi^2$ error in the density between the snapshot and model. The side-on projection has the least error since the input image is closest to this. The face-on projection shows some regions with larger errors, particularly in the low-density region surrounding the disc and the spiral features. We also note that the outer parts of the disc show larger errors in the side-on and end-on projections since the density in the snapshot itself is quite low in these regions. }
    \label{fig:projections}
\end{figure*}

In order to qualitatively check the goodness of the density fit, we now compare various projections of the N-body snapshot (projections not used by IMFIT) with the same projections of the deprojected model. These projections qualitatively illustrate the validity of our 3D model and can be useful for identifying where the errors are coming from. The left column of figure \ref{fig:projections} shows the face-on, side-on, and end-on projections of the N-body snapshot (recall that the image provided to IMFIT has the disc edge-on but the bar angle $\psi=45^\circ$, an orientation not shown in this figure).
Each projected image is constructed by simply projecting the N-body snapshot after rotating it through the appropriate angles. 

The middle column shows the same three projections of the deprojected model.  We can see that all three projections of the deprojected model match the N-body snapshot reasonably well. Naturally, there are several features in the snapshot that cannot be reproduced by an analytical density model, in particular the spiral arms  and the ``ansae'' at the ends of the bar which are visible in a significant fraction of barred galaxies \citep{Martinez2007}. This is not surprising since neither the spiral arms nor ``ansae'' are  visible in the edge-on projection of the galaxy, and therefore reproducing these features would not be possible. Adding more components to the model such as spiral arms or ``ansae'' can be undertaken in future when we focus on non-edge-on discs.

The right-hand column of figure \ref{fig:projections} shows the $\chi^2$ error map between the deprojected model and the projected snapshot. We opt to show the $\chi^2$ map instead of the residual (percentage difference) map since the latter is dominated by the low-density regions in the outer regions of the galaxy. These low-density regions are expected to contribute little to the gravitational potential and forces. On the other hand, the $\chi^2$ map is heavily biased towards the high-density regions, where even a small error can result in a large $\chi^2$. 

The discrepancies between the model and snapshot are most clearly visible in the face-on projection of the $\chi^2$ map. The main regions with a high $\chi^2$ are the center, the ansae of the bar, and the underdense ``hole'' surrounding the bar. The errors are the lowest in the side-on projection, since this projection is the closest to the input image (which was projected at $\psi=45^\circ$). In addition, we note that we use the \textit{symmetric} snapshot for constructing the input projected image which is used for deprojection. By contrast, the deprojected model density is compared with the \textit{unsymmetric} (original) snapshot. This results in additional errors, most notably, the bent nature of the disc is evident in the side-on and edge-on projections. We use the original (unsymmetrized) snapshot as an input to orbit-superposition dynamical modelling, even though these density models themselves are symmetric by construction, therefore it makes sense to compare the symmetric deprojected density model to the original snapshot.

We further quantify the quality of the deprojected model as follows. We discretize the space and bin the particles of the N-body snapshot using the same method as in section \ref{subsec:3dfit}. We then calculate the density of the model in the same bins, and compare these binned IMFIT-recovered deprojected model densities ($\rho_\mathrm{dep}$) with the corresponding snapshot densities ($\rho_\mathrm{true}$). 

\begin{figure*}
    \centering
    \includegraphics[height=0.35\textwidth,trim={0 0 93 0},clip]{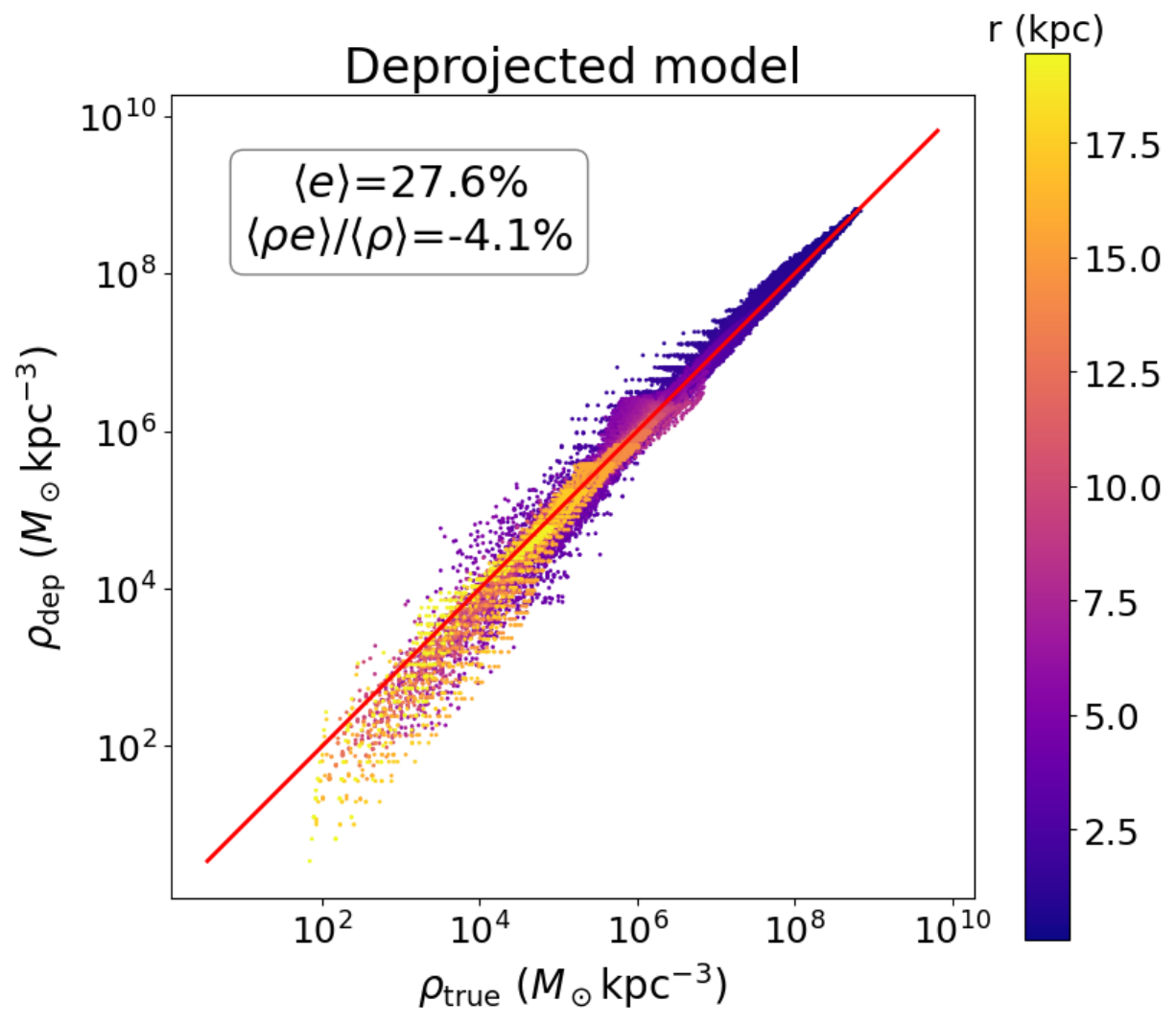} 
    \includegraphics[height=0.35\textwidth,trim={81 0 0 0},clip]{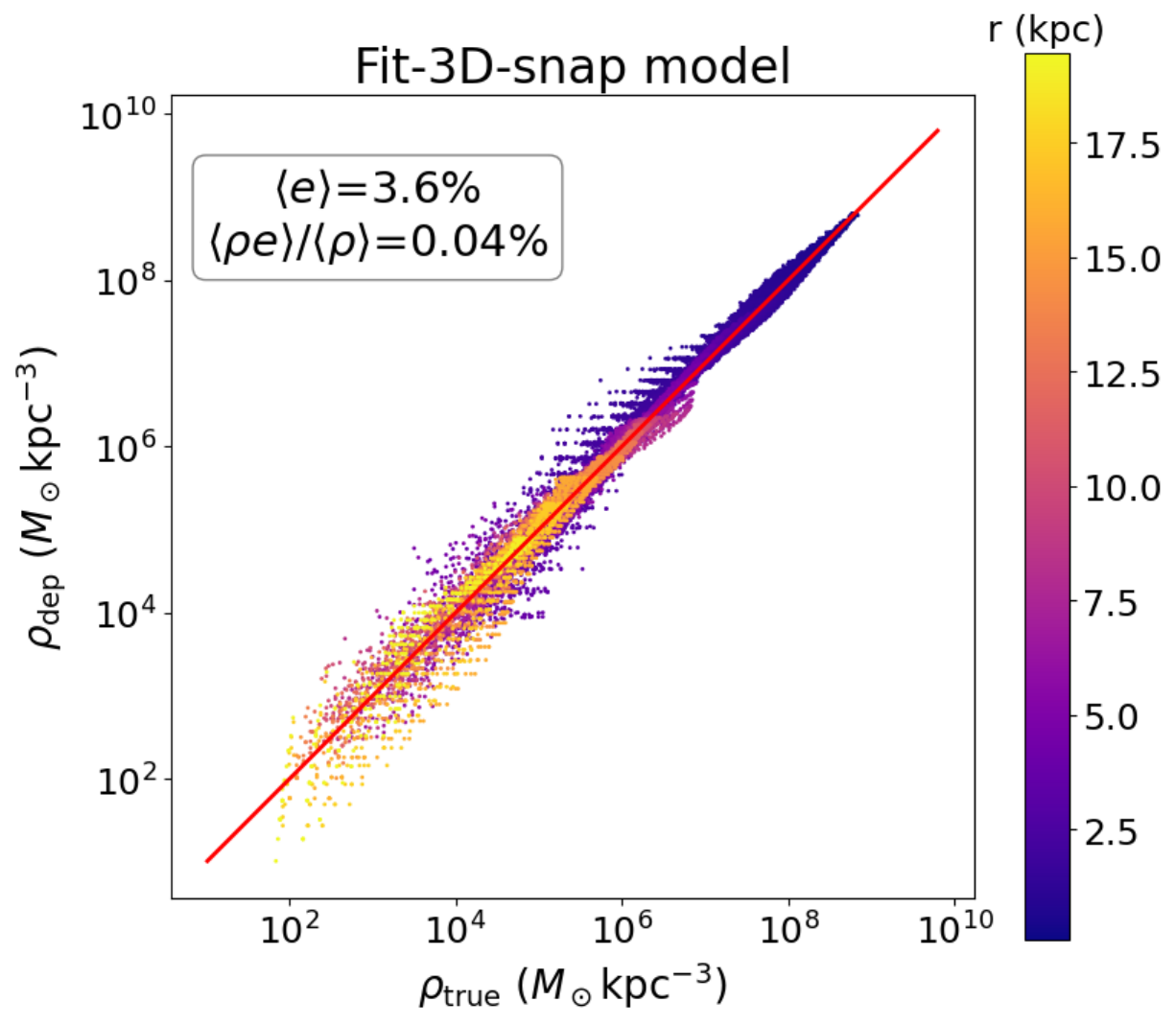}
    \includegraphics[height=0.35\textwidth]{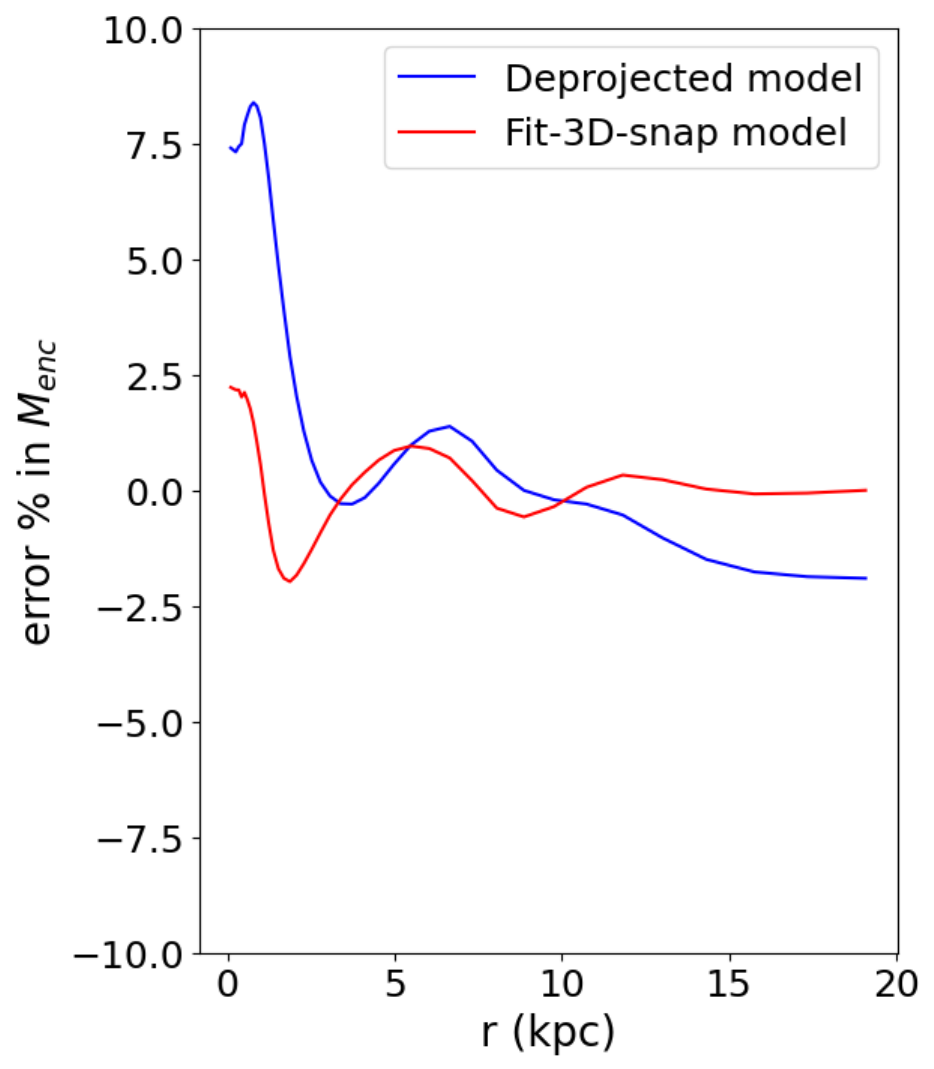}
    \caption{Left: binned snapshot density ($\rho_\mathrm{true}$) vs. deprojected model density ($\rho_\mathrm{dep}$). The red line indicates $y=x$, and the average unweighted and weighted errors ($\langle e \rangle$ and $\langle \rho e \rangle/ \langle \rho \rangle$ respectively) are indicated. The points are colour coded by the distance of the bin center from the galaxy center. We can see that while there is scatter particularly in the low-density outer regions, the inner regions of the model match the snapshot very well. Middle: similar plot for the fit-3D-snap model ($\rho_\mathrm{true}$ vs. $\rho_\mathrm{3d-fit}$). The scatter around $y=x$ is lower than that for the deprojected model, although there is still a non-zero amount of scatter. Right: error in enclosed mass (model - snapshot) vs. $r$ for both the deprojected model and the fit-3D-snap model.}
    \label{fig:3D_fit}
\end{figure*}

Figure \ref{fig:3D_fit}~(left) shows   $\rho_\mathrm{dep}$ versus $\rho_\mathrm{true}$ in each of 19200 bins. The points are coloured by the distance of each bin center from the center of the galaxy. The red line corresponds to $y=x$, so all the points would lie on this line if the deprojected model recovered by IMFIT perfectly matched the snapshot. We can see that there are deviations from the red line, particularly in the low-density outer disc regions as these regions are dominated by Poisson noise.  These deviations are fairly symmetric about $y=x$ (although there is a slight underestimation of the density by the deprojected model). There is also a set of points at around 10 kpc where the model overestimates the density, which corresponds to the underdense region surrounding the bar. We see that the inner high-density regions of the bulge and bar are fairly well reproduced by the model, and the average unweighted and density-weighted errors $\langle e \rangle$ and $\langle \rho e \rangle / \langle \rho \rangle$ (listed above the figure) are both reasonably low.  

As discussed in section~\ref{subsec:3dfit}, since the multi-component 3D parameteric model described in section~\ref{sec:components} may not perfectly describe the 3D density distribution of the N-body snapshot, we also considered the fit-3D-snap model obtained by directly fitting the 3D parametric model to the 3D snapshot.  The middle panel of figure \ref{fig:3D_fit} shows the densities obtained from fit-3D-snap model ($\rho_\mathrm{3D-fit}$) in the same bins vs. true snapshot density $\rho_\mathrm{true}$. It is immediately clear that the scatter around the $y=x$ line is lower for the fit-3D-snap model compared to the deprojected model. The weighted and unweighted errors are significantly smaller. However, features of the snapshot which are not reproduced even in the fit-3D-snap model like the spiral arms and the puffiness of the outer disc are evident. 

Figure \ref{fig:3D_fit} (right) shows the percentage error in the enclosed mass ($M_{\rm enc}(<r)$) between the model and the snapshot for the deprojected model (blue) and fit-3D-snap model (orange).   Since the fit-3D-snap model is obtained without deprojection, the enclosed mass shows considerably less error than the deprojected model (blue).  In particular we find that the deprojected model has a maximum error of $\sim 7.5 \%$ in enclosed mass, where the density is overestimated in the underdense ring surrounding the bar. The total mass of the galaxy is underestimated in the deprojected model by $\sim 2 \%$. On the other hand, the fit-3D-snap model has a total mass nearly equal to the snapshot, and the maximum error in enclosed mass is only $\sim 2.5 \%$. However it is clear from the orange curve that our 3D parametric model does not provide a perfect fit even when directly fitting the snapshot.

\subsubsection{Gravitational potential and forces}
In order to calculate the gravitational potential from a given density distribution, we numerically solve the Poisson equation 
\begin{equation}
    \nabla^2 \Phi=4 \pi G \rho \ .
\end{equation}
We use AGAMA's \texttt{CylSpline} expansion tool, which expands the potential as a sum of azimuthal Fourier harmonics in $\phi$, computes it on a 2D grid in $\{R,z\}$ for each harmonic, and interpolates in the entire space. This potential solver can take as input either the N-body snapshot (which we take as the ground truth) or the analytical parametric density profile. We use \texttt{CylSpline} instead of the more popular multipole expansion in order to accurately calculate the potential of the flattened disc, for which multipole expansion yields large errors. 

\begin{figure*}
        \includegraphics[width=\textwidth]{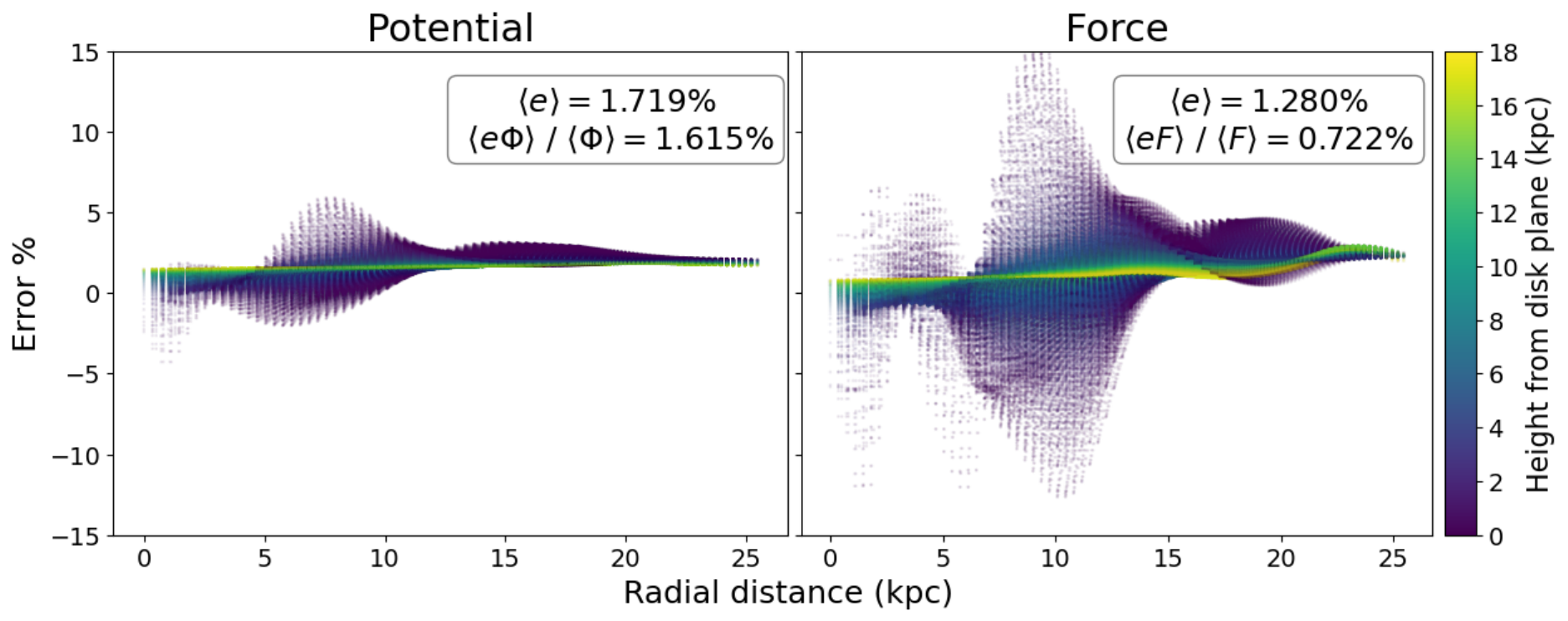}
    \caption{Left: percentage error in the gravitational potential between the deprojected model and snapshot  $ \left[ 100 \times \left( \Phi_{\rm model}-\Phi_{\rm snap} \right) / \Phi_{\rm snap} \right]$ as calculated by the \texttt{CylSpline}  expansion tool in Agama, plotted versus the radial distance from the center of the galaxy. The average unweighted and weighted errors ($\langle e \rangle$ and $\langle x e \rangle/ \langle x \rangle$ respectively) are indicated. The points are coloured according to the vertical distance from the disc plane. Right: the same plot, but for the gravitational force magnitude. It is evident from the figure that both the gravitational potential and force is recovered to high accuracy in the deprojected model. The average error decreases at large radial distance and large vertical height, as at large distances the exact density distribution becomes unimportant and it is the total mass that matters. The error is the highest at moderate radial distance and small vertical height, which is the location of the spiral arms of the galaxy (features that are not included in the deprojected model).}
    \label{fig:pot_forces}
\end{figure*}

The left panel of figure \ref{fig:pot_forces} shows the percentage error in the gravitational potential between the model and the snapshot as a function of the radial distance from the center of the galaxy. The points are colour coded by the vertical distance from the disc plane. As we go further away from the galactic center (both radially and vertically), the error in the potential decreases. This is because at large distances the gravitational potential is insensitive to the detailed density distribution and instead mostly depends on the enclosed mass. The $\sim 2$ percent error at large distances corresponds to the difference in the total mass between the model and snapshot. The potential error is larger in the inner regions; however, the maximum error is still $\lesssim 5$ percent. 

Once the gravitational potential is calculated, the corresponding gravitational force and its magnitude are obtained as 
\begin{equation}
    \vec{F} = -\nabla \Phi \quad \quad F = | \vec{F} | = \sqrt{F_x^2+F_y^2+F_z^2}
\end{equation}
The right panel of figure \ref{fig:pot_forces} shows the percentage error in magnitude of the gravitational force in a similar manner to the left panel. There is a slightly larger distribution of errors due to the fact that the force is the derivative of potential. However, the errors are still reasonably low throughout the model. 

In both the potential and force panels, the largest errors occur at moderate radial distance in the plane of the disc. These errors are attributed to spiral arms and the low-density hole surrounding the bar, which are features not well reproduced in the deprojected model.

The results of this section show that if the orientation of the BP/X bulge is known the 3D density can be recovered reasonably well, and the corresponding gravitational potential and forces are in good agreement with those of the original snapshot. In section \ref{sec:results_forstand}, we further test the accuracy of dynamical properties of the deprojected model by integrating orbits and constructing orbit-superposition models that fit the projected kinematics, in addition to the surface density, and provide constraints on the bar angle and pattern speed.

\subsection{Degeneracy in bar angle from IMFIT image deprojection}
\label{subsec:degen}
In section \ref{sec:results_trueorient}, we  fixed the inclination of the disc $i$ and the bar angle $\psi$ at their true values ($90^\circ$ and $45^\circ$ respectively) and showed how the 2D projection, 3D model projections, 3D densities, potential and forces were recovered. The value of $\theta$ (position angle) can be estimated from photometric data reasonably accurately, and the inclination of the disc can be estimated using the shape of the disc. However  especially for an edge-on disc the angle $\psi$ of the bar to the line-of-sight is difficult to determine. We now assess how well $\psi$ can be recovered by IMFIT from photometric deprojection alone. 

For this experiment we provided the same input image to IMFIT as in section \ref{sec:results_2dfit} ($\psi_{\rm true}=45^\circ$) but  we varied the value of $\psi$ used by IMFIT to fit the image between $0^\circ$ (side-on) to $90^\circ$ (end-on) and construct a best-fit model for each value of $\psi$. 

\begin{figure*}
    \centering
    \includegraphics[width=\textwidth]{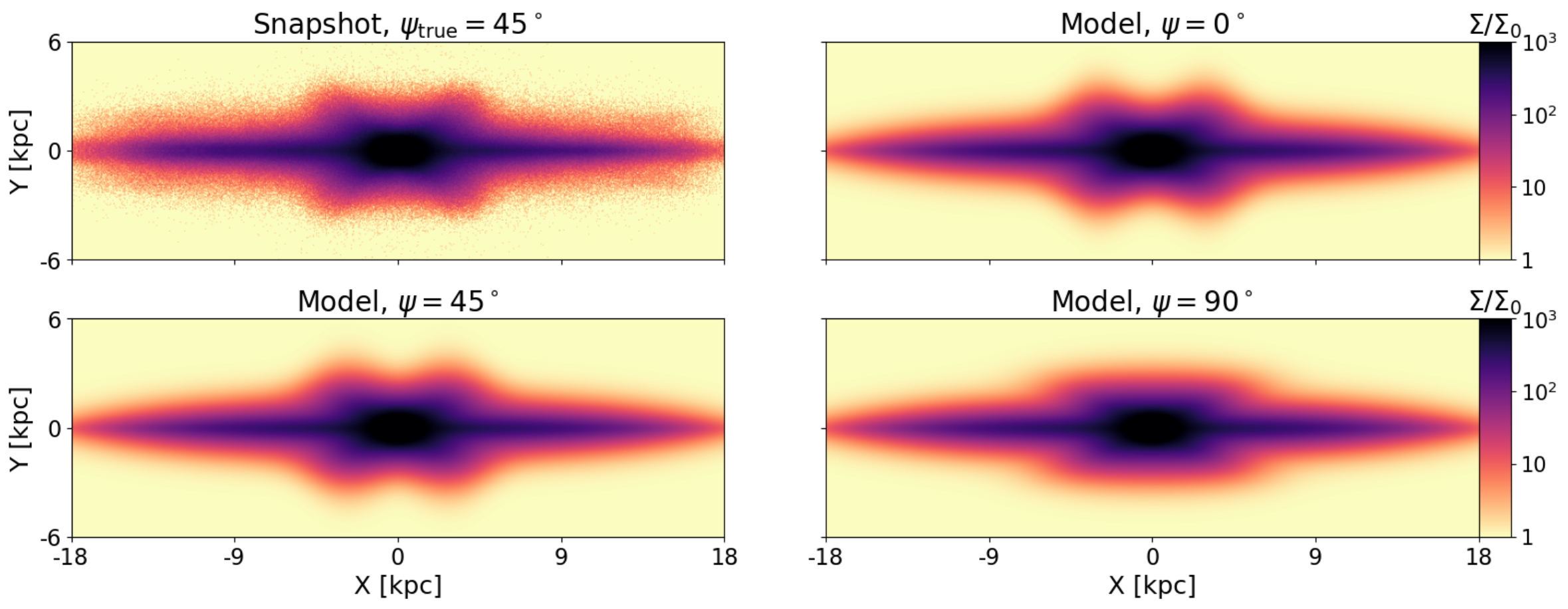}
    \caption{Top left: input image, which is the N body snapshot projected at an angle $\psi_{\rm true} = 45^\circ$. All other panels: projections of the best-fit model assuming a particular value of $\psi$. The models $\psi=0^\circ$ and $\psi=45^\circ$ are both able to model the BP/X shape well, and only $\psi=90^\circ$ shows significant deviation. }
    \label{fig:psi_snapshots}
\end{figure*}

Figure \ref{fig:psi_snapshots} shows the best-fit images for some values of $\psi$ that we explore. We can see that both $\psi=0^\circ$ and $\psi=45^\circ$ are qualitatively very similar to the projected snapshot. At $\psi=90^\circ$, the peanut shape is no longer present because it is viewed end-on. However, the general shape of the disc, bar, and bulge are reasonably well reproduced in all models. This degeneracy arises because a bar of length $l$ as viewed side-on will have the same projected length as a bar of length $l/\cos(\psi)$ as viewed from an angle $\psi$.

\begin{figure}
    \centering
    \includegraphics[width=\columnwidth]{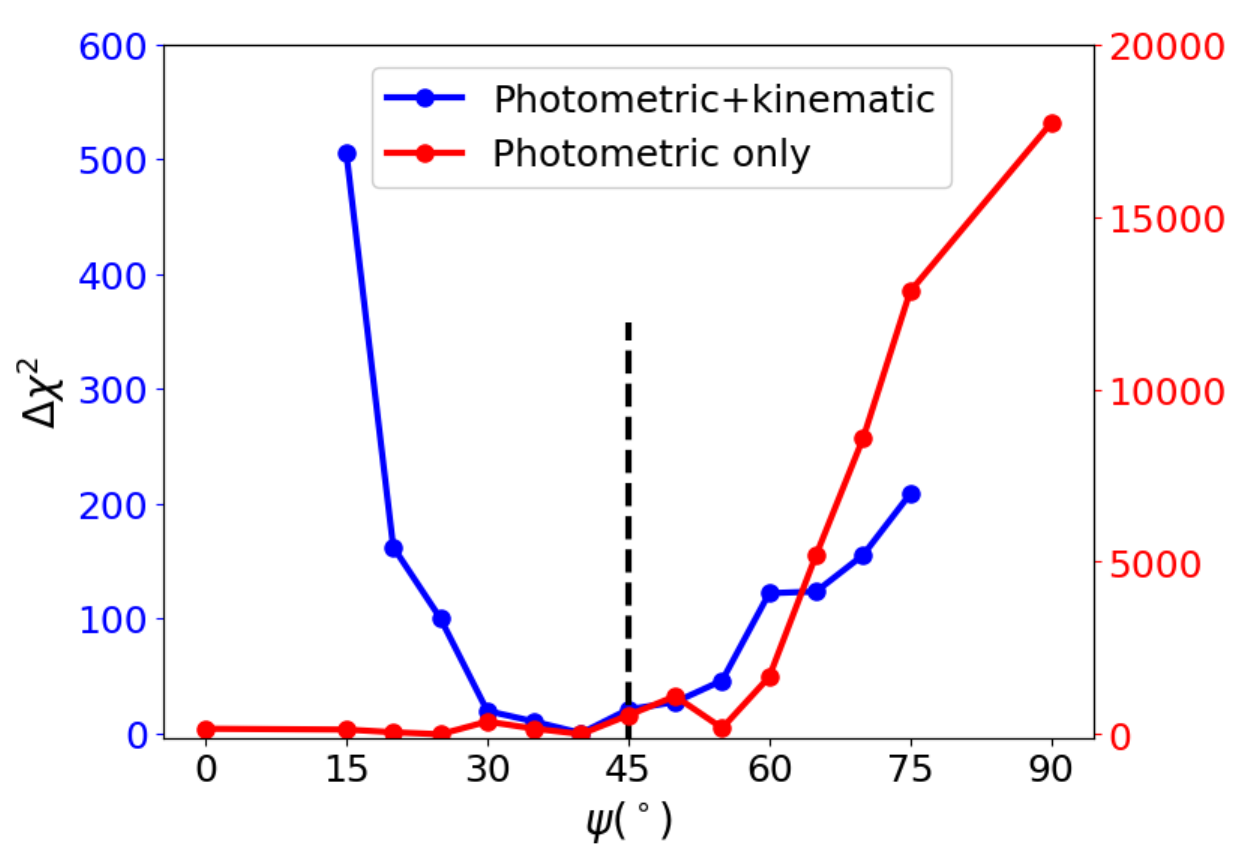}
    \caption{$\Delta \chi^2$ values vs. bar angle $\psi$. The red curve shows $\Delta \chi^2$ values for the deprojected (IMFIT) models with the bar angle $\psi$ as a free parameter. The true value of $\psi$ is $45^\circ$, shown as the black dashed line. The models with $\psi \lesssim 60^\circ$ have statistically similar $\Delta \chi^2$ values, making them indistinguishable with photometric modelling alone. The blue curve shows the values as obtained from dynamical modelling with FORSTAND (discussed in section \ref{sec:results_forstand}). With the addition of kinematic data, we can see that the degeneracy is resolved and $\psi$ is constrained.}
    \label{fig:psi_chisqr}
\end{figure}

Figure \ref{fig:psi_chisqr} shows the resulting  $\Delta \chi^2$ values ($=\chi^2 - \chi^2_{\rm min}$, where we use the \textit{total} $\chi^2$, not the reduced value) for each of the fits, plotted against the (fixed) input value of $\psi$. The plot shows the $\Delta \chi^2$ from photometric modelling alone in red and from dynamical modelling (photometric+kinematic data) in blue (which we discuss in section \ref{sec:results_forstand}). The curve of $\Delta \chi^2$ for photometric data shows a near plateau for all models with $\psi \lesssim 60^\circ$, and we can only rule out models with $\psi \gtrsim 60^\circ$. It is therefore clear that photometric data alone is insufficient to constrain the value of $\psi$. In section~\ref{sec:results_forstand}, we show that this degeneracy can be resolved via dynamical modelling with FORSTAND using photometric+kinematic data.

\section{Results: Schwarzschild modelling with Forstand}
\label{sec:results_forstand}
\subsection{Overview of Code}
\label{sec:forstand_overview}
We used the Schwarzschild orbit-superposition code FORSTAND \citep{VV20}, which is built on top of the AGAMA stellar-dynamical toolbox. FORSTAND generates self-consistent models that are constrained by the surface brightness distribution and 2D kinematic maps provided by the user. We used the 2D Voronoi binned kinematics maps and 2D surface brightness distributions described in section~\ref{sec:nbody}.  While the current version of FORSTAND is capable of building triaxial galaxy models that can simulate bars, it has so far only been tested on mock data where the true 3D density of the bar was provided as input. Here we test how well our deprojected 3D model recovered by IMFIT is able to reproduce the ``observed’’ mock kinematics and photometry of the snapshot, as well as how accurately the bar pattern speed $\Omega$, bar angle $\psi$, and central SMBH mass $M_{\rm BH}$ can be determined.

\subsubsection{Construction of gravitational potential}
For each choice of bar angle $\psi$, we obtain a 3D stellar luminosity distribution using IMFIT. This luminosity distribution is converted to a mass distribution by multiplying with a fiducial mass/light ratio $\Upsilon_*$ which we arbitrarily set to unity. For observed galaxies this fiducial value can be set using the colour of the stellar population \citep[e.g.][]{bell_dejong_2001}. The gravitational potential of the galaxy is composed of three parts:
\begin{enumerate}
    \item The stellar gravitational potential, constructed from parametric multi-component 3D density profile using the \texttt{CylSpline} Poisson solver in AGAMA. In order to quantify any discrepancies due to the inherent limitations of deprojection, we  run the FORSTAND using the 3D density distributions obtained from each of the following:
    \begin{enumerate}
        \item IMFIT (i.e. the deprojected model)
        \item from fitting the multi-component parametric model in Section~\ref{sec:components} directly  to the N-body snapshot (i.e. the fit-3D-snap model)
        \item the true 3D density of the snapshot without fitting to our analytic profile (i.e. the true density model).
    \end{enumerate}
    \item An NFW dark matter halo with scale radius $r_s=18 \ \rm{kpc}$ and asymptotic circular velocity $v_c=180 \ \rm{km/s}$ (the true parameters of the dark matter halo in the N-body simulation)
    \item A central SMBH, represented as a Plummer potential with scale radius $\sim 10^{-4}$ kpc. The true mass of the SMBH in the snapshot is $7.5 \times 10^7 M_\odot$.
\end{enumerate} 

Since our goal is to test whether the deprojection method outlined in section \ref{sec:imfit} can generate realistic orbits, we fix the parameters of the dark matter halo and do not vary them across runs.

\subsubsection{Construction of orbit library}
A large number of orbital initial conditions ($\sim 20\,000$) are randomly drawn from the stellar density profile, with their velocities assigned from an axisymmetric Jeans model. 
The orbits are integrated in the given potential for 100 orbital times, and the spatial density of each orbit is recorded on a grid in $R,z$ and expanded into Fourier harmonics in $\phi$. We note that the orbits are integrated in the frame rotating with angular velocity $\Omega$ (where $\Omega$ is an input parameter) and in the total potential of the galaxy which includes a contribution from the SMBH. Therefore a different orbit library must be constructed for each value of $\Omega$ and $M_{\rm BH}$. 

\subsubsection{Construction of mock IFU kinematical data}
\begin{figure*}
    \centering
    \includegraphics[width=\textwidth]{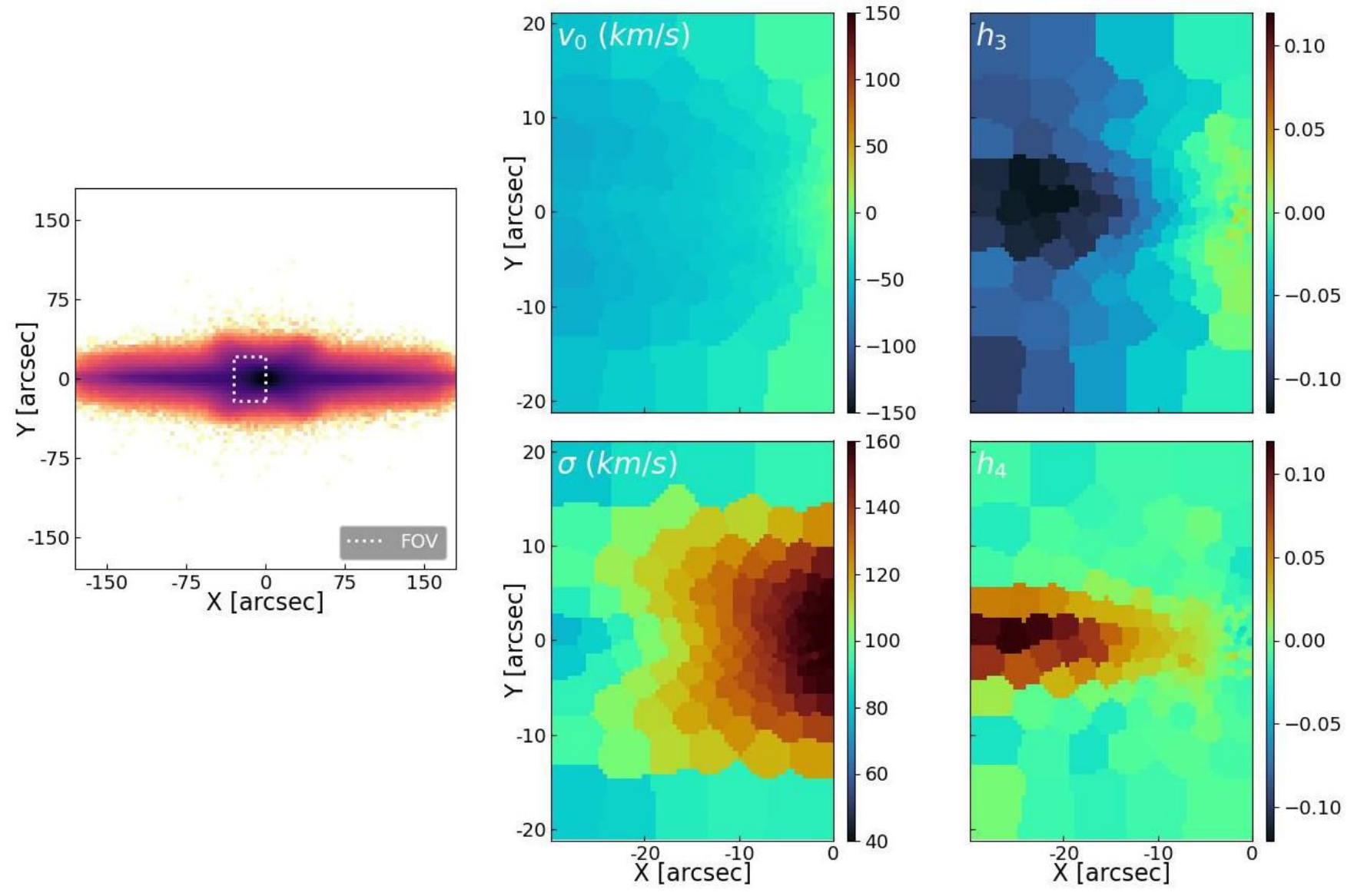}
    \caption{Left: field of view of the snapshot that is used for generating mock photometric and kinematic data, to be used as the input to FORSTAND. Other panels: maps of GH coefficients $V$ (top center), $\sigma$ (bottom center), $h_3$ (top right), and $h_4$ (bottom right). While only the first four coefficients are shown, the code also uses $h_5$ and $h_6$ in the fitting.} 
    \label{fig:kinematic_maps}
\end{figure*}

We construct mock IFU data using the same projection ($i=90^\circ$ and $\psi=45^\circ$) of the N-body snapshot. The snapshot is placed at a distance of 20 Mpc, where $1''$ $\approx$ 100 pc. Since the snapshot contains a central supermassive black hole  (represented as a softened point mass with softening parameter $\sim$ 33 pc), we construct two kinematic datasets: a low resolution (LR) dataset covering the entire bar region of the galaxy, and a high resolution (HR) dataset focused on the central region where the SMBH dominates the potential. 
The LR dataset has a field of view of $1'$ and a resolution (pixel size) of $0.46''$ (corresponding to a field of view and resolution of 5.8 kpc and 45 pc respectively). The HR dataset has a field of view of $7.5''$ and a resolution of $0.042''$ (corresponding to a field of view and resolution of 0.72 kpc and 4 pc respectively). We use a Gaussian point spread function (PSF) with width equal to the pixel size for each dataset. Figure \ref{fig:kinematic_maps} shows the field-of-view and maps of the first four GH coefficients ($V$, $\sigma$, $h_3$, and $h_4$) for the input snapshot. 

Both the kinematic datasets are constructed with the disk edge-on and with bar position angle $\psi=45^\circ$. Since the edge-on galaxy is nearly symmetric about the $Y=0$ plane, we only use half of the sky plane ($X<0$) in the modelling (where $X$ and $Y$ refer to the sky coordinates). We have verified that using the full sky plane results in similar results. Since the galaxy is not symmetric about the $Y=0$ plane due to the bending of the disc, we use the entire $X<0$ half-plane and not just a single quadrant. The galactic center is located at right-edge of the kinematic map.

We use the Voronoi binning method \citep{Cappellari2003} to group the pixels into apertures with a target signal-to-noise ratio. The pixels are binned into 150 apertures for the LR dataset and 50 apertures for the HR dataset (corresponding to a S/N value of $\sim$ 120 for LR and $\sim$ 75 for HR). The LOSVDs in each aperture are computed directly from the snapshot, and are expanded into a Gauss-Hermite series using 6 GH moments (the code can be configured to use higher-order moments, as advocated by \citealt{Quenneville_2021}). Errors are assigned to each GH coefficient by bootstrapping over random subsets of particles in the snapshot, and are therefore determined by the Poisson noise ($\mathrm{err} \propto \sqrt{I}$, where $I$ is the amount of light in the bin, which is roughly constant). Because we are using a high-resolution N-body snapshot, these errors are lower than the amount of noise that we can expect from real IFU data. The error-free (Poisson noise only) models are presented here; we discuss the effect of adding a realistic amount of noise in Appendix~\ref{sec:appendix}.

Changing the overall mass normalization of the entire model (i.e., adjusting $\Upsilon_*$, $M_{\rm BH}$ and the dark matter halo mass) is equivalent to rescaling the velocity axis of the model LOSVD by $\sqrt{\Upsilon}$, thus can be performed without reintegrating the orbits. Therefore, each orbit library is reused multiple times, scanning the range of $\Upsilon_*$ values in multiplicative steps of $2^{0.05}$, until the minimum of $\chi^2$ is found and bracketed from both ends. In total, we ran over a thousand realizations of orbit libraries, each one typically reused for $\sim 10$ values of $\Upsilon_*$.

\subsubsection{Fitting the mock IFU kinematics}
For each orbit, the line-of-sight kinematic contributions within the field of view of the mock IFU is recorded on an intermediate 3D datacube, which is then convolved with the instrumental PSF, rebinned onto the same Voronoi bins as used for the mock kinematics, and converted to the GH moments. These conversions are performed using B-splines for the representation of this intermediate 3D datacube \citep{VV20}. The code then determines the orbit weights that minimize the deviation between the 3D density model, its 2D projection and the observed kinematic datacube.

For each Voronoi bin, the mock LOSVD is expanded in terms of Gauss-Hermite coefficients. The first six coefficients ($V$, $\sigma$, $h_3$, $h_4$, $h_5$, and $h_6$) are taken as constraints in the optimization problem ($V$ and $\sigma$ are converted to $h_1$ and $h_2$, as explained in Section~2.6 of the above paper). For a set of $N_{\rm o}$ orbits and $N_{\rm c}$ constraints, we minimize the objective function
\begin{equation}
\label{eq:obj_func}
    \mathcal{Q} = \sum_{n=1}^{N_{\rm c}} \left( \frac{ \sum_{i=1}^{N_{\rm o}} w_i u_{in} - U_n}{\epsilon_{U_n}} \right)^2 + \lambda N_{\rm o}^{-1} \sum_{i=1}^{N_{\rm o}} \left( \frac{w_i}{\tilde w_i} \right)^2
\end{equation}
where $w_i$ are the orbital weights, $u_{in}$ is the contribution of the $i^{\rm th}$ orbit to the $n^{\rm th}$ constraint, $U_n$ are the constraint values, and $\epsilon_{U_n}$ are the errors/uncertainties on the constraints. The second term in equation \ref{eq:obj_func} represents the regularization term, where we penalize orbital weights that deviate far from their priors $\tilde w_i$ (which we take as equal weights for every orbit). We use a value of $\lambda=10$ in our modelling. 

The orbital weights must satisfy the density (mass) constraint in each Voronoi bin. Since we are attempting to fit a smooth analytical model to a discrete N-body snapshot which is noisy in low-density regions, we assign a formal error of 10\% in satisfying the mass constraints. This is larger than the tolerance parameters used in other works (e.g. 1\% in \citealt{Tahmasebzadeh2022}). While we find that we can still recover the quantities of interest for smaller values of the tolerance parameter, this results in a more noisy orbital weight distribution.

The goodness of fit between the data and the model is measured by the $\chi^2$ value, which is composed of contributions from the density constraints, two kinematic constraints, and regularization:
\begin{equation}
\label{eq:chisqr}
    \chi^2 = \chi_{\rm dens}^2+\chi_{\rm kin,lr}^2+\chi_{\rm kin,hr}^2+\chi_{\rm reg}^2
\end{equation}
In all of our models, the total $\chi^2$ is dominated by $\chi_{\rm kin,lr}^2$. We find that if the value of $\Delta \chi^2 = \chi^2 - \chi^2_{\rm min}$ is taken as the formal statistical uncertainties in the best-fit parameters (e.g. $\Delta \chi^2 = 2.3,6.2,11.8,$ etc. corresponding to $1\sigma$, $2\sigma$, $3\sigma$ respectively for 2 degrees of freedom), the resulting posteriors are unreasonably tight. This has been attributed to the large number of ``hidden'' degrees of freedom in the model, since we select the best fit orbital weights instead of marginalizing over them (see \citealt{Magorrian2006}). Several different alternatives have been proposed in order to relate $\Delta \chi^2$ with the uncertainty levels \citep[e.g.][]{vdb2008,Zhu2018,Lipka2021}. In our analysis, we use the $\Delta \chi^2$ values in order to quote the posteriors around the best-fit parameters; but we note that our confidence intervals require a more rigorous statistical analysis.  

\subsection{Recovery of bar pattern speed and mass-to-light ratio}
We first attempt to recover the ``large scale'' parameters of the galaxy, namely the stellar mass-to-light ratio $\Upsilon_{*}$ and the bar pattern speed $\Omega$. For these runs, since $\chi^2$ is dominated by the contribution from the LR kinematic dataset, the mass of the BH at $\Upsilon_{*}=1$ is kept constant at the true value. As explained earlier, changing $\Upsilon_{*}$ implies a proportional change in all other mass components, including $M_{\rm BH}$, but as the range of variation of $\Upsilon_{*}$ is small (typically $\pm 10\%$), we ignore the BH mass variation in this section. In any case, reasonable values of $M_{\rm BH}$ do not effect the measurement of the large scale parameters. The true value of $\Omega$ is calculated by measuring the rotation speed of the moment of inertia tensor of the system. We note that the in the N-body simulation, the "true" value of $\Omega$ itself is oscillating with time, with amplitude $\sim 0.5$ km/s/kpc at the time of the final snapshot. Hereafter, $\Omega_{\rm true}$ refers to the instantaneous value in the snapshot, which is $\sim 15$ km/s/kpc, and the oscillation amplitude is quoted as error bars.

\begin{figure*}
    \centering
    \begin{tabular}{c|c}
    \includegraphics[width=0.49\textwidth]{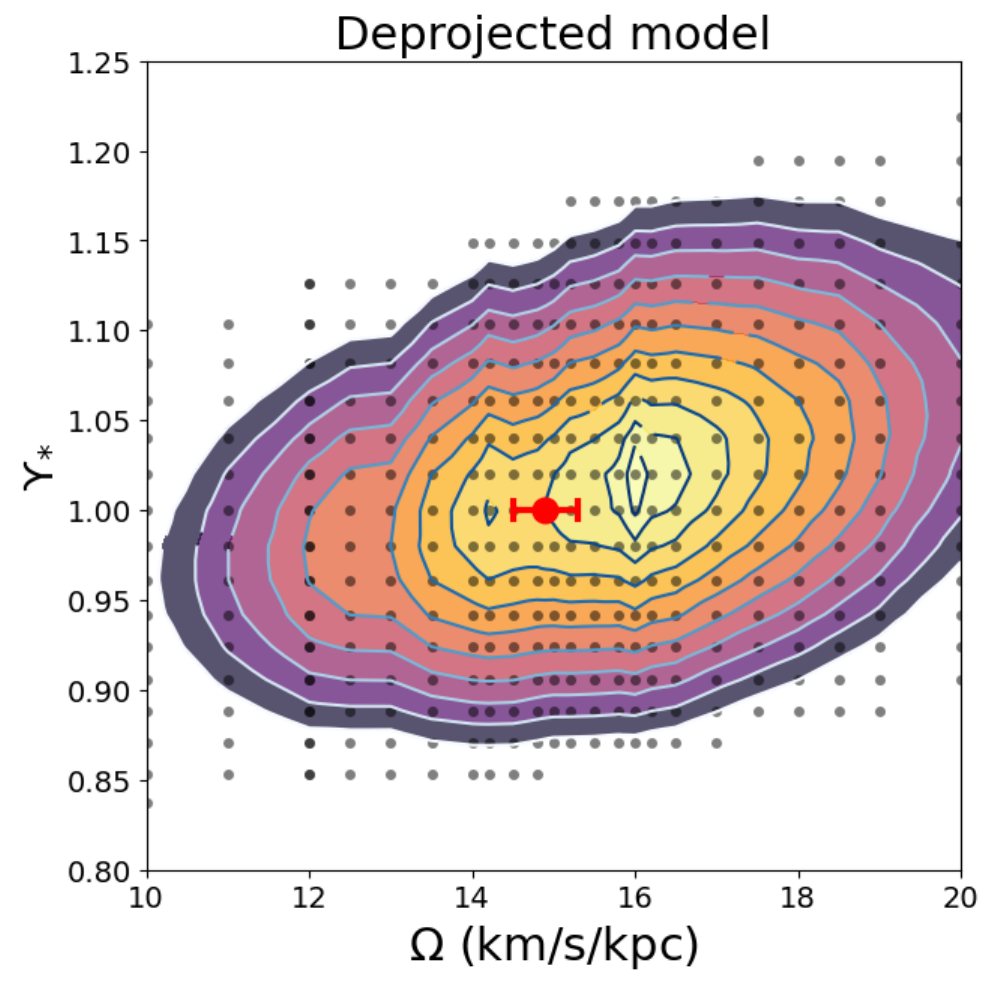}
    \includegraphics[width=0.49\textwidth]{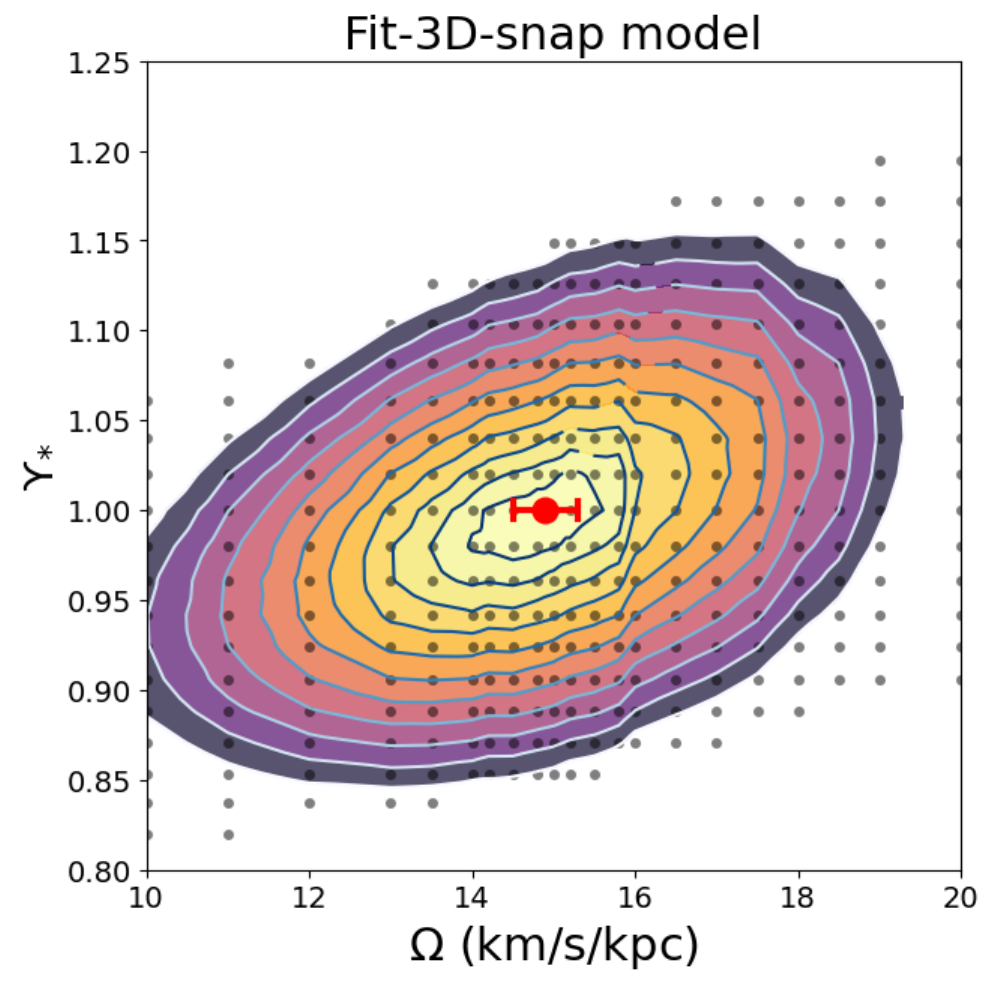}\\
    \includegraphics[width=0.49\textwidth]{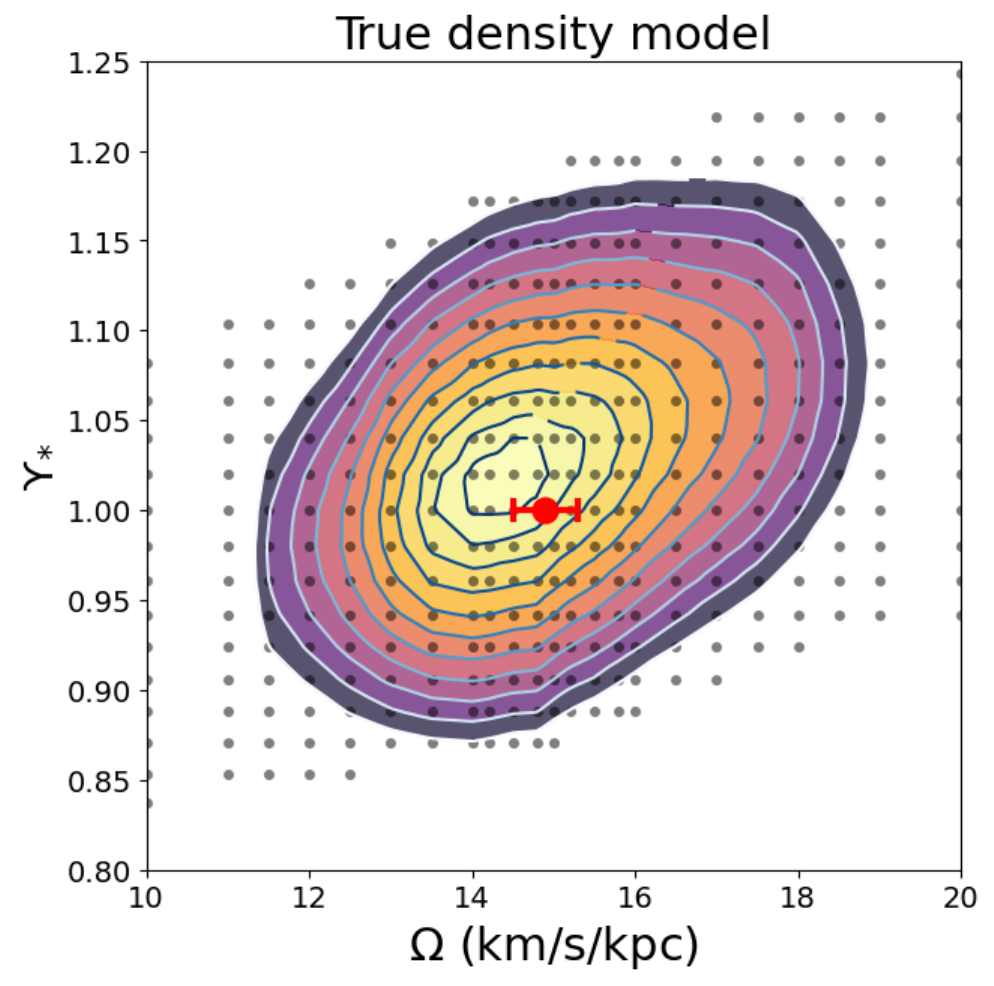}
    \includegraphics[width=0.49\textwidth]{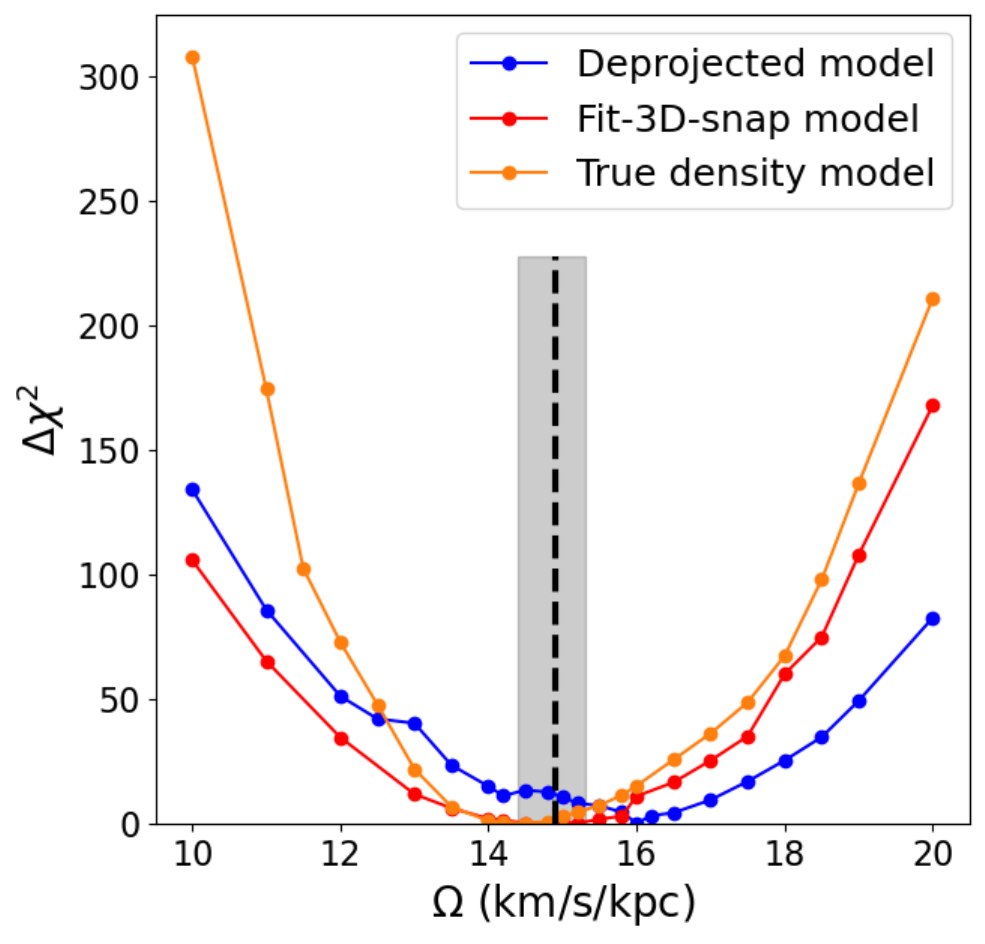}
    \end{tabular}
    \caption{Top left: contours of $\Delta \chi^2$ in the $\Omega-\Upsilon_*$ plane for the deprojected model. The contour levels show $\Delta \chi^2 = 2.3,6.2,11.8$, etc. Gray points represent parameter values tested, and the red dot denotes the true values of the parameters ($\Omega_{\rm true} = 15$ km/s/kpc and $\Upsilon_*=1$), with the error bars denoting the amplitude of oscillation of $\Omega_{\rm true}$. Top right: similar contours for the fit-3D-snap model. Bottom left: similar contours for the true density model. Bottom right: one-dimensional cuts through $\Upsilon_*$. The true value of $\Omega$ is indicated by the black dashed line, with the shaded region denoting the oscillation in $\Omega$. We can see that the value of $\Omega$ is reasonably tightly constrained between $\sim 10 \%$ of its true value for the deprojected model and is almost perfectly recovered by the fit-3D-snap model and true density model. As the density estimation of the snapshot becomes more accurate, the constraints in $\Omega$ become tighter.} 
    \label{fig:omega_contours.pdf}
\end{figure*}

Figure \ref{fig:omega_contours.pdf} shows the results. The top left panel shows contours of $\Delta \chi^2$ in the $\Upsilon_{*}-\Omega$ plane for the deprojected model. The true projection angles of the snapshot ($i=90^\circ$ and $\psi=45^\circ$) are assumed. Gray dots represent values of parameters tested, and the red dot marks the true values of the parameters. The middle panel shows the same for the fit-3D-snap model. We can see that in both models, $\Upsilon_{*}$ and $\Omega$ are well recovered. The parameters of the best-fit model are in excellent agreement with the true values, with deviations $\lesssim 10\%$ for $\Omega$ and just a couple percent for $\Upsilon_*$. When the fit-3D-snap model is used in FORSTAND (instead of the deprojected model), the parameter recovery is nearly perfect and constraints are slightly tighter, as seen in the top right panel of figure \ref{fig:omega_contours.pdf}. This is because the fit-3D-snap model provides a better density estimate to the input snapshot. The constraints are even tighter when the true density model is used (bottom left panel). The small discrepancy between the best-fit and true values may be due to the oscillating nature of $\Omega_{\rm true}$ over time. Nevertheless, even with the deprojected model, the best-fit estimates of $\Omega$ and $\Upsilon_{*}$ are within $\sim 10\%$ of the true values. 

The bottom right panel of figure \ref{fig:omega_contours.pdf} shows the values of $\Delta \chi^2$ marginalized over $\Upsilon_{*}$ for the three models. It is clear from this figure that fitting the large-scale LOSVD and photometry via deprojection tightly constrains the bar pattern speed. We can also see that as the estimation of the density/potential of the snapshot becomes more accurate (deprojected $<$ fit-3D-snap $<$ true density), the constraints on $\Omega$ from the $\Delta \chi^2$ values become tighter. However as discussed earlier, the relationship between $\Delta \chi^2$ and the confidence intervals in Schwarzschild modelling requires a more detailed study.

\begin{figure*}
    \centering
    \includegraphics[width=\textwidth]{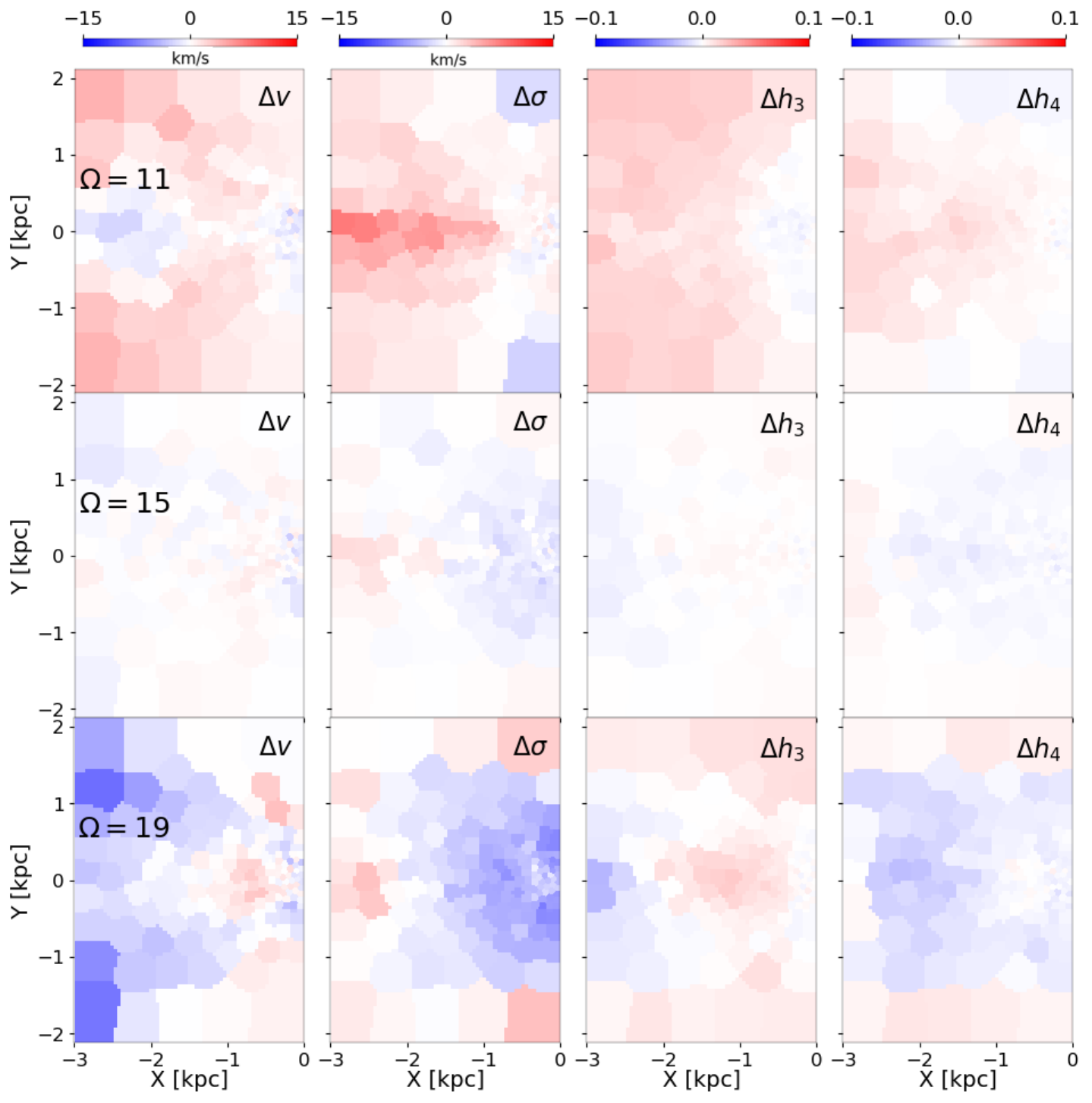}
    \caption{Errors in the GH coefficients (model-data) for three models with different $\Omega$ using the true density of the snapshot. The values of $\Upsilon_*$ and $M_{\rm BH}$ are fixed at their true values (1 and $7.5 \times 10^7 M_\odot$ respectively). The top panel where $\Omega$ is underestimated shows that the projected LOSVD is more radial (positive $\Delta h_4$) and has higher dispersion (positive $\Delta \sigma$). The opposite is true for the bottom panel where $\Omega$ is overestimated. The central panel where $\Omega$ is at its true value shows minimum error in the GH coefficients. } 
    \label{fig:ghm_errors}
\end{figure*}

It is instructive to analyze the features in the kinematic maps of the models which allow us to constrain $\Omega$. Figure \ref{fig:ghm_errors} shows the errors (model - data) in the first four GH coefficients for different models using the true density distribution. Three values of $\Omega=11,15,19$ km/s/kpc are shown, and the values of $\Upsilon_*$ and $M_{\rm BH}$ are fixed at their true values. The center panel with $\Omega=\Omega_{\rm true} = 15$ km/s/kpc fits the data very well and therefore shows the least error. When $\Omega$ is decreased to 11 km/s/kpc (top panel), the underestimation of $\Omega$ leads to less tangential orbits as seen from an inertial frame of reference compared to the snapshot. When projected along the edge-on line of sight, this shows up in the LOSVDs as higher $h_4$ coefficients, i.e. positive values in the $\Delta h_4$ map. These orbits also contribute to a large $\sigma$ when projected, and therefore lead to positive $\Delta \sigma$ values. The opposite is true when $\Omega$ is overestimated, as seen in the bottom panel with $\Omega=19$ km/s/kpc. The large pattern speed leads to more tangential orbits in the model as viewed from an inertial frame compared to the snapshot. The edge-on projected LOSVD therefore is more tangential (negative $\Delta h_4$) and has lower dispersion (negative $\Delta \sigma$). 

\subsection{Recovery of bar angle}
We now consider variation in the bar angle $\psi$ in our deprojected model. We saw in section \ref{subsec:degen} that photometric deprojection produces near identical projected fits to the input image for $\psi \lesssim 60^\circ$. Here, we construct a deprojected model for values of $\psi$ between $0^\circ$ and $75^\circ$. We then run FORSTAND for each of these deprojected models, using each deprojected density for the 3D mass distribution and to derive the stellar component of gravitational potential. The kinematic constraints are held fixed across runs. Therefore we investigate how well the code fits the kinematics of the original snapshot using the various deprojected density distributions, in order to determine the bar's orientation angle $\psi$.

\begin{figure}
    \centering\includegraphics[width=\columnwidth]{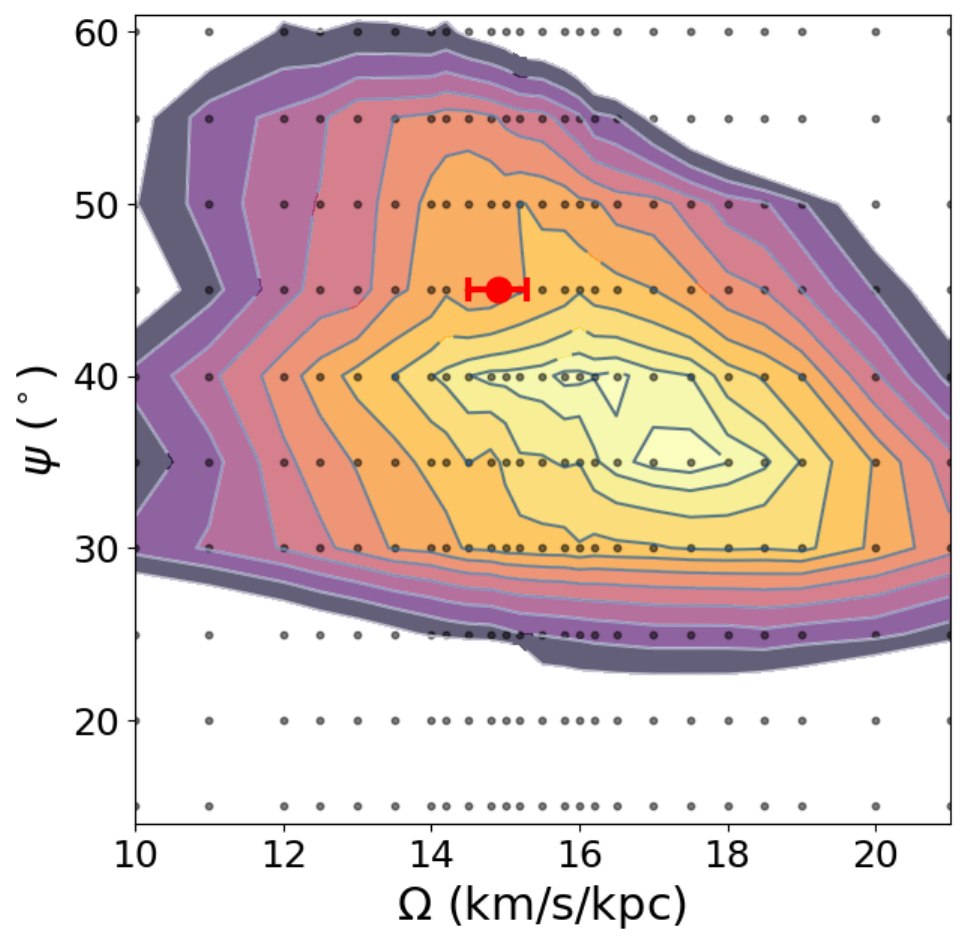}
    \caption{$\Delta \chi^2$ contours in the $\psi-\Omega$ plane when the deprojected models for different $\psi$ are used in Schwarzschild modelling.  The contour levels show $\Delta \chi^2 = 2.3, 6.2, 11.8$, etc. Gray points indicate parameter values tested, and the red dot indicates the true values $\psi_{\rm true}=45^\circ$ and $\Omega_{\rm true}=$ 15 km/s/kpc with the error bar denoting the amplitude of oscillation of $\Omega$.
    }\label{fig:psi_omega_contours}
\end{figure}

Figure \ref{fig:psi_omega_contours} shows the $\Delta \chi^2$ contours in the $\psi-\Omega$ plane. We see that both the bar angle and the pattern speed are reasonably well constrained by the models. The best-fit values of the parameters are within $\sim$ 10\% of the true values. When marginalized over all other parameters, the curve of $\Delta \chi^2$ vs. $\psi$, shown as the blue curve in figure \ref{fig:psi_chisqr}, shows a clear minimum around the true value of $\psi$. It is clear from figures \ref{fig:psi_chisqr} and \ref{fig:psi_omega_contours} that with the addition of kinematic data, the degeneracy in $\psi$ seen with photometric modelling alone (red curve in Figure~\ref{fig:psi_chisqr}) is broken, and the true value is recovered to $\sim 10\%$ accuracy. In principle, this method can also be used to determine the inclination angle of the disc $i$. We defer this exploration for a future study.

\subsection{Recovery of black hole mass}
Since the black hole dominates the potential of the galaxy only within the inner $\sim 100$ pc, it is reasonable to first estimate the values of $\Omega$ and $\Upsilon_{*}$ using a fiducial value of $M_{\rm BH}$. Once these parameters have been recovered to reasonable accuracy, we now focus our attention on the recovery of $M_{\rm BH}$. However, the large-scale parameters themselves are not perfectly recovered and have uncertainty themselves. Therefore, we run a grid of models across $\Omega$, $\Upsilon_{*}$, and $M_{\rm BH}$, and obtain the best-fit values of $\Omega$ and $\Upsilon_*$. We place a flat prior on $\Omega$ and $\Upsilon_*$ between $\pm 1$ km/s/kpc and $\pm 0.05$ around the best fit values in $\Omega$ and $\Upsilon_*$ respectively. Using this prior, we then marginalize over $\Omega$ and $\Upsilon_{*}$. 

\begin{figure*}
    \centering
    \includegraphics[width=\textwidth]{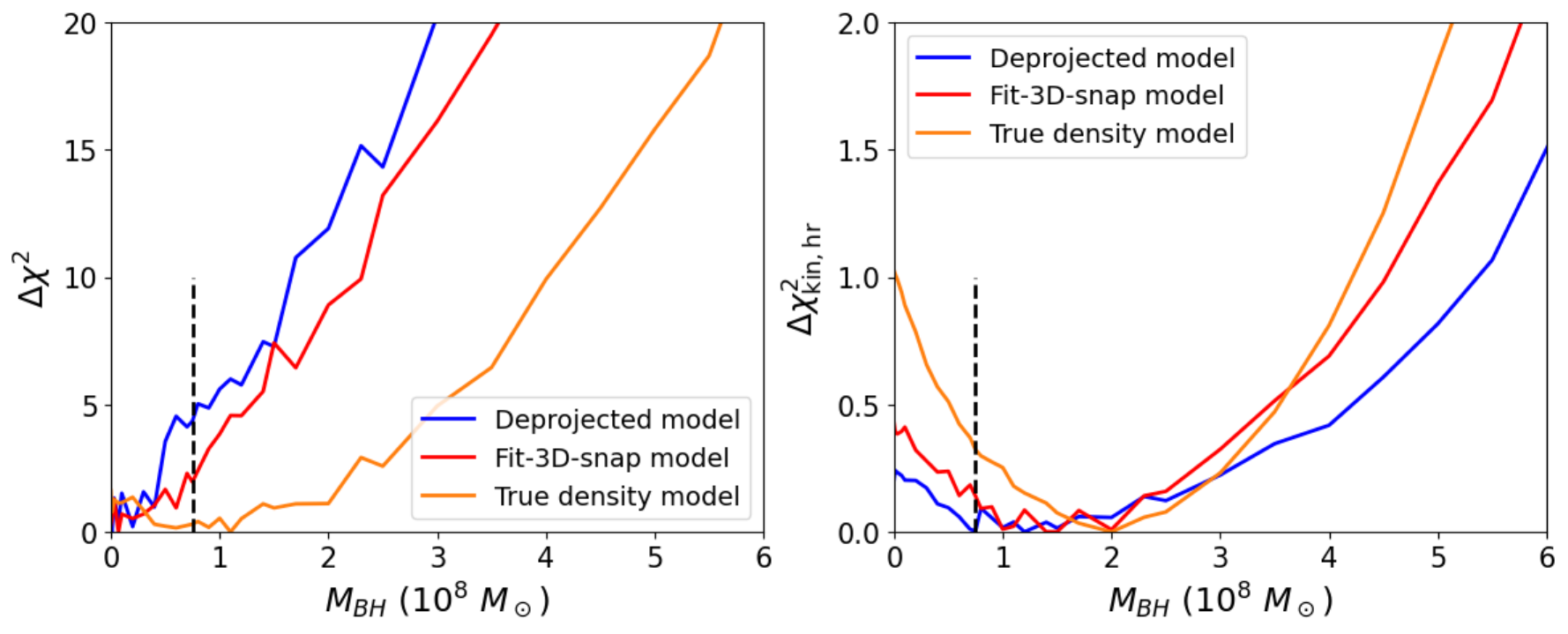}
    \caption{Left: $\Delta \chi^2$ vs. $M_{\rm BH}$ for the deprojected, fit-3D-snap, and true density models (this is the total $\chi^2$ from the density, kinematic, and regularization constraints). The true value of $M_{\rm BH}$ is marked with the black dashed line. The deprojected and fit-3D-snap models are unable to recover the correct value of $M_{\rm BH}$, and the true density model shows only a very shallow minimum around the true value. Right: similar curves but for $\Delta \chi^2_{\rm kin,hr}$, i.e. the contribution from only the high-resolution kinematics. All three models show a minimum in $\Delta \chi^2_{\rm kin,hr}$ around the true value of $M_{\rm BH}$. }
    \label{fig:mbh}
\end{figure*}

The left panel of figure \ref{fig:mbh} shows the values of $\Delta \chi^2$ vs. $M_{\rm BH}$ for the deprojected, fit-3D-snap, and true density models. The black dashed line shows the true value of $M_{\rm BH} = 7.5 \times 10^7 M_\odot$. It is clear from the plot that neither the deprojected model nor the fit-3D-snap model can recover $M_{\rm BH}$. The fit-3D-snap model shows a broad plateau in $\Delta \chi^2$ for $M_{\rm BH} \lesssim 10^8 M_\odot$. The deprojected model also shows a plateau, but it only extends until $\sim 4 \times 10^7 M_\odot$. Both of these models favor smaller SMBH masses, although we cannot draw any conclusions about the exact value. 

The discrepancy originates from the fact that the enclosed mass within the inner $\sim$ 1 kpc is
overestimated in both the deprojected and the fit-3D-snap model (see figure \ref{fig:3D_fit}). While the large-scale parameters are mostly insensitive to this, it becomes important when trying to measure $M_{\rm BH}$. The code compensates for this extra mass by preferring a lower $M_{\rm BH}$. This is evident from the fact that the true density model does indeed show a minimum around the true value of $M_{\rm BH}$ (orange line in figure \ref{fig:mbh}). However, this minimum is both too noisy and too shallow confidently rule out lower values of $M_{\rm BH}$. Therefore, we can only obtain upper limits for $M_{\rm BH}$ from $\Delta \chi^2$ values. 

Instead of using the total $\chi^2$ values (defined in equation \ref{eq:chisqr}), we now instead look at $\chi^2_{\rm kin,hr}$, i.e. the contribution from only the high-resolution central kinematics. Since this kinematic dataset focuses on the radius of influence of the SMBH, we expect that it will be able to better trace the effect of the SMBH. The right panel of figure \ref{fig:mbh} shows $\Delta \chi^2_{\rm kin,hr}$ vs. $M_{\rm BH}$ for the three models. Note the different y-axis scale. We see that all three models show a minimum around the true $M_{\rm BH}$. The depth of the minimum gets shallower as the density estimation gets less accurate (true density $>$ fit-3D-snap $>$ deprojected). While this minimum is present for all models and an upper limit can be established, with $\Delta \chi^2_{\rm kin,hr} \lesssim 1$ it is difficult to rule out lower values of $M_{\rm BH}$ from this data, especially for the deprojected and fit-3D-snap models. We further discuss the recovery of $M_{\rm BH}$ and the dependence of the measured value on the other large-scale parameters in section \ref{subsec:calc_mbh}.

\section{Discussion}
\label{sec:discussion}
\subsection{Coevolution of bars and BP/X bulges}
The strength of the BP/X bulge in a barred galaxy is correlated with various dynamical properties and hence is an indirect tracer of the evolution history of the galaxy. This is especially relevant for edge-on discs where limited additional information is available regarding their structure. Specifically, there has been overwhelming evidence that the the BP/X bulge is a part of the bar itself and not a separate component \citep{Kuijken1995,Bureau1999,Bureau1999_2,Laurikainen2016} and is correlated with large-scale gas kinematics \citep{Athanassoula2002} and the bar strength \citep{Bureau2005}. Moreover, various studies have shown that a strong BP/X shape is a characteristic feature of a buckling instability \citep{Raha_91,Debattista2004,Martinez-Valpuesta06, Lokas2019,Collier2020}. However recent work has shown that not all BP/X bulges are necessarily due to buckling \citep{Quillen14}, and the presence of a mid-plane asymmetry (i.e. bending) can be used to identify a BP/X bulge formed from a recent buckling event \citep{Cuomo_etal_2023}. Therefore although the bent disc shows up merely as an error in our photometric fits \ref{fig:projections}, this may contain valuable information regarding its dynamical history.

Notably, \citet{Wheeler2023} study the coevolution of bars and SMBHs using N-body simulations and find strong differences between an early growing and a late growing SMBH. Contrary to expectation from previous studies, early growing SMBHs can strengthen bars and suppress bar buckling, both of which lead to a stronger BP/X shape in the central region. They applied the model described in this paper to fit the 3D densities of their snapshots (i.e. fit-3D-snap models) and found a strong correlation between the bar amplitude and the peanut parameters $R_{\rm pea}$ and $h_{\rm pea} = A_{\rm pea}+z_0$. This shows that the strength of the BP/X shape can be used to indirectly determine bar strength in edge-on galaxies. While further work is required to determine if there is a correlation between the BP/X bulge and the evolution of the SMBH itself, the existence of these correlations provide strong evidence for the coupling between the large-scale and small-scale dynamics. 

\subsection{Measuring bar pattern speeds}
Since a significant fraction of disc galaxies host bars, a complete model of an edge-on galaxy requires the determination of the presence or absence of bars, and if present, requires a calculation of the pattern speed. The commonly used Tremaine--Weinberg method for measuring pattern speed is inapplicable for edge-on galaxies. Therefore, alternative methods such as Schwarzschild modelling are required. While most Schwarzschild codes use MGE for deprojecting the observed photometry, this is not suitable for edge-on bars, especially if a strong BP/X is present.

Barred galaxies are often classified by their ratio of the corotation radius to the bar length, i.e. $\mathcal{R} = R_{\rm CR} / R_{\rm bar}$. Bars with $1.0 \leq \mathcal{R} \leq 1.4$ are classified as "fast", whereas bars with $\mathcal{R} \geq 1.4$ are "slow". Theoretically, bars should slow down over time due to dynamical friction. Analytical calculations and simulations predict that most bars should be "slow" (\citealt{Hernquist1992, Debattista2000, Athanassoula2003_2, Roshan2021}, but see \citealt{Fragkoudi2021} for a different perspective). On the other hand, observations have revealed a large number of "fast" bars \citep{Rautiainen2008, Aguerri2015, Guo2019, Williams2021}. In addition, bars should not survive beyond corotation due to instability in the bar-supporting ($x_1$) orbits \citep{Contopoulos1980}, however there have been several observed cases of such "ultrafast" bars \citep{Aguerri2015, Guo2019, Cuomo2019}. As emphasized, the primary method of calculating pattern speed has been the Tremaine--Weinberg method, and therefore measurements using alternative techniques such as dynamical modelling may prove useful in addressing this discrepancy.

Recent integral field spectroscopy (IFS) surveys such as CALIFA \citep{Sanchez2012} and MANGA \citep{Bundy2015} have revealed samples of barred galaxies with BP/X bulges \citep{Kruk2019}. When combined with high-resolution IFS data from MUSE surveys such as TIMER \citep{Gadotti2019}, PHANGS \citep{Emsellem2022} or Composite Bulges Survey \citep{Erwin2021}, these galaxies have the potential to reveal a great deal of information about the formation and evolution of bars. The upcoming GECKOS survey \citep{vandesande2023} focuses exclusively on edge-on galaxies and therefore will likely expand our sample of BP/X bulges with IFU kinematics. The deprojection method presented here may be an important modelling tool for this survey. 

\subsection{Recovery of black hole masses in barred galaxies\label{subsec:calc_mbh}}

Section \ref{sec:results_forstand} showed that the calculation of $M_{\rm BH}$ can be sensitive to the mass profile of the model. While the total $\Delta \chi^2$ values were not able to accurately recover $M_{\rm BH}$, the values of $\Delta \chi^2_{\rm kin,hr}$ showed a minimum around the true value, although this minimum may not be significant enough to rule out lower values. The depth of the $\Delta \chi^2_{\rm kin,hr}$ valley was highest for the true density model and lowest for the deprojected model. Since this arises from discrepancies in the density/enclosed mass profiles, it may be alleviated by more accurately modelling the mass profile of the galaxy. Figure \ref{fig:mbh}  showed that deviations of even a few percent in $M_{\rm enc}(<r)$ can significantly alter the $\Delta \chi^2$ vs. $M_{\rm BH}$ curves. Therefore we expect to require $\lesssim 1 \%$ error in the central mass profile of galaxies in order to achieve the same results as the true density model.

\begin{figure*}
    \centering
    \includegraphics[width=\textwidth]{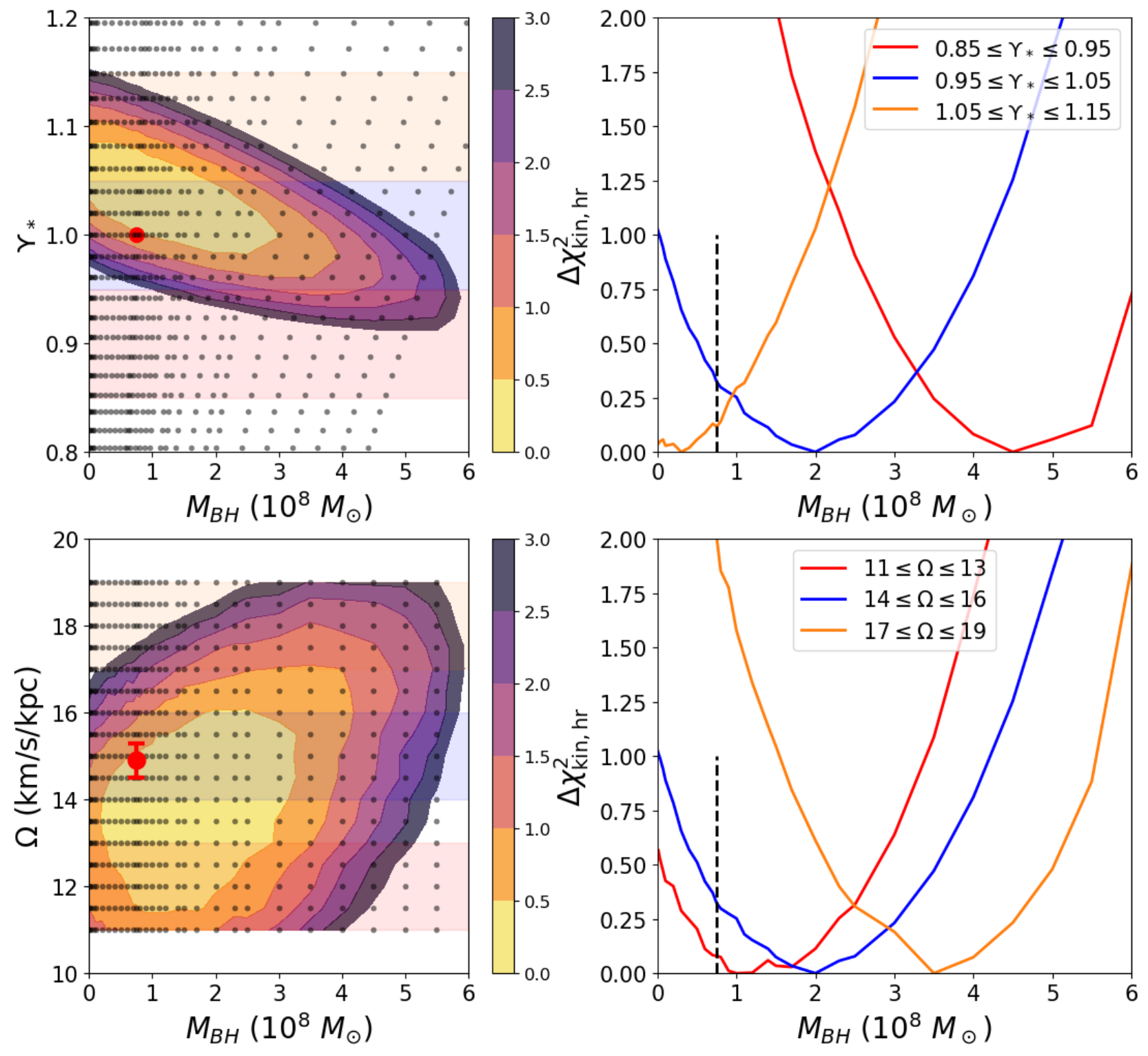}
    \caption{Left panels: $\Delta \chi^2_{\rm kin,kr}$ contours in the $\Upsilon_*$--$M_{\rm BH}$ plane (top) and the $\Omega$-$M_{\rm BH}$ plane (bottom) for the true density model ($\Omega$ is marginalized between $15 \pm 1$ km/s/kpc and $\Upsilon_*$ between $1 \pm 0.05$ for the top and bottom plots respectively). There is a negative correlation between $M_{\rm BH}$ and $\Upsilon_*$ and a positive correlation between $M_{\rm BH}$ and $\Omega$. The shaded regions indicate the regions over which we marginalize to produce one-dimensional cuts in $\Delta \chi^2_{\rm kin,hr}$ vs. $M_{\rm BH}$ for different ranges in $\Upsilon_*$ (top right) and $\Omega$ (bottom right). See discussion in text.}
    \label{fig:mbh_var}
\end{figure*}

We also note that in figure \ref{fig:mbh} we have marginalized around the best-fit large-scale parameters ($\Omega$ and $\Upsilon_*$). These parameters were recovered from the total $\chi^2$ values and not $\chi^2_{\rm kin,hr}$. Deviations of either of these parameters from the true values can bias the $M_{\rm BH}$ measurement. In order to test this, in figure \ref{fig:mbh_var} we analyze the variation of the $\Delta \chi^2_{\rm kin,hr}$ vs. $M_{\rm BH}$ curves when we marginalize over different ranges in $\Omega$ and $\Upsilon_*$. We use the true density model for this so that there are no biases/discrepancies due to deprojection or fitting to the analytic density profile.

The top left panel of figure \ref{fig:mbh_var} shows $\Delta \chi^2_{\rm kin,hr}$ contours in the $\Upsilon_*$-$M_{\rm BH}$ plane ($\Omega$ is marginalized over $15 \pm 1$ km/s/kpc). Note that the dots do not form a perfectly rectangular grid because $M_{\rm BH}$ is rescaled for every $\Upsilon_*$. The true values are indicated by the red dot. We can clearly see that around the true values, there is a negative correlation between $\Upsilon_*$ and $M_{\rm BH}$, as is evident from the tilted $\Delta \chi^2$ contours. The top right panel shows one-dimensional cuts in $\Delta \chi^2$ when we marginalize over different ranges in $\Upsilon_*$. As we marginalize over higher values in $\Upsilon_*$ and thereby increasing the stellar mass in the central region, the best-fit value of $M_{\rm BH}$ decreases in order to satisfy the constraints. This is essentially the same reason as the underestimation of $M_{\rm BH}$ by the deprojected and fit-3D-snap models in figure \ref{fig:mbh}. Since the mass of central SMBHs in local galaxies is of the order $\sim 10^{-3} \times $ the bulge mass, overestimating the enclosed mass within the bulge by even a few percent can lead to errors in the small-scale dynamics and hence a biased black hole mass.

While the above issue is seen when modelling galaxies of all morphologies, the issue of coupling between measuring the pattern speed of the bar and the black hole mass is unique to barred galaxies. A  higher value of $\Omega$ will result in more tangentially biased orbits as viewed from the inertial frame of an observer outside the bar. This will manifest in the orbits having more negative $h_4$ values, and tangential orbits contribute little to the high central velocity dispersion values (see figure \ref{fig:ghm_errors}). Therefore the mass in this region (either the mass of the black hole or the mass-to-light ratio of the stars) needs to be raised, resulting in it being biased toward higher values. The bottom left panel of figure \ref{fig:mbh_var} shows the $\Delta \chi^2$ contours in the $M_{\rm BH}$--$\Omega$ plane ($\Upsilon_*$ is marginalized over $ 1 \pm 0.05$). The contours are slightly tilted, showing a positive correlation between $M_{\rm BH}$ and $\Omega$. When we marginalize over different ranges of $\Omega$ (bottom right panel), higher $\Omega$ ranges result in higher values of $M_{\rm BH}$. This shows that even if we have the density of the galaxy exactly correct, errors may still arise from the incorrect estimation of $\Omega$. \citet{Brown2013} and \citet{Onken2014} argued that if a bar is ignored and modeled as an axisymmetric galaxy, a similar overestimate of the black hole mass results for the same reason. We see here that even when the bar is correctly modeled, if the pattern speed is over/under estimated, the black hole mass will also be correspondingly over/under estimated. 

Therefore it is important to use both large-scale (low resolution) and a small-scale (high resolution) datasets in order to measure $M_{\rm BH}$. For axisymmetric models, the ability of the Schwarzschild technique itself to self-consistently recover $\Upsilon_{*}$ and $M_{\rm BH}$ is limited if the sphere of influence of the BH is not well resolved \citep{Valluri2004,Cretton2004}, and there are no reasons to expect that the situation may be better in triaxial systems. We also note that the the large-scale parameters $\Omega$ and $\Upsilon_{*}$ are not completely independent either. Larger values of $\Omega$ give a lower (more negative) Jacobi energy, and therefore a deeper effective potential well. This can manifest as either a larger $M_{\rm BH}$ as explained above, or a larger $\Upsilon_{*}$. We can see that the $\Delta \chi^2$ contours in the $\Omega-\Upsilon_{*}$ plane are not circular but rather tilted with a positive correlation (Figure~\ref{fig:omega_contours.pdf}). This effect is further explored in detail by \citet{Koda2002}.

Finally, the main goal of this paper was to accurately model the BP/X shape of the bar, and we were still able to obtain loose constraints on $M_{\rm BH}$ using the deprojected model. We modelled the central bulge with a simple Einasto profile and only explored a few alternative density profiles. Future work focusing on the inner $\sim 1$ kpc may be needed to better model the density close to the radius of influence of the black hole, which can provide tighter constraints on $M_{\rm BH}$. 

\section{Conclusion}
\label{sec:conclusion}
We have developed a new parametric method to model bars with a peanut/X-shaped bulge (section \ref{sec:components}). This density model offers great flexibility to model bars of varying strengths and shapes (figure \ref{fig:bar_models}). We have tested the applicability of this model with N-body snapshots of a barred galaxy. Our important findings are: 
\begin{itemize}
    \item The parametric model provides an excellent fit to the density distribution snapshot when fit directly to the 3D density distribution of the snapshot (which we call the fit-3D-snap model, figure \ref{fig:3D_fit}). 
    \item We project the snapshot along the edge-on inclination, with a viewing angle such that the BP/X shape is clearly visible. This projected image is then used as an input to IMFIT, and we reconstruct the 3D shape using our parametric model (which we call the deprojected model), varying its parameters to best fit the projected image. We find that the overall features of the snapshot are well reproduced in the deprojected model, with minor discrepancies (e.g. lack of spiral arms and ansae of the bar which are not modelled). We quantify the goodness of fit by comparing the binned densities of the snapshot and model and find that the high-density inner regions are in good agreement (see figures \ref{fig:2D_fits},\ref{fig:projections},\ref{fig:3D_fit}). 
    \item From the deprojected density, we  calculate the gravitational potential and forces of the model and compare them to the original snapshot. The potential agrees to within a maximum error of $\sim 5\%$ and the force to within $\sim 15 \%$ in all parts of the galaxy, with the average errors at the level of 1--2\% (figure \ref{fig:pot_forces}).
    \item We use the deprojected density and the fit-3D-snap density to construct a stellar dynamical model of the galaxy using the Schwarzschild orbit-superposition technique with the FORSTAND code. This stands in contrast to the recent studies by \citet{Tahmasebzadeh2022}, who used a MGE parametrization of the bar density, or to \citet{deNicola2020,deNicola2022}, who developed another deprojection method and demonstrated the good recovery of viewing angles with the Schwarzschild method, but for triaxial elliptical galaxies rather than barred discs.
    For both the deprojected and fit-3D-snap models, we are able to recover with good accuracy the ``large-scale'' properties of the galaxy: the bar pattern speed, mass-to-light ratio, and orientation angle of the bar (see figures \ref{fig:omega_contours.pdf},\ref{fig:psi_omega_contours}). 
    \item We attempt to use the high-resolution kinematics in the inner region to recover the mass of the central SMBH. While the total $\chi^2$ of the models can provide upper constraints at best, the value of $\chi^2_{\rm kin,hr}$ (i.e. the contribution from the high-resolution dataset) is a better tracer of $M_{\rm BH}$, and can provide an upper limit and a weak lower limit to the black hole mass. We explored the sensitivity of the best-fit $M_{\rm BH}$ to the large-scale parameters $\Omega$ and $\Upsilon_{*}$. In particular, we find that underestimation of the mass-to-light ratio and/or overestimation of the pattern speed of the bar leads to higher calculated black hole masses (see figures \ref{fig:mbh},\ref{fig:mbh_var}). 
\end{itemize}
Although the method presented in this paper has only been applied to N-body snapshots so far, we are confident that it can be easily adapted to model real galaxies. This is the first method that recovers the density distribution in BP/X bars and uses it in fitting self-consistent dynamical models to the photometric/kinematic data of edge-on galaxies, where traditional methods may fail.  We hope that this will prove useful in illuminating the dynamics and evolution of barred galaxies. 



\section{Acknowledgements}
The authors thank Behzad Tahmasebzadeh and Leandro Beraldo e Silva for helpful and illuminating discussions. SD, MV, \& VW gratefully acknowledge funding from the National Science Foundation (grants NSF-AST-1515001, NSF-AST-2009122) and the Space Telescope Science Institute (grant JWST-ERS-01364.002-A).
\section{Data Availability}
Software: AGAMA \citep{Vasiliev2019}, IMFIT \citep{erwin2015}, matplotlib \citep{Hunter2007}, numpy \citep{Harris2020}, scipy \citep{Virtanen2020}.\\
The N-body snapshot used in this paper (the final snapshot of model BB1) can be found at \url{https://zenodo.org/record/8230972}.

\bibliographystyle{mnras}
\bibliography{references} 

\begin{thebibliography}{}
\makeatletter
\relax
\def\mn@urlcharsother{\let\do\@makeother \do\$\do\&\do\#\do\^\do\_\do\%\do\~}
\def\mn@doi{\begingroup\mn@urlcharsother \@ifnextchar [ {\mn@doi@} {\mn@doi@[]}}
\def\mn@doi@[#1]#2{\def\@tempa{#1}\ifx\@tempa\@empty \href {http://dx.doi.org/#2} {doi:#2}\else \href {http://dx.doi.org/#2} {#1}\fi \endgroup}
\def\mn@eprint#1#2{\mn@eprint@#1:#2::\@nil}
\def\mn@eprint@arXiv#1{\href {http://arxiv.org/abs/#1} {{\tt arXiv:#1}}}
\def\mn@eprint@dblp#1{\href {http://dblp.uni-trier.de/rec/bibtex/#1.xml} {dblp:#1}}
\def\mn@eprint@#1:#2:#3:#4\@nil{\def\@tempa {#1}\def\@tempb {#2}\def\@tempc {#3}\ifx \@tempc \@empty \let \@tempc \@tempb \let \@tempb \@tempa \fi \ifx \@tempb \@empty \def\@tempb {arXiv}\fi \@ifundefined {mn@eprint@\@tempb}{\@tempb:\@tempc}{\expandafter \expandafter \csname mn@eprint@\@tempb\endcsname \expandafter{\@tempc}}}

\bibitem[\protect\citeauthoryear{{Abbott}, {Valluri}, {Shen}  \& {Debattista}}{{Abbott} et~al.}{2017}]{Abbott17}
{Abbott} C.~G.,  {Valluri} M.,  {Shen} J.,   {Debattista} V.~P.,  2017, \mn@doi [\mnras] {10.1093/mnras/stx1262}, \href {https://ui.adsabs.harvard.edu/abs/2017MNRAS.470.1526A} {470, 1526}

\bibitem[\protect\citeauthoryear{{Aguerri}, {M{\'e}ndez-Abreu}  \& {Corsini}}{{Aguerri} et~al.}{2009}]{Aguerri2009}
{Aguerri} J.~A.~L.,  {M{\'e}ndez-Abreu} J.,   {Corsini} E.~M.,  2009, \mn@doi [\aap] {10.1051/0004-6361:200810931}, \href {https://ui.adsabs.harvard.edu/abs/2009A&A...495..491A} {495, 491}

\bibitem[\protect\citeauthoryear{{Aguerri} et~al.,}{{Aguerri} et~al.}{2015}]{Aguerri2015}
{Aguerri} J.~A.~L.,  et~al., 2015, \mn@doi [\aap] {10.1051/0004-6361/201423383}, \href {https://ui.adsabs.harvard.edu/abs/2015A&A...576A.102A} {576, A102}

\bibitem[\protect\citeauthoryear{{Anderson}, {Debattista}, {Erwin}, {Liddicott}, {Deg}  \& {Beraldo e Silva}}{{Anderson} et~al.}{2022}]{Anderson_etal_2022}
{Anderson} S.~R.,  {Debattista} V.~P.,  {Erwin} P.,  {Liddicott} D.~J.,  {Deg} N.,   {Beraldo e Silva} L.,  2022, \mn@doi [\mnras] {10.1093/mnras/stac913}, \href {https://ui.adsabs.harvard.edu/abs/2022MNRAS.513.1642A} {513, 1642}

\bibitem[\protect\citeauthoryear{{Athanassoula}}{{Athanassoula}}{2003}]{Athanassoula2003_2}
{Athanassoula} E.,  2003, \mn@doi [\mnras] {10.1046/j.1365-8711.2003.06473.x}, \href {https://ui.adsabs.harvard.edu/abs/2003MNRAS.341.1179A} {341, 1179}

\bibitem[\protect\citeauthoryear{{Athanassoula} \& {Misiriotis}}{{Athanassoula} \& {Misiriotis}}{2002}]{Athanassoula2002}
{Athanassoula} E.,  {Misiriotis} A.,  2002, \mn@doi [\mnras] {10.1046/j.1365-8711.2002.05028.x}, \href {https://ui.adsabs.harvard.edu/abs/2002MNRAS.330...35A} {330, 35}

\bibitem[\protect\citeauthoryear{{Athanassoula}, {Morin}, {Wozniak}, {Puy}, {Pierce}, {Lombard}  \& {Bosma}}{{Athanassoula} et~al.}{1990}]{Athanassoula1990}
{Athanassoula} E.,  {Morin} S.,  {Wozniak} H.,  {Puy} D.,  {Pierce} M.~J.,  {Lombard} J.,   {Bosma} A.,  1990, \mnras, \href {https://ui.adsabs.harvard.edu/abs/1990MNRAS.245..130A} {245, 130}

\bibitem[\protect\citeauthoryear{{Barazza}, {Jogee}  \& {Marinova}}{{Barazza} et~al.}{2008}]{Barazza2008}
{Barazza} F.~D.,  {Jogee} S.,   {Marinova} I.,  2008, \mn@doi [\apj] {10.1086/526510}, \href {https://ui.adsabs.harvard.edu/abs/2008ApJ...675.1194B} {675, 1194}

\bibitem[\protect\citeauthoryear{{Barnes} \& {Sellwood}}{{Barnes} \& {Sellwood}}{2003}]{Barnes2003}
{Barnes} E.~I.,  {Sellwood} J.~A.,  2003, \mn@doi [\aj] {10.1086/346142}, \href {https://ui.adsabs.harvard.edu/abs/2003AJ....125.1164B} {125, 1164}

\bibitem[\protect\citeauthoryear{{Bell} \& {de Jong}}{{Bell} \& {de Jong}}{2001}]{bell_dejong_2001}
{Bell} E.~F.,  {de Jong} R.~S.,  2001, \mn@doi [\apj] {10.1086/319728}, \href {https://ui.adsabs.harvard.edu/abs/2001ApJ...550..212B} {550, 212}

\bibitem[\protect\citeauthoryear{{Bendinelli}}{{Bendinelli}}{1991}]{Bendinelli1991}
{Bendinelli} O.,  1991, \mn@doi [\apj] {10.1086/169595}, \href {https://ui.adsabs.harvard.edu/abs/1991ApJ...366..599B} {366, 599}

\bibitem[\protect\citeauthoryear{{Blitz} \& {Spergel}}{{Blitz} \& {Spergel}}{1991}]{Blitz_spergel_1991}
{Blitz} L.,  {Spergel} D.~N.,  1991, \mn@doi [\apj] {10.1086/170535}, \href {https://ui.adsabs.harvard.edu/abs/1991ApJ...379..631B} {379, 631}

\bibitem[\protect\citeauthoryear{{Borodina}, {Williams}, {Sormani}, {Meidt}  \& {Schinnerer}}{{Borodina} et~al.}{2023}]{Borodina2023}
{Borodina} O.,  {Williams} T.~G.,  {Sormani} M.~C.,  {Meidt} S.,   {Schinnerer} E.,  2023, \mn@doi [\mnras] {10.1093/mnras/stad2068}, \href {https://ui.adsabs.harvard.edu/abs/2023MNRAS.tmp.1991B} {524, 3437}

\bibitem[\protect\citeauthoryear{{Brown}, {Valluri}, {Shen}  \& {Debattista}}{{Brown} et~al.}{2013}]{Brown2013}
{Brown} J.~S.,  {Valluri} M.,  {Shen} J.,   {Debattista} V.~P.,  2013, \mn@doi [\apj] {10.1088/0004-637X/778/2/151}, \href {https://ui.adsabs.harvard.edu/abs/2013ApJ...778..151B} {778, 151}

\bibitem[\protect\citeauthoryear{{Bundy} et~al.,}{{Bundy} et~al.}{2015}]{Bundy2015}
{Bundy} K.,  et~al., 2015, \mn@doi [\apj] {10.1088/0004-637X/798/1/7}, \href {https://ui.adsabs.harvard.edu/abs/2015ApJ...798....7B} {798, 7}

\bibitem[\protect\citeauthoryear{{Bureau} \& {Athanassoula}}{{Bureau} \& {Athanassoula}}{1999}]{Bureau1999}
{Bureau} M.,  {Athanassoula} E.,  1999, \mn@doi [\apj] {10.1086/307675}, \href {https://ui.adsabs.harvard.edu/abs/1999ApJ...522..686B} {522, 686}

\bibitem[\protect\citeauthoryear{{Bureau} \& {Athanassoula}}{{Bureau} \& {Athanassoula}}{2005}]{Bureau2005}
{Bureau} M.,  {Athanassoula} E.,  2005, \mn@doi [\apj] {10.1086/430056}, \href {https://ui.adsabs.harvard.edu/abs/2005ApJ...626..159B} {626, 159}

\bibitem[\protect\citeauthoryear{{Bureau} \& {Freeman}}{{Bureau} \& {Freeman}}{1999}]{Bureau1999_2}
{Bureau} M.,  {Freeman} K.~C.,  1999, \mn@doi [\aj] {10.1086/300922}, \href {https://ui.adsabs.harvard.edu/abs/1999AJ....118..126B} {118, 126}

\bibitem[\protect\citeauthoryear{{Cappellari}}{{Cappellari}}{2008}]{Cappellari2008}
{Cappellari} M.,  2008, \mn@doi [\mnras] {10.1111/j.1365-2966.2008.13754.x}, \href {https://ui.adsabs.harvard.edu/abs/2008MNRAS.390...71C} {390, 71}

\bibitem[\protect\citeauthoryear{{Cappellari} \& {Copin}}{{Cappellari} \& {Copin}}{2003}]{Cappellari2003}
{Cappellari} M.,  {Copin} Y.,  2003, \mn@doi [\mnras] {10.1046/j.1365-8711.2003.06541.x}, \href {https://ui.adsabs.harvard.edu/abs/2003MNRAS.342..345C} {342, 345}

\bibitem[\protect\citeauthoryear{{Collier}}{{Collier}}{2020}]{Collier2020}
{Collier} A.,  2020, \mn@doi [\mnras] {10.1093/mnras/stz3625}, \href {https://ui.adsabs.harvard.edu/abs/2020MNRAS.492.2241C} {492, 2241}

\bibitem[\protect\citeauthoryear{{Combes}, {Debbasch}, {Friedli}  \& {Pfenniger}}{{Combes} et~al.}{1990}]{Combes1990}
{Combes} F.,  {Debbasch} F.,  {Friedli} D.,   {Pfenniger} D.,  1990, \aap, \href {https://ui.adsabs.harvard.edu/abs/1990A&A...233...82C} {233, 82}

\bibitem[\protect\citeauthoryear{{Contopoulos}}{{Contopoulos}}{1980}]{Contopoulos1980}
{Contopoulos} G.,  1980, \aap, \href {https://ui.adsabs.harvard.edu/abs/1980A&A....81..198C} {81, 198}

\bibitem[\protect\citeauthoryear{{Cretton} \& {Emsellem}}{{Cretton} \& {Emsellem}}{2004}]{Cretton2004}
{Cretton} N.,  {Emsellem} E.,  2004, \mn@doi [\mnras] {10.1111/j.1365-2966.2004.07374.x}, \href {https://ui.adsabs.harvard.edu/abs/2004MNRAS.347L..31C} {347, L31}

\bibitem[\protect\citeauthoryear{{Cuomo} et~al.,}{{Cuomo} et~al.}{2019}]{Cuomo2019}
{Cuomo} V.,  et~al., 2019, \mn@doi [\mnras] {10.1093/mnras/stz1943}, \href {https://ui.adsabs.harvard.edu/abs/2019MNRAS.488.4972C} {488, 4972}

\bibitem[\protect\citeauthoryear{{Cuomo} et~al.,}{{Cuomo} et~al.}{2023}]{Cuomo_etal_2023}
{Cuomo} V.,  et~al., 2023, \mn@doi [\mnras] {10.1093/mnras/stac3047}, \href {https://ui.adsabs.harvard.edu/abs/2023MNRAS.518.2300C} {518, 2300}

\bibitem[\protect\citeauthoryear{{Debattista} \& {Sellwood}}{{Debattista} \& {Sellwood}}{2000}]{Debattista2000}
{Debattista} V.~P.,  {Sellwood} J.~A.,  2000, \mn@doi [\apj] {10.1086/317148}, \href {https://ui.adsabs.harvard.edu/abs/2000ApJ...543..704D} {543, 704}

\bibitem[\protect\citeauthoryear{{Debattista}, {Gerhard}  \& {Sevenster}}{{Debattista} et~al.}{2002}]{Debattista2002}
{Debattista} V.~P.,  {Gerhard} O.,   {Sevenster} M.~N.,  2002, \mn@doi [\mnras] {10.1046/j.1365-8711.2002.05500.x}, \href {https://ui.adsabs.harvard.edu/abs/2002MNRAS.334..355D} {334, 355}

\bibitem[\protect\citeauthoryear{{Debattista}, {Carollo}, {Mayer}  \& {Moore}}{{Debattista} et~al.}{2004}]{Debattista2004}
{Debattista} V.~P.,  {Carollo} C.~M.,  {Mayer} L.,   {Moore} B.,  2004, \mn@doi [\apjl] {10.1086/386332}, \href {https://ui.adsabs.harvard.edu/abs/2004ApJ...604L..93D} {604, L93}

\bibitem[\protect\citeauthoryear{{Debattista}, {Ness}, {Gonzalez}, {Freeman}, {Zoccali}  \& {Minniti}}{{Debattista} et~al.}{2017a}]{Debattista2017}
{Debattista} V.~P.,  {Ness} M.,  {Gonzalez} O.~A.,  {Freeman} K.,  {Zoccali} M.,   {Minniti} D.,  2017a, \mn@doi [\mnras] {10.1093/mnras/stx947}, \href {https://ui.adsabs.harvard.edu/abs/2017MNRAS.469.1587D} {469, 1587}

\bibitem[\protect\citeauthoryear{{Debattista}, {Ness}, {Gonzalez}, {Freeman}, {Zoccali}  \& {Minniti}}{{Debattista} et~al.}{2017b}]{Debattista_etal_2017}
{Debattista} V.~P.,  {Ness} M.,  {Gonzalez} O.~A.,  {Freeman} K.,  {Zoccali} M.,   {Minniti} D.,  2017b, \mn@doi [\mnras] {10.1093/mnras/stx947}, \href {https://ui.adsabs.harvard.edu/abs/2017MNRAS.469.1587D} {469, 1587}

\bibitem[\protect\citeauthoryear{{Debattista}, {Liddicott}, {Khachaturyants}  \& {Beraldo e Silva}}{{Debattista} et~al.}{2020}]{Debattista_etal_2020}
{Debattista} V.~P.,  {Liddicott} D.~J.,  {Khachaturyants} T.,   {Beraldo e Silva} L.,  2020, \mn@doi [\mnras] {10.1093/mnras/staa2568}, \href {https://ui.adsabs.harvard.edu/abs/2020MNRAS.498.3334D} {498, 3334}

\bibitem[\protect\citeauthoryear{{Einasto}}{{Einasto}}{1965}]{Einasto}
{Einasto} J.,  1965, Trudy Astrofizicheskogo Instituta Alma-Ata, \href {https://ui.adsabs.harvard.edu/abs/1965TrAlm...5...87E} {5, 87}

\bibitem[\protect\citeauthoryear{{Emsellem}, {Monnet}  \& {Bacon}}{{Emsellem} et~al.}{1994a}]{Emsellem1994}
{Emsellem} E.,  {Monnet} G.,   {Bacon} R.,  1994a, \aap, \href {https://ui.adsabs.harvard.edu/abs/1994A&A...285..723E} {285, 723}

\bibitem[\protect\citeauthoryear{{Emsellem}, {Monnet}, {Bacon}  \& {Nieto}}{{Emsellem} et~al.}{1994b}]{Emsellem1994_2}
{Emsellem} E.,  {Monnet} G.,  {Bacon} R.,   {Nieto} J.~L.,  1994b, \aap, \href {https://ui.adsabs.harvard.edu/abs/1994A&A...285..739E} {285, 739}

\bibitem[\protect\citeauthoryear{{Emsellem} et~al.,}{{Emsellem} et~al.}{2022}]{Emsellem2022}
{Emsellem} E.,  et~al., 2022, \mn@doi [\aap] {10.1051/0004-6361/202141727}, \href {https://ui.adsabs.harvard.edu/abs/2022A&A...659A.191E} {659, A191}

\bibitem[\protect\citeauthoryear{{Erwin}}{{Erwin}}{2015}]{erwin2015}
{Erwin} P.,  2015, \mn@doi [\apj] {10.1088/0004-637X/799/2/226}, \href {https://ui.adsabs.harvard.edu/abs/2015ApJ...799..226E} {799, 226}

\bibitem[\protect\citeauthoryear{{Erwin}}{{Erwin}}{2018}]{Erwin2018}
{Erwin} P.,  2018, \mn@doi [\mnras] {10.1093/mnras/stx3117}, \href {https://ui.adsabs.harvard.edu/abs/2018MNRAS.474.5372E} {474, 5372}

\bibitem[\protect\citeauthoryear{{Erwin} \& {Debattista}}{{Erwin} \& {Debattista}}{2013}]{Erwin2013}
{Erwin} P.,  {Debattista} V.~P.,  2013, \mn@doi [\mnras] {10.1093/mnras/stt385}, \href {https://ui.adsabs.harvard.edu/abs/2013MNRAS.431.3060E} {431, 3060}

\bibitem[\protect\citeauthoryear{{Erwin} \& {Debattista}}{{Erwin} \& {Debattista}}{2016}]{Erwin_Debattista_2016}
{Erwin} P.,  {Debattista} V.~P.,  2016, \mn@doi [\apjl] {10.3847/2041-8205/825/2/L30}, \href {https://ui.adsabs.harvard.edu/abs/2016ApJ...825L..30E} {825, L30}

\bibitem[\protect\citeauthoryear{{Erwin} \& {Debattista}}{{Erwin} \& {Debattista}}{2017}]{Erwin_Debattista2017}
{Erwin} P.,  {Debattista} V.~P.,  2017, \mn@doi [\mnras] {10.1093/mnras/stx620}, \href {https://ui.adsabs.harvard.edu/abs/2017MNRAS.468.2058E} {468, 2058}

\bibitem[\protect\citeauthoryear{{Erwin} et~al.,}{{Erwin} et~al.}{2021}]{Erwin2021}
{Erwin} P.,  et~al., 2021, \mn@doi [\mnras] {10.1093/mnras/stab126}, \href {https://ui.adsabs.harvard.edu/abs/2021MNRAS.502.2446E} {502, 2446}

\bibitem[\protect\citeauthoryear{{Eskridge} et~al.,}{{Eskridge} et~al.}{2002}]{eskridge_etal_02}
{Eskridge} P.~B.,  et~al., 2002, \apjs, 143, 73

\bibitem[\protect\citeauthoryear{{Font}, {Beckman}, {Epinat}, {Fathi}, {Guti{\'e}rrez}  \& {Hernandez}}{{Font} et~al.}{2011}]{Font2011}
{Font} J.,  {Beckman} J.~E.,  {Epinat} B.,  {Fathi} K.,  {Guti{\'e}rrez} L.,   {Hernandez} O.,  2011, \mn@doi [\apjl] {10.1088/2041-8205/741/1/L14}, \href {https://ui.adsabs.harvard.edu/abs/2011ApJ...741L..14F} {741, L14}

\bibitem[\protect\citeauthoryear{{Fragkoudi}, {Athanassoula}, {Bosma}  \& {Iannuzzi}}{{Fragkoudi} et~al.}{2015}]{Fragkoudi2015}
{Fragkoudi} F.,  {Athanassoula} E.,  {Bosma} A.,   {Iannuzzi} F.,  2015, \mn@doi [\mnras] {10.1093/mnras/stv537}, \href {https://ui.adsabs.harvard.edu/abs/2015MNRAS.450..229F} {450, 229}

\bibitem[\protect\citeauthoryear{{Fragkoudi}, {Di Matteo}, {Haywood}, {G{\'o}mez}, {Combes}, {Katz}  \& {Semelin}}{{Fragkoudi} et~al.}{2017}]{Fragkoudi2017}
{Fragkoudi} F.,  {Di Matteo} P.,  {Haywood} M.,  {G{\'o}mez} A.,  {Combes} F.,  {Katz} D.,   {Semelin} B.,  2017, \mn@doi [\aap] {10.1051/0004-6361/201630244}, \href {https://ui.adsabs.harvard.edu/abs/2017A&A...606A..47F} {606, A47}

\bibitem[\protect\citeauthoryear{{Fragkoudi}, {Grand}, {Pakmor}, {Springel}, {White}, {Marinacci}, {Gomez}  \& {Navarro}}{{Fragkoudi} et~al.}{2021}]{Fragkoudi2021}
{Fragkoudi} F.,  {Grand} R.~J.~J.,  {Pakmor} R.,  {Springel} V.,  {White} S.~D.~M.,  {Marinacci} F.,  {Gomez} F.~A.,   {Navarro} J.~F.,  2021, \mn@doi [\aap] {10.1051/0004-6361/202140320}, \href {https://ui.adsabs.harvard.edu/abs/2021A&A...650L..16F} {650, L16}

\bibitem[\protect\citeauthoryear{{Freeman}}{{Freeman}}{1970}]{Freeman1970}
{Freeman} K.~C.,  1970, \mn@doi [\apj] {10.1086/150474}, \href {https://ui.adsabs.harvard.edu/abs/1970ApJ...160..811F} {160, 811}

\bibitem[\protect\citeauthoryear{{Gadotti}}{{Gadotti}}{2009}]{Gadotti2009}
{Gadotti} D.~A.,  2009, in Chaos in Astronomy. p.~159 (\mn@eprint {arXiv} {0802.0495}), \mn@doi{10.1007/978-3-540-75826-6\_15}

\bibitem[\protect\citeauthoryear{{Gadotti}, {Athanassoula}, {Carrasco}, {Bosma}, {de Souza}  \& {Recillas}}{{Gadotti} et~al.}{2007}]{Gadotti2007}
{Gadotti} D.~A.,  {Athanassoula} E.,  {Carrasco} L.,  {Bosma} A.,  {de Souza} R.~E.,   {Recillas} E.,  2007, \mn@doi [\mnras] {10.1111/j.1365-2966.2007.12295.x}, \href {https://ui.adsabs.harvard.edu/abs/2007MNRAS.381..943G} {381, 943}

\bibitem[\protect\citeauthoryear{{Gadotti} et~al.,}{{Gadotti} et~al.}{2019}]{Gadotti2019}
{Gadotti} D.~A.,  et~al., 2019, \mn@doi [\mnras] {10.1093/mnras/sty2666}, \href {https://ui.adsabs.harvard.edu/abs/2019MNRAS.482..506G} {482, 506}

\bibitem[\protect\citeauthoryear{{Garc{\'\i}a-G{\'o}mez}, {Athanassoula}  \& {Barber{\`a}}}{{Garc{\'\i}a-G{\'o}mez} et~al.}{2002}]{GarciaGomez2002}
{Garc{\'\i}a-G{\'o}mez} C.,  {Athanassoula} E.,   {Barber{\`a}} C.,  2002, \mn@doi [\aap] {10.1051/0004-6361:20020460}, \href {https://ui.adsabs.harvard.edu/abs/2002A&A...389...68G} {389, 68}

\bibitem[\protect\citeauthoryear{{Garma-Oehmichen}, {Cano-D{\'\i}az}, {Hern{\'a}ndez-Toledo}, {Aquino-Ort{\'\i}z}, {Valenzuela}, {Aguerri}, {S{\'a}nchez}  \& {Merrifield}}{{Garma-Oehmichen} et~al.}{2020}]{Oehmichen2020}
{Garma-Oehmichen} L.,  {Cano-D{\'\i}az} M.,  {Hern{\'a}ndez-Toledo} H.,  {Aquino-Ort{\'\i}z} E.,  {Valenzuela} O.,  {Aguerri} J.~A.~L.,  {S{\'a}nchez} S.~F.,   {Merrifield} M.,  2020, \mn@doi [\mnras] {10.1093/mnras/stz3101}, \href {https://ui.adsabs.harvard.edu/abs/2020MNRAS.491.3655G} {491, 3655}

\bibitem[\protect\citeauthoryear{{Garma-Oehmichen} et~al.,}{{Garma-Oehmichen} et~al.}{2022}]{Oehmichen2022}
{Garma-Oehmichen} L.,  et~al., 2022, \mn@doi [\mnras] {10.1093/mnras/stac3069}, \href {https://ui.adsabs.harvard.edu/abs/2022MNRAS.517.5660G} {517, 5660}

\bibitem[\protect\citeauthoryear{{Gerhard} \& {Binney}}{{Gerhard} \& {Binney}}{1996}]{Gerhard1996}
{Gerhard} O.~E.,  {Binney} J.~J.,  1996, \mn@doi [\mnras] {10.1093/mnras/279.3.993}, \href {https://ui.adsabs.harvard.edu/abs/1996MNRAS.279..993G} {279, 993}

\bibitem[\protect\citeauthoryear{{Guo}, {Mao}, {Athanassoula}, {Li}, {Ge}, {Long}, {Merrifield}  \& {Masters}}{{Guo} et~al.}{2019}]{Guo2019}
{Guo} R.,  {Mao} S.,  {Athanassoula} E.,  {Li} H.,  {Ge} J.,  {Long} R.~J.,  {Merrifield} M.,   {Masters} K.,  2019, \mn@doi [\mnras] {10.1093/mnras/sty2715}, \href {https://ui.adsabs.harvard.edu/abs/2019MNRAS.482.1733G} {482, 1733}

\bibitem[\protect\citeauthoryear{{H{\"a}fner}, {Evans}, {Dehnen}  \& {Binney}}{{H{\"a}fner} et~al.}{2000}]{Hafner2000}
{H{\"a}fner} R.,  {Evans} N.~W.,  {Dehnen} W.,   {Binney} J.,  2000, \mn@doi [\mnras] {10.1046/j.1365-8711.2000.03242.x}, \href {https://ui.adsabs.harvard.edu/abs/2000MNRAS.314..433H} {314, 433}

\bibitem[\protect\citeauthoryear{{Harris} et~al.,}{{Harris} et~al.}{2020}]{Harris2020}
{Harris} C.~R.,  et~al., 2020, \mn@doi [\nat] {10.1038/s41586-020-2649-2}, \href {https://ui.adsabs.harvard.edu/abs/2020Natur.585..357H} {585, 357}

\bibitem[\protect\citeauthoryear{{Hernquist} \& {Weinberg}}{{Hernquist} \& {Weinberg}}{1992}]{Hernquist1992}
{Hernquist} L.,  {Weinberg} M.~D.,  1992, \mn@doi [\apj] {10.1086/171975}, \href {https://ui.adsabs.harvard.edu/abs/1992ApJ...400...80H} {400, 80}

\bibitem[\protect\citeauthoryear{{Hunter}}{{Hunter}}{2007}]{Hunter2007}
{Hunter} J.~D.,  2007, \mn@doi [Computing in Science and Engineering] {10.1109/MCSE.2007.55}, \href {https://ui.adsabs.harvard.edu/abs/2007CSE.....9...90H} {9, 90}

\bibitem[\protect\citeauthoryear{{Jethwa}, {Thater}, {Maindl}  \& {Van de Ven}}{{Jethwa} et~al.}{2020}]{Jethwa2020}
{Jethwa} P.,  {Thater} S.,  {Maindl} T.,   {Van de Ven} G.,  2020, {DYNAMITE: DYnamics, Age and Metallicity Indicators Tracing Evolution}, Astrophysics Source Code Library, record ascl:2011.007 (\mn@eprint {ascl} {2011.007})

\bibitem[\protect\citeauthoryear{{Knapen}}{{Knapen}}{1999}]{Knapen99}
{Knapen} J.~H.,  1999, in {Beckman} J.~E.,  {Mahoney} T.~J.,  eds,  Astronomical Society of the Pacific Conference Series Vol. 187, The Evolution of Galaxies on Cosmological Timescales. pp 72--87 (\mn@eprint {} {astro-ph/9907290})

\bibitem[\protect\citeauthoryear{{Kochanek} \& {Rybicki}}{{Kochanek} \& {Rybicki}}{1996}]{Kochanek1996}
{Kochanek} C.~S.,  {Rybicki} G.~B.,  1996, \mn@doi [\mnras] {10.1093/mnras/280.4.1257}, \href {https://ui.adsabs.harvard.edu/abs/1996MNRAS.280.1257K} {280, 1257}

\bibitem[\protect\citeauthoryear{{Koda} \& {Wada}}{{Koda} \& {Wada}}{2002}]{Koda2002}
{Koda} J.,  {Wada} K.,  2002, \mn@doi [\aap] {10.1051/0004-6361:20021461}, \href {https://ui.adsabs.harvard.edu/abs/2002A&A...396..867K} {396, 867}

\bibitem[\protect\citeauthoryear{{Kormendy} \& {Kennicutt}}{{Kormendy} \& {Kennicutt}}{2004}]{Kormendy2004}
{Kormendy} J.,  {Kennicutt} Robert~C. J.,  2004, \mn@doi [\araa] {10.1146/annurev.astro.42.053102.134024}, \href {https://ui.adsabs.harvard.edu/abs/2004ARA&A..42..603K} {42, 603}

\bibitem[\protect\citeauthoryear{{Krajnovi{\'c}} et~al.,}{{Krajnovi{\'c}} et~al.}{2015}]{Krajnovic2015}
{Krajnovi{\'c}} D.,  et~al., 2015, \mn@doi [\mnras] {10.1093/mnras/stv958}, \href {https://ui.adsabs.harvard.edu/abs/2015MNRAS.452....2K} {452, 2}

\bibitem[\protect\citeauthoryear{{Kruk}, {Erwin}, {Debattista}  \& {Lintott}}{{Kruk} et~al.}{2019}]{Kruk2019}
{Kruk} S.~J.,  {Erwin} P.,  {Debattista} V.~P.,   {Lintott} C.,  2019, \mn@doi [\mnras] {10.1093/mnras/stz2877}, \href {https://ui.adsabs.harvard.edu/abs/2019MNRAS.490.4721K} {490, 4721}

\bibitem[\protect\citeauthoryear{{Kuijken} \& {Dubinski}}{{Kuijken} \& {Dubinski}}{1995}]{Kuijken_Dubinski_1995}
{Kuijken} K.,  {Dubinski} J.,  1995, \mn@doi [\mnras] {10.1093/mnras/277.4.1341}, \href {https://ui.adsabs.harvard.edu/abs/1995MNRAS.277.1341K} {277, 1341}

\bibitem[\protect\citeauthoryear{{Kuijken} \& {Merrifield}}{{Kuijken} \& {Merrifield}}{1995}]{Kuijken1995}
{Kuijken} K.,  {Merrifield} M.~R.,  1995, \mn@doi [\apjl] {10.1086/187824}, \href {https://ui.adsabs.harvard.edu/abs/1995ApJ...443L..13K} {443, L13}

\bibitem[\protect\citeauthoryear{{Lablanche} et~al.,}{{Lablanche} et~al.}{2012}]{Lablanche2012}
{Lablanche} P.-Y.,  et~al., 2012, \mn@doi [\mnras] {10.1111/j.1365-2966.2012.21343.x}, \href {https://ui.adsabs.harvard.edu/abs/2012MNRAS.424.1495L} {424, 1495}

\bibitem[\protect\citeauthoryear{{Laurikainen} \& {Salo}}{{Laurikainen} \& {Salo}}{2016}]{Laurikainen2016}
{Laurikainen} E.,  {Salo} H.,  2016, in {Laurikainen} E.,  {Peletier} R.,   {Gadotti} D.,  eds,  Astrophysics and Space Science Library Vol. 418, Galactic Bulges. p.~77 (\mn@eprint {arXiv} {1505.00590}), \mn@doi{10.1007/978-3-319-19378-6_4}

\bibitem[\protect\citeauthoryear{{Laurikainen}, {Salo}, {Buta}  \& {Knapen}}{{Laurikainen} et~al.}{2011}]{Laurikainen_etal_2011}
{Laurikainen} E.,  {Salo} H.,  {Buta} R.,   {Knapen} J.~H.,  2011, \mn@doi [\mnras] {10.1111/j.1365-2966.2011.19283.x}, \href {https://ui.adsabs.harvard.edu/abs/2011MNRAS.418.1452L} {418, 1452}

\bibitem[\protect\citeauthoryear{{Li}, {Ho}, {Barth}  \& {Peng}}{{Li} et~al.}{2011}]{Li2011}
{Li} Z.-Y.,  {Ho} L.~C.,  {Barth} A.~J.,   {Peng} C.~Y.,  2011, \mn@doi [\apjs] {10.1088/0067-0049/197/2/22}, \href {https://ui.adsabs.harvard.edu/abs/2011ApJS..197...22L} {197, 22}

\bibitem[\protect\citeauthoryear{{Lipka} \& {Thomas}}{{Lipka} \& {Thomas}}{2021}]{Lipka2021}
{Lipka} M.,  {Thomas} J.,  2021, \mn@doi [\mnras] {10.1093/mnras/stab1092}, \href {https://ui.adsabs.harvard.edu/abs/2021MNRAS.504.4599L} {504, 4599}

\bibitem[\protect\citeauthoryear{{{\L}okas}}{{{\L}okas}}{2019}]{Lokas2019}
{{\L}okas} E.~L.,  2019, \mn@doi [\aap] {10.1051/0004-6361/201936056}, \href {https://ui.adsabs.harvard.edu/abs/2019A&A...629A..52L} {629, A52}

\bibitem[\protect\citeauthoryear{{Long} \& {Mao}}{{Long} \& {Mao}}{2010}]{Long2010}
{Long} R.~J.,  {Mao} S.,  2010, \mn@doi [\mnras] {10.1111/j.1365-2966.2010.16438.x}, \href {https://ui.adsabs.harvard.edu/abs/2010MNRAS.405..301L} {405, 301}

\bibitem[\protect\citeauthoryear{{Long}, {Mao}, {Shen}  \& {Wang}}{{Long} et~al.}{2013}]{Long2013}
{Long} R.~J.,  {Mao} S.,  {Shen} J.,   {Wang} Y.,  2013, \mn@doi [\mnras] {10.1093/mnras/sts285}, \href {https://ui.adsabs.harvard.edu/abs/2013MNRAS.428.3478L} {428, 3478}

\bibitem[\protect\citeauthoryear{{Magorrian}}{{Magorrian}}{1999}]{Magorrain1999}
{Magorrian} J.,  1999, \mn@doi [\mnras] {10.1046/j.1365-8711.1999.02135.x}, \href {https://ui.adsabs.harvard.edu/abs/1999MNRAS.302..530M} {302, 530}

\bibitem[\protect\citeauthoryear{{Magorrian}}{{Magorrian}}{2006}]{Magorrian2006}
{Magorrian} J.,  2006, \mn@doi [\mnras] {10.1111/j.1365-2966.2006.11054.x}, \href {https://ui.adsabs.harvard.edu/abs/2006MNRAS.373..425M} {373, 425}

\bibitem[\protect\citeauthoryear{{Marinova} \& {Jogee}}{{Marinova} \& {Jogee}}{2007a}]{marinova_jogee_07}
{Marinova} I.,  {Jogee} S.,  2007a, \mn@doi [\apj] {10.1086/512355}, \href {http://adsabs.harvard.edu/abs/2007ApJ...659.1176M} {659, 1176}

\bibitem[\protect\citeauthoryear{{Marinova} \& {Jogee}}{{Marinova} \& {Jogee}}{2007b}]{Irina2007}
{Marinova} I.,  {Jogee} S.,  2007b, \mn@doi [\apj] {10.1086/512355}, \href {https://ui.adsabs.harvard.edu/abs/2007ApJ...659.1176M} {659, 1176}

\bibitem[\protect\citeauthoryear{{Martinez-Valpuesta}, {Shlosman}  \& {Heller}}{{Martinez-Valpuesta} et~al.}{2006}]{Martinez-Valpuesta06}
{Martinez-Valpuesta} I.,  {Shlosman} I.,   {Heller} C.,  2006, \mn@doi [\apj] {10.1086/498338}, \href {https://ui.adsabs.harvard.edu/abs/2006ApJ...637..214M} {637, 214}

\bibitem[\protect\citeauthoryear{{Martinez-Valpuesta}, {Knapen}  \& {Buta}}{{Martinez-Valpuesta} et~al.}{2007}]{Martinez2007}
{Martinez-Valpuesta} I.,  {Knapen} J.~H.,   {Buta} R.,  2007, \mn@doi [\aj] {10.1086/522205}, \href {https://ui.adsabs.harvard.edu/abs/2007AJ....134.1863M} {134, 1863}

\bibitem[\protect\citeauthoryear{{Merrell}, {Vasiliev}, {Bentz}, {Valluri}  \& {Onken}}{{Merrell} et~al.}{2023}]{Merrell2023}
{Merrell} K.~A.,  {Vasiliev} E.,  {Bentz} M.~C.,  {Valluri} M.,   {Onken} C.~A.,  2023, \mn@doi [\apj] {10.3847/1538-4357/acc4bc}, \href {https://ui.adsabs.harvard.edu/abs/2023ApJ...949...13M} {949, 13}

\bibitem[\protect\citeauthoryear{{Miller} \& {van Dokkum}}{{Miller} \& {van Dokkum}}{2021}]{Miller2021}
{Miller} T.~B.,  {van Dokkum} P.,  2021, \mn@doi [\apj] {10.3847/1538-4357/ac2b30}, \href {https://ui.adsabs.harvard.edu/abs/2021ApJ...923..124M} {923, 124}

\bibitem[\protect\citeauthoryear{{Monnet}, {Bacon}  \& {Emsellem}}{{Monnet} et~al.}{1992}]{Monnet1992}
{Monnet} G.,  {Bacon} R.,   {Emsellem} E.,  1992, \aap, \href {https://ui.adsabs.harvard.edu/abs/1992A&A...253..366M} {253, 366}

\bibitem[\protect\citeauthoryear{{Navarro}, {Eke}  \& {Frenk}}{{Navarro} et~al.}{1996}]{Navarro_etal_1996}
{Navarro} J.~F.,  {Eke} V.~R.,   {Frenk} C.~S.,  1996, \mn@doi [\mnras] {10.1093/mnras/283.3.L72}, \href {https://ui.adsabs.harvard.edu/abs/1996MNRAS.283L..72N} {283, L72}

\bibitem[\protect\citeauthoryear{{Ness} \& {Lang}}{{Ness} \& {Lang}}{2016}]{Ness2016}
{Ness} M.,  {Lang} D.,  2016, \mn@doi [\aj] {10.3847/0004-6256/152/1/14}, \href {https://ui.adsabs.harvard.edu/abs/2016AJ....152...14N} {152, 14}

\bibitem[\protect\citeauthoryear{{Neureiter} et~al.,}{{Neureiter} et~al.}{2021}]{Neureiter2021}
{Neureiter} B.,  et~al., 2021, \mn@doi [\mnras] {10.1093/mnras/staa3014}, \href {https://ui.adsabs.harvard.edu/abs/2021MNRAS.500.1437N} {500, 1437}

\bibitem[\protect\citeauthoryear{{Onken} et~al.,}{{Onken} et~al.}{2014}]{Onken2014}
{Onken} C.~A.,  et~al., 2014, \mn@doi [\apj] {10.1088/0004-637X/791/1/37}, \href {https://ui.adsabs.harvard.edu/abs/2014ApJ...791...37O} {791, 37}

\bibitem[\protect\citeauthoryear{{Palicio} et~al.,}{{Palicio} et~al.}{2018}]{Palicio2018}
{Palicio} P.~A.,  et~al., 2018, \mn@doi [\mnras] {10.1093/mnras/sty1156}, \href {https://ui.adsabs.harvard.edu/abs/2018MNRAS.478.1231P} {478, 1231}

\bibitem[\protect\citeauthoryear{{Palmer}}{{Palmer}}{1994}]{palmer1994}
{Palmer} P.~L.,  1994, \mn@doi [\mnras] {10.1093/mnras/266.3.697}, \href {https://ui.adsabs.harvard.edu/abs/1994MNRAS.266..697P} {266, 697}

\bibitem[\protect\citeauthoryear{{Pfenniger} \& {Friedli}}{{Pfenniger} \& {Friedli}}{1991}]{Pfenniger_Friedli_1991}
{Pfenniger} D.,  {Friedli} D.,  1991, \aap, \href {https://ui.adsabs.harvard.edu/abs/1991A&A...252...75P} {252, 75}

\bibitem[\protect\citeauthoryear{{Pfenniger}, {Saha}  \& {Wu}}{{Pfenniger} et~al.}{2023}]{Pfenniger2023}
{Pfenniger} D.,  {Saha} K.,   {Wu} Y.-T.,  2023, \mn@doi [\aap] {10.1051/0004-6361/202245463}, \href {https://ui.adsabs.harvard.edu/abs/2023A&A...673A..36P} {673, A36}

\bibitem[\protect\citeauthoryear{{Pi{\~n}ol-Ferrer}, {Fathi}, {Carignan}, {Font}, {Hernandez}, {Karlsson}  \& {van de Ven}}{{Pi{\~n}ol-Ferrer} et~al.}{2014}]{Ferrer2014}
{Pi{\~n}ol-Ferrer} N.,  {Fathi} K.,  {Carignan} C.,  {Font} J.,  {Hernandez} O.,  {Karlsson} R.,   {van de Ven} G.,  2014, \mn@doi [\mnras] {10.1093/mnras/stt2162}, \href {https://ui.adsabs.harvard.edu/abs/2014MNRAS.438..971P} {438, 971}

\bibitem[\protect\citeauthoryear{{Picaud} \& {Robin}}{{Picaud} \& {Robin}}{2004}]{Picaud2004}
{Picaud} S.,  {Robin} A.~C.,  2004, \mn@doi [\aap] {10.1051/0004-6361:20041218}, \href {https://ui.adsabs.harvard.edu/abs/2004A&A...428..891P} {428, 891}

\bibitem[\protect\citeauthoryear{{Pilawa}, {Liepold}, {Delgado Andrade}, {Walsh}, {Ma}, {Quenneville}, {Greene}  \& {Blakeslee}}{{Pilawa} et~al.}{2022}]{Pilawa2022}
{Pilawa} J.~D.,  {Liepold} C.~M.,  {Delgado Andrade} S.~C.,  {Walsh} J.~L.,  {Ma} C.-P.,  {Quenneville} M.~E.,  {Greene} J.~E.,   {Blakeslee} J.~P.,  2022, \mn@doi [\apj] {10.3847/1538-4357/ac58fd}, \href {https://ui.adsabs.harvard.edu/abs/2022ApJ...928..178P} {928, 178}

\bibitem[\protect\citeauthoryear{{Portail}, {Wegg}, {Gerhard}  \& {Martinez-Valpuesta}}{{Portail} et~al.}{2015a}]{Portail2015}
{Portail} M.,  {Wegg} C.,  {Gerhard} O.,   {Martinez-Valpuesta} I.,  2015a, \mn@doi [\mnras] {10.1093/mnras/stv058}, \href {https://ui.adsabs.harvard.edu/abs/2015MNRAS.448..713P} {448, 713}

\bibitem[\protect\citeauthoryear{{Portail}, {Wegg}, {Gerhard}  \& {Martinez-Valpuesta}}{{Portail} et~al.}{2015b}]{Portail2015b}
{Portail} M.,  {Wegg} C.,  {Gerhard} O.,   {Martinez-Valpuesta} I.,  2015b, \mn@doi [\mnras] {10.1093/mnras/stv058}, \href {https://ui.adsabs.harvard.edu/abs/2015MNRAS.448..713P} {448, 713}

\bibitem[\protect\citeauthoryear{{Portail}, {Gerhard}, {Wegg}  \& {Ness}}{{Portail} et~al.}{2017}]{Portail2017}
{Portail} M.,  {Gerhard} O.,  {Wegg} C.,   {Ness} M.,  2017, \mn@doi [\mnras] {10.1093/mnras/stw2819}, \href {https://ui.adsabs.harvard.edu/abs/2017MNRAS.465.1621P} {465, 1621}

\bibitem[\protect\citeauthoryear{{Quenneville}, {Liepold}  \& {Ma}}{{Quenneville} et~al.}{2021}]{Quenneville_2021}
{Quenneville} M.~E.,  {Liepold} C.~M.,   {Ma} C.-P.,  2021, \mn@doi [\apjs] {10.3847/1538-4365/abe6a0}, \href {https://ui.adsabs.harvard.edu/abs/2021ApJS..254...25Q} {254, 25}

\bibitem[\protect\citeauthoryear{{Quillen}}{{Quillen}}{2002}]{Quillen02}
{Quillen} A.~C.,  2002, \mn@doi [\aj] {10.1086/341753}, \href {https://ui.adsabs.harvard.edu/abs/2002AJ....124..722Q} {124, 722}

\bibitem[\protect\citeauthoryear{{Quillen}, {Minchev}, {Sharma}, {Qin}  \& {Di Matteo}}{{Quillen} et~al.}{2014}]{Quillen14}
{Quillen} A.~C.,  {Minchev} I.,  {Sharma} S.,  {Qin} Y.-J.,   {Di Matteo} P.,  2014, \mn@doi [\mnras] {10.1093/mnras/stt1972}, \href {https://ui.adsabs.harvard.edu/abs/2014MNRAS.437.1284Q} {437, 1284}

\bibitem[\protect\citeauthoryear{{Raha}, {Sellwood}, {James}  \& {Kahn}}{{Raha} et~al.}{1991}]{Raha_91}
{Raha} N.,  {Sellwood} J.~A.,  {James} R.~A.,   {Kahn} F.~D.,  1991, \mn@doi [\nat] {10.1038/352411a0}, \href {https://ui.adsabs.harvard.edu/abs/1991Natur.352..411R} {352, 411}

\bibitem[\protect\citeauthoryear{{Rattenbury}, {Mao}, {Sumi}  \& {Smith}}{{Rattenbury} et~al.}{2007}]{Rattenbury2007}
{Rattenbury} N.~J.,  {Mao} S.,  {Sumi} T.,   {Smith} M.~C.,  2007, \mn@doi [\mnras] {10.1111/j.1365-2966.2007.11843.x}, \href {https://ui.adsabs.harvard.edu/abs/2007MNRAS.378.1064R} {378, 1064}

\bibitem[\protect\citeauthoryear{{Rautiainen}, {Salo}  \& {Laurikainen}}{{Rautiainen} et~al.}{2008}]{Rautiainen2008}
{Rautiainen} P.,  {Salo} H.,   {Laurikainen} E.,  2008, \mn@doi [\mnras] {10.1111/j.1365-2966.2008.13522.x}, \href {https://ui.adsabs.harvard.edu/abs/2008MNRAS.388.1803R} {388, 1803}

\bibitem[\protect\citeauthoryear{{Robin}, {Marshall}, {Schultheis}  \& {Reyl{\'e}}}{{Robin} et~al.}{2012}]{Robin2012}
{Robin} A.~C.,  {Marshall} D.~J.,  {Schultheis} M.,   {Reyl{\'e}} C.,  2012, \mn@doi [\aap] {10.1051/0004-6361/201116512}, \href {https://ui.adsabs.harvard.edu/abs/2012A&A...538A.106R} {538, A106}

\bibitem[\protect\citeauthoryear{{Roshan}, {Ghafourian}, {Kashfi}, {Banik}, {Haslbauer}, {Cuomo}, {Famaey}  \& {Kroupa}}{{Roshan} et~al.}{2021}]{Roshan2021}
{Roshan} M.,  {Ghafourian} N.,  {Kashfi} T.,  {Banik} I.,  {Haslbauer} M.,  {Cuomo} V.,  {Famaey} B.,   {Kroupa} P.,  2021, \mn@doi [\mnras] {10.1093/mnras/stab2553}, \href {https://ui.adsabs.harvard.edu/abs/2021MNRAS.508..926R} {508, 926}

\bibitem[\protect\citeauthoryear{Rybicki}{Rybicki}{1987}]{Rybicki1987}
Rybicki G.~B.,  1987, in De~Zeeuw T.,  ed., Structure and Dynamics of Elliptical Galaxies. Springer Netherlands, Dordrecht, pp 397--398

\bibitem[\protect\citeauthoryear{{S{\'a}nchez} et~al.,}{{S{\'a}nchez} et~al.}{2012}]{Sanchez2012}
{S{\'a}nchez} S.~F.,  et~al., 2012, \mn@doi [\aap] {10.1051/0004-6361/201117353}, \href {https://ui.adsabs.harvard.edu/abs/2012A&A...538A...8S} {538, A8}

\bibitem[\protect\citeauthoryear{{Sanders}, {Smith}  \& {Evans}}{{Sanders} et~al.}{2019}]{Sanders2019}
{Sanders} J.~L.,  {Smith} L.,   {Evans} N.~W.,  2019, \mn@doi [\mnras] {10.1093/mnras/stz1827}, \href {https://ui.adsabs.harvard.edu/abs/2019MNRAS.488.4552S} {488, 4552}

\bibitem[\protect\citeauthoryear{{Schwarzschild}}{{Schwarzschild}}{1979}]{Schwarzschild1979}
{Schwarzschild} M.,  1979, \mn@doi [\apj] {10.1086/157282}, \href {https://ui.adsabs.harvard.edu/abs/1979ApJ...232..236S} {232, 236}

\bibitem[\protect\citeauthoryear{{Sellwood}}{{Sellwood}}{2014}]{GALAXY}
{Sellwood} J.~A.,  2014, arXiv e-prints, \href {https://ui.adsabs.harvard.edu/abs/2014arXiv1406.6606S} {p. arXiv:1406.6606}

\bibitem[\protect\citeauthoryear{{Sellwood} \& {Athanassoula}}{{Sellwood} \& {Athanassoula}}{1986}]{Sellwood1986}
{Sellwood} J.~A.,  {Athanassoula} E.,  1986, \mn@doi [\mnras] {10.1093/mnras/221.2.195}, \href {https://ui.adsabs.harvard.edu/abs/1986MNRAS.221..195S} {221, 195}

\bibitem[\protect\citeauthoryear{{Sellwood} \& {Gerhard}}{{Sellwood} \& {Gerhard}}{2020}]{Sellwood_Gerhard_2020}
{Sellwood} J.~A.,  {Gerhard} O.,  2020, \mn@doi [\mnras] {10.1093/mnras/staa1336}, \href {https://ui.adsabs.harvard.edu/abs/2020MNRAS.495.3175S} {495, 3175}

\bibitem[\protect\citeauthoryear{{Sellwood} \& {Wilkinson}}{{Sellwood} \& {Wilkinson}}{1993}]{Sellwood1993}
{Sellwood} J.~A.,  {Wilkinson} A.,  1993, \mn@doi [Reports on Progress in Physics] {10.1088/0034-4885/56/2/001}, \href {https://ui.adsabs.harvard.edu/abs/1993RPPh...56..173S} {56, 173}

\bibitem[\protect\citeauthoryear{{S\'ersic}}{{S\'ersic}}{1968}]{Sersic}
{S\'ersic} J.~L.,  1968, {Atlas de Galaxias Australes}.
Observatorio Astronomico Cordoba

\bibitem[\protect\citeauthoryear{{Sheth} et~al.,}{{Sheth} et~al.}{2008}]{Sheth08}
{Sheth} K.,  et~al., 2008, \mn@doi [\apj] {10.1086/524980}, \href {http://adsabs.harvard.edu/abs/2008ApJ...675.1141S} {675, 1141}

\bibitem[\protect\citeauthoryear{{Smirnov} \& {Savchenko}}{{Smirnov} \& {Savchenko}}{2020}]{Smirnov}
{Smirnov} A.~A.,  {Savchenko} S.~S.,  2020, \mn@doi [\mnras] {10.1093/mnras/staa2892}, \href {https://ui.adsabs.harvard.edu/abs/2020MNRAS.499..462S} {499, 462}

\bibitem[\protect\citeauthoryear{{Sormani}, {Gerhard}, {Portail}, {Vasiliev}  \& {Clarke}}{{Sormani} et~al.}{2022}]{Sormani2022}
{Sormani} M.~C.,  {Gerhard} O.,  {Portail} M.,  {Vasiliev} E.,   {Clarke} J.,  2022, \mn@doi [\mnras] {10.1093/mnrasl/slac046}, \href {https://ui.adsabs.harvard.edu/abs/2022MNRAS.514L...1S} {514, L1}

\bibitem[\protect\citeauthoryear{{Syer} \& {Tremaine}}{{Syer} \& {Tremaine}}{1996}]{Syer1996}
{Syer} D.,  {Tremaine} S.,  1996, \mn@doi [\mnras] {10.1093/mnras/282.1.223}, \href {https://ui.adsabs.harvard.edu/abs/1996MNRAS.282..223S} {282, 223}

\bibitem[\protect\citeauthoryear{{Tahmasebzadeh}, {Zhu}, {Shen}, {Gerhard}  \& {Qin}}{{Tahmasebzadeh} et~al.}{2021}]{Tahmasebzadeh2021}
{Tahmasebzadeh} B.,  {Zhu} L.,  {Shen} J.,  {Gerhard} O.,   {Qin} Y.,  2021, \mn@doi [\mnras] {10.1093/mnras/stab3002}, \href {https://ui.adsabs.harvard.edu/abs/2021MNRAS.508.6209T} {508, 6209}

\bibitem[\protect\citeauthoryear{{Tahmasebzadeh}, {Zhu}, {Shen}, {Gerhard}  \& {van de Ven}}{{Tahmasebzadeh} et~al.}{2022}]{Tahmasebzadeh2022}
{Tahmasebzadeh} B.,  {Zhu} L.,  {Shen} J.,  {Gerhard} O.,   {van de Ven} G.,  2022, \mn@doi [\apj] {10.3847/1538-4357/ac9df6}, \href {https://ui.adsabs.harvard.edu/abs/2022ApJ...941..109T} {941, 109}

\bibitem[\protect\citeauthoryear{{Thater}, {Krajnovi{\'c}}, {Cappellari}, {Davis}, {de Zeeuw}, {McDermid}  \& {Sarzi}}{{Thater} et~al.}{2019}]{Thater2019}
{Thater} S.,  {Krajnovi{\'c}} D.,  {Cappellari} M.,  {Davis} T.~A.,  {de Zeeuw} P.~T.,  {McDermid} R.~M.,   {Sarzi} M.,  2019, \mn@doi [\aap] {10.1051/0004-6361/201834808}, \href {https://ui.adsabs.harvard.edu/abs/2019A&A...625A..62T} {625, A62}

\bibitem[\protect\citeauthoryear{{Thater} et~al.,}{{Thater} et~al.}{2022}]{Thater2022}
{Thater} S.,  et~al., 2022, \mn@doi [\aap] {10.1051/0004-6361/202243926}, \href {https://ui.adsabs.harvard.edu/abs/2022A&A...667A..51T} {667, A51}

\bibitem[\protect\citeauthoryear{{Thomas}, {Saglia}, {Bender}, {Erwin}  \& {Fabricius}}{{Thomas} et~al.}{2014}]{Thomas2014}
{Thomas} J.,  {Saglia} R.~P.,  {Bender} R.,  {Erwin} P.,   {Fabricius} M.,  2014, \mn@doi [\apj] {10.1088/0004-637X/782/1/39}, \href {https://ui.adsabs.harvard.edu/abs/2014ApJ...782...39T} {782, 39}

\bibitem[\protect\citeauthoryear{{Tremaine} \& {Weinberg}}{{Tremaine} \& {Weinberg}}{1984}]{Tremaine_Weinberg_1984}
{Tremaine} S.,  {Weinberg} M.~D.,  1984, \mn@doi [\apjl] {10.1086/184292}, \href {https://ui.adsabs.harvard.edu/abs/1984ApJ...282L...5T} {282, L5}

\bibitem[\protect\citeauthoryear{{Valluri}, {Merritt}  \& {Emsellem}}{{Valluri} et~al.}{2004}]{Valluri2004}
{Valluri} M.,  {Merritt} D.,   {Emsellem} E.,  2004, \mn@doi [\apj] {10.1086/380896}, \href {https://ui.adsabs.harvard.edu/abs/2004ApJ...602...66V} {602, 66}

\bibitem[\protect\citeauthoryear{{Vasiliev}}{{Vasiliev}}{2019}]{Vasiliev2019}
{Vasiliev} E.,  2019, \mn@doi [\mnras] {10.1093/mnras/sty2672}, \href {https://ui.adsabs.harvard.edu/abs/2019MNRAS.482.1525V} {482, 1525}

\bibitem[\protect\citeauthoryear{{Vasiliev} \& {Athanassoula}}{{Vasiliev} \& {Athanassoula}}{2015}]{Vasiliev2015}
{Vasiliev} E.,  {Athanassoula} E.,  2015, \mn@doi [\mnras] {10.1093/mnras/stv805}, \href {https://ui.adsabs.harvard.edu/abs/2015MNRAS.450.2842V} {450, 2842}

\bibitem[\protect\citeauthoryear{{Vasiliev} \& {Valluri}}{{Vasiliev} \& {Valluri}}{2020a}]{VV20_conference}
{Vasiliev} E.,  {Valluri} M.,  2020a, in {Valluri} M.,  {Sellwood} J.~A.,  eds,  Proceedings of the International Astronomical Union Vol. 353, Galactic Dynamics in the Era of Large Surveys. Cambridge University Press, pp 176--183, \mn@doi{10.1017/S1743921319008706}

\bibitem[\protect\citeauthoryear{{Vasiliev} \& {Valluri}}{{Vasiliev} \& {Valluri}}{2020b}]{VV20}
{Vasiliev} E.,  {Valluri} M.,  2020b, \mn@doi [\apj] {10.3847/1538-4357/ab5fe0}, \href {https://ui.adsabs.harvard.edu/abs/2020ApJ...889...39V} {889, 39}

\bibitem[\protect\citeauthoryear{{Virtanen} et~al.,}{{Virtanen} et~al.}{2020}]{Virtanen2020}
{Virtanen} P.,  et~al., 2020, \mn@doi [Nature Methods] {10.1038/s41592-019-0686-2}, \href {https://ui.adsabs.harvard.edu/abs/2020NatMe..17..261V} {17, 261}

\bibitem[\protect\citeauthoryear{{Walsh}, {van den Bosch}, {Barth}  \& {Sarzi}}{{Walsh} et~al.}{2012}]{Walsh2012}
{Walsh} J.~L.,  {van den Bosch} R. C.~E.,  {Barth} A.~J.,   {Sarzi} M.,  2012, \mn@doi [\apj] {10.1088/0004-637X/753/1/79}, \href {https://ui.adsabs.harvard.edu/abs/2012ApJ...753...79W} {753, 79}

\bibitem[\protect\citeauthoryear{{Wang}, {Zhao}, {Mao}  \& {Rich}}{{Wang} et~al.}{2012}]{Wang2012}
{Wang} Y.,  {Zhao} H.,  {Mao} S.,   {Rich} R.~M.,  2012, \mn@doi [\mnras] {10.1111/j.1365-2966.2012.22063.x}, \href {https://ui.adsabs.harvard.edu/abs/2012MNRAS.427.1429W} {427, 1429}

\bibitem[\protect\citeauthoryear{{Wang}, {Mao}, {Long}  \& {Shen}}{{Wang} et~al.}{2013}]{Wang2013}
{Wang} Y.,  {Mao} S.,  {Long} R.~J.,   {Shen} J.,  2013, \mn@doi [\mnras] {10.1093/mnras/stt1537}, \href {https://ui.adsabs.harvard.edu/abs/2013MNRAS.435.3437W} {435, 3437}

\bibitem[\protect\citeauthoryear{{Wegg} \& {Gerhard}}{{Wegg} \& {Gerhard}}{2013}]{Wegg2013}
{Wegg} C.,  {Gerhard} O.,  2013, \mn@doi [\mnras] {10.1093/mnras/stt1376}, \href {https://ui.adsabs.harvard.edu/abs/2013MNRAS.435.1874W} {435, 1874}

\bibitem[\protect\citeauthoryear{{Wegg}, {Gerhard}  \& {Portail}}{{Wegg} et~al.}{2015}]{Wegg2015}
{Wegg} C.,  {Gerhard} O.,   {Portail} M.,  2015, \mn@doi [\mnras] {10.1093/mnras/stv745}, \href {https://ui.adsabs.harvard.edu/abs/2015MNRAS.450.4050W} {450, 4050}

\bibitem[\protect\citeauthoryear{{Wheeler}, {Valluri}, {Beraldo e Silva}, {Dattathri}  \& {Debattista}}{{Wheeler} et~al.}{2023}]{Wheeler2023}
{Wheeler} V.,  {Valluri} M.,  {Beraldo e Silva} L.,  {Dattathri} S.,   {Debattista} V.~P.,  2023, ApJ submitted

\bibitem[\protect\citeauthoryear{{Widrow} \& {Dubinski}}{{Widrow} \& {Dubinski}}{2005}]{Widrow_Dubinski_2005}
{Widrow} L.~M.,  {Dubinski} J.,  2005, \mn@doi [\apj] {10.1086/432710}, \href {https://ui.adsabs.harvard.edu/abs/2005ApJ...631..838W} {631, 838}

\bibitem[\protect\citeauthoryear{{Widrow}, {Pym}  \& {Dubinski}}{{Widrow} et~al.}{2008}]{Widrow_etal_2008}
{Widrow} L.~M.,  {Pym} B.,   {Dubinski} J.,  2008, \mn@doi [\apj] {10.1086/587636}, \href {https://ui.adsabs.harvard.edu/abs/2008ApJ...679.1239W} {679, 1239}

\bibitem[\protect\citeauthoryear{{Williams} et~al.,}{{Williams} et~al.}{2021}]{Williams2021}
{Williams} T.~G.,  et~al., 2021, \mn@doi [\aj] {10.3847/1538-3881/abe243}, \href {https://ui.adsabs.harvard.edu/abs/2021AJ....161..185W} {161, 185}

\bibitem[\protect\citeauthoryear{{Xiang} et~al.,}{{Xiang} et~al.}{2021}]{Xiang2021}
{Xiang} K.~M.,  et~al., 2021, \mn@doi [\apj] {10.3847/1538-4357/abdab5}, \href {https://ui.adsabs.harvard.edu/abs/2021ApJ...909..125X} {909, 125}

\bibitem[\protect\citeauthoryear{{Yoshino} \& {Yamauchi}}{{Yoshino} \& {Yamauchi}}{2015}]{Yoshino2015}
{Yoshino} A.,  {Yamauchi} C.,  2015, \mn@doi [\mnras] {10.1093/mnras/stu2249}, \href {https://ui.adsabs.harvard.edu/abs/2015MNRAS.446.3749Y} {446, 3749}

\bibitem[\protect\citeauthoryear{{Zhao}}{{Zhao}}{1996}]{Zhao1996}
{Zhao} H.,  1996, \mn@doi [\mnras] {10.1093/mnras/283.1.149}, \href {https://ui.adsabs.harvard.edu/abs/1996MNRAS.283..149Z} {283, 149}

\bibitem[\protect\citeauthoryear{{Zhu} et~al.,}{{Zhu} et~al.}{2018}]{Zhu2018}
{Zhu} L.,  et~al., 2018, \mn@doi [\mnras] {10.1093/mnras/stx2409}, \href {https://ui.adsabs.harvard.edu/abs/2018MNRAS.473.3000Z} {473, 3000}

\bibitem[\protect\citeauthoryear{{Zou}, {Shen}, {Bureau}  \& {Li}}{{Zou} et~al.}{2019}]{Zou2019}
{Zou} Y.,  {Shen} J.,  {Bureau} M.,   {Li} Z.-Y.,  2019, \mn@doi [\apj] {10.3847/1538-4357/ab3f34}, \href {https://ui.adsabs.harvard.edu/abs/2019ApJ...884...23Z} {884, 23}

\bibitem[\protect\citeauthoryear{{de Lorenzi}, {Debattista}, {Gerhard}  \& {Sambhus}}{{de Lorenzi} et~al.}{2007}]{deLorenzi2007}
{de Lorenzi} F.,  {Debattista} V.~P.,  {Gerhard} O.,   {Sambhus} N.,  2007, \mn@doi [\mnras] {10.1111/j.1365-2966.2007.11434.x}, \href {https://ui.adsabs.harvard.edu/abs/2007MNRAS.376...71D} {376, 71}

\bibitem[\protect\citeauthoryear{{de Nicola}, {Saglia}, {Thomas}, {Dehnen}  \& {Bender}}{{de Nicola} et~al.}{2020}]{deNicola2020}
{de Nicola} S.,  {Saglia} R.~P.,  {Thomas} J.,  {Dehnen} W.,   {Bender} R.,  2020, \mn@doi [\mnras] {10.1093/mnras/staa1703}, \href {https://ui.adsabs.harvard.edu/abs/2020MNRAS.496.3076D} {496, 3076}

\bibitem[\protect\citeauthoryear{{de Nicola}, {Neureiter}, {Thomas}, {Saglia}  \& {Bender}}{{de Nicola} et~al.}{2022}]{deNicola2022}
{de Nicola} S.,  {Neureiter} B.,  {Thomas} J.,  {Saglia} R.~P.,   {Bender} R.,  2022, \mn@doi [\mnras] {10.1093/mnras/stac2852}, \href {https://ui.adsabs.harvard.edu/abs/2022MNRAS.517.3445D} {517, 3445}

\bibitem[\protect\citeauthoryear{{den Brok}, {Krajnovi{\'c}}, {Emsellem}, {Brinchmann}  \& {Maseda}}{{den Brok} et~al.}{2021}]{denbrok2021}
{den Brok} M.,  {Krajnovi{\'c}} D.,  {Emsellem} E.,  {Brinchmann} J.,   {Maseda} M.,  2021, \mn@doi [\mnras] {10.1093/mnras/stab2852}, \href {https://ui.adsabs.harvard.edu/abs/2021MNRAS.508.4786D} {508, 4786}

\bibitem[\protect\citeauthoryear{{van de Sande}, {Fraser-McKelvie}, {Fisher}, {Martig}, {Hayden}  \& {the GECKOS Survey collaboration}}{{van de Sande} et~al.}{2023}]{vandesande2023}
{van de Sande} J.,  {Fraser-McKelvie} A.,  {Fisher} D.~B.,  {Martig} M.,  {Hayden} M.~R.,   {the GECKOS Survey collaboration} 2023, \mn@doi [arXiv e-prints] {10.48550/arXiv.2306.00059}, \href {https://ui.adsabs.harvard.edu/abs/2023arXiv230600059V} {p. arXiv:2306.00059}

\bibitem[\protect\citeauthoryear{{van den Bosch}}{{van den Bosch}}{1997}]{vdb1997}
{van den Bosch} F.~C.,  1997, \mn@doi [\mnras] {10.1093/mnras/287.3.543}, \href {https://ui.adsabs.harvard.edu/abs/1997MNRAS.287..543V} {287, 543}

\bibitem[\protect\citeauthoryear{{van den Bosch} \& {de Zeeuw}}{{van den Bosch} \& {de Zeeuw}}{2010}]{vdb2010}
{van den Bosch} R. C.~E.,  {de Zeeuw} P.~T.,  2010, \mn@doi [\mnras] {10.1111/j.1365-2966.2009.15832.x}, \href {https://ui.adsabs.harvard.edu/abs/2010MNRAS.401.1770V} {401, 1770}

\bibitem[\protect\citeauthoryear{{van den Bosch}, {Jaffe}  \& {van der Marel}}{{van den Bosch} et~al.}{1998}]{vdb1998}
{van den Bosch} F.~C.,  {Jaffe} W.,   {van der Marel} R.~P.,  1998, \mn@doi [\mnras] {10.1046/j.1365-8711.1998.01069.x}, \href {https://ui.adsabs.harvard.edu/abs/1998MNRAS.293..343V} {293, 343}

\bibitem[\protect\citeauthoryear{{van den Bosch}, {van de Ven}, {Verolme}, {Cappellari}  \& {de Zeeuw}}{{van den Bosch} et~al.}{2008}]{vdb2008}
{van den Bosch} R.~C.~E.,  {van de Ven} G.,  {Verolme} E.~K.,  {Cappellari} M.,   {de Zeeuw} P.~T.,  2008, \mn@doi [\mnras] {10.1111/j.1365-2966.2008.12874.x}, \href {https://ui.adsabs.harvard.edu/abs/2008MNRAS.385..647V} {385, 647}

\bibitem[\protect\citeauthoryear{{van der Kruit}}{{van der Kruit}}{1988}]{Kruit1988}
{van der Kruit} P.~C.,  1988, \aap, \href {https://ui.adsabs.harvard.edu/abs/1988A&A...192..117V} {192, 117}

\makeatother
\end{thebibliography}




\appendix

\section{FORSTAND models with added noise}  \label{sec:appendix}
\begin{figure}
    \centering
    \includegraphics[width=\columnwidth]{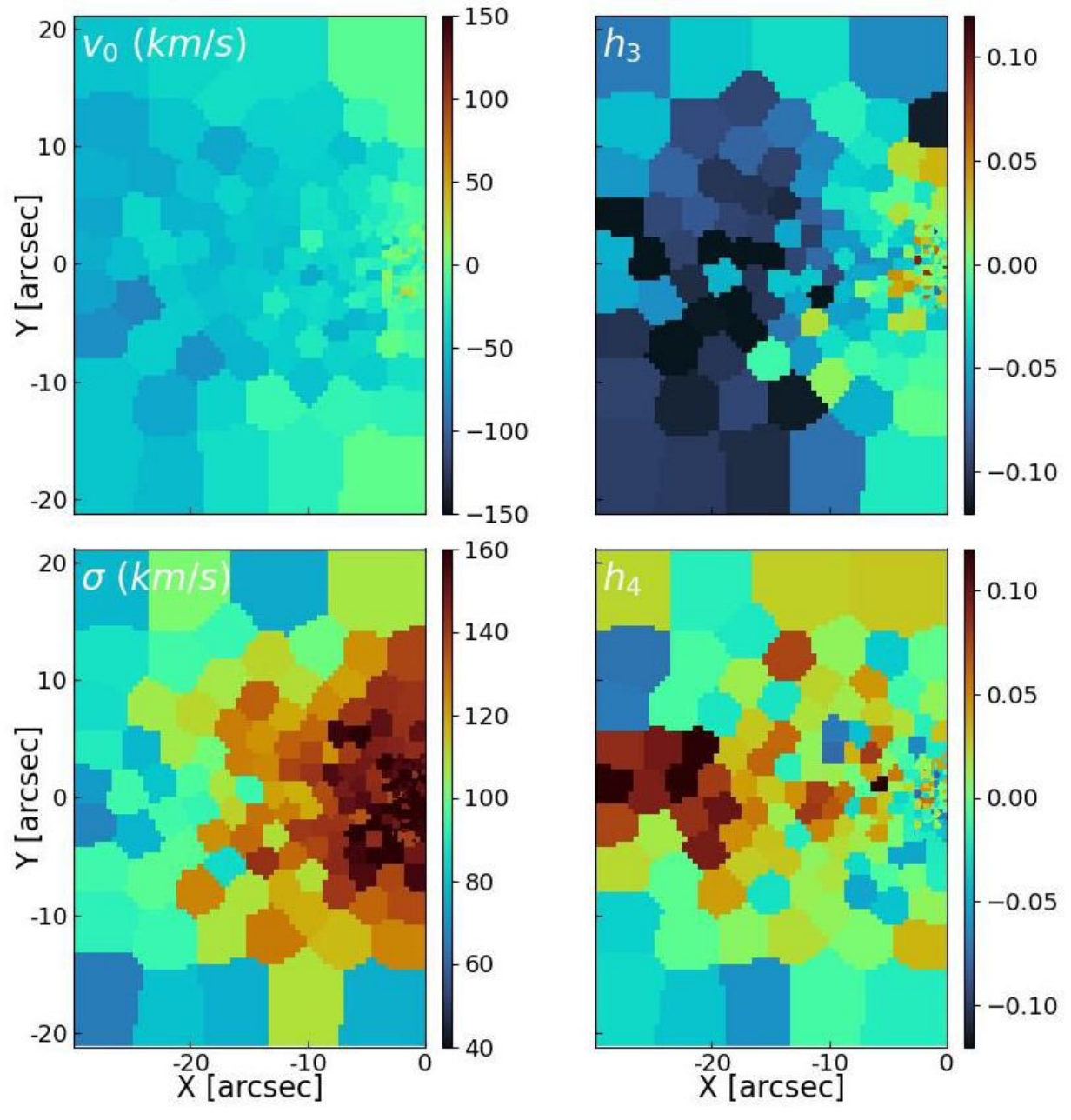}
    \caption{An example kinematic map with added noise, with only the first four GH moment maps shown. The values of $V$ and $\sigma$ are perturbed by random noise of amplitude $10$ km/s, whereas $h_3-h_6$ are perturbed by random noise of amplitude $\pm 0.03$.}
    \label{fig:noisy_maps}
\end{figure}

Here we present the results of dynamical modelling with FORSTAND with added noise. The pixels of the input snapshot are grouped using Voronoi binning as before, and the LOSVDs are computed in each bin. Each LOSVD is expanded into a Gauss-Hermite series up to 6$^{\rm th}$ order. Then, each of the GH coefficients are perturbed by a random value within a fixed amplitude. For $V$ and $\sigma$, the noise amplitude is $10$ km/s, and for $h_3-h_6$ the noise amplitude is $0.03$. These noise amplitudes are slightly higher than, but of the same order as, the noise expected from MUSE \citep[e.g.][]{Krajnovic2015, denbrok2021, Thater2022}. Figure \ref{fig:noisy_maps} shows an example of an input noisy kinematic map. 

\begin{figure}
    \centering
    \includegraphics[width=\columnwidth]{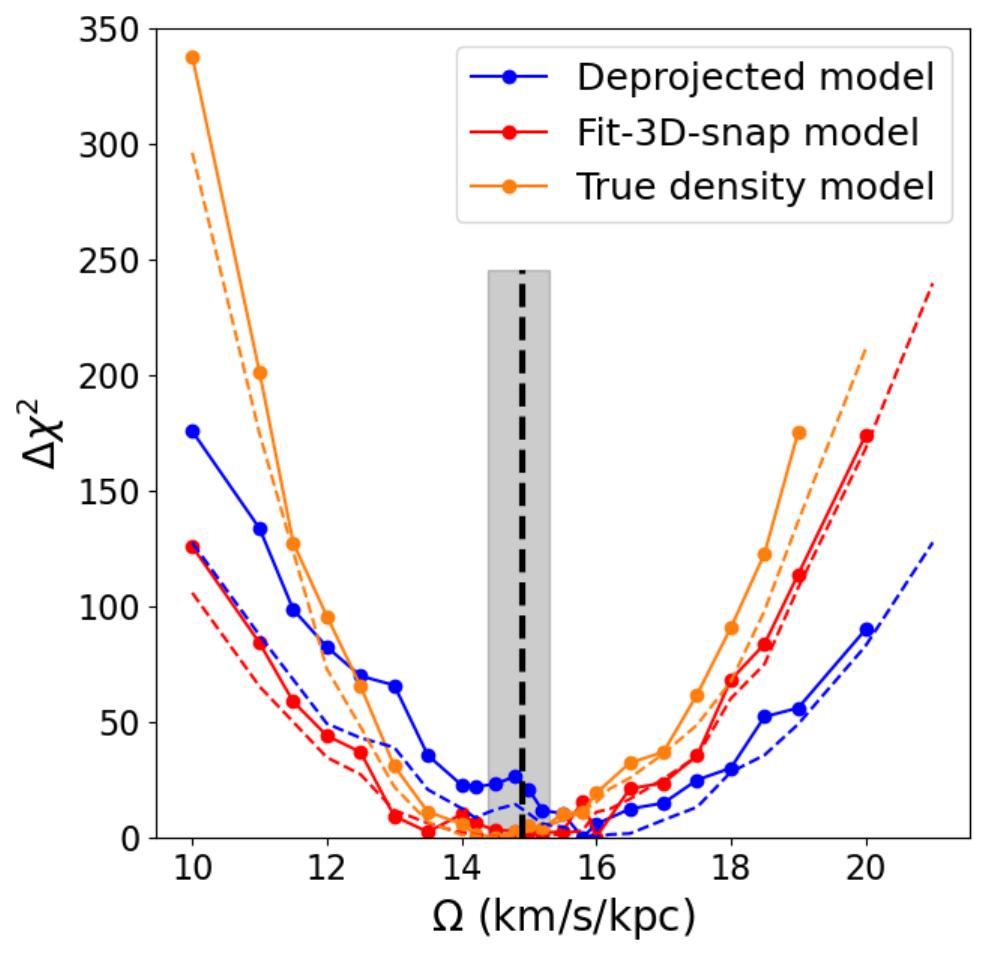}
    \caption{One-dimensional marginalized $\Delta \chi^2$ vs. $\Omega$ curves for noisy input data (solid lines), with the corresponding curves for Poisson noise only models as dashed lines. It is clear that even with the addition of noise, the recovery of $\Omega$ is robust. }
    \label{fig:omega_noisy}
\end{figure}

We then run FORSTAND on the noisy input data to recover the quantities of interest ($\Omega$, $\Upsilon_*$, $\psi$, $M_{\rm BH}$). Figure \ref{fig:omega_noisy} shows the marginalized one dimensional curves of $\Delta \chi^2$ vs. $\Omega$ for the deprojected, fit-3D-snap, and true density models with added noise as solid curves. For reference, the error-free (Poisson noise only) models are shown as dashed curves. We can clearly see that even with the added noise, the value of $\Omega$ is recovered to reasonable accuracy. Although we do not show them here, the analogues of figures \ref{fig:omega_contours.pdf} and  \ref{fig:psi_omega_contours} show that all three large-scale parameters ($\Omega$, $\Upsilon_*$, and $\psi$) can be recovered even with noisy input data. 

\begin{figure*}
    \centering
    \includegraphics[width=\textwidth]{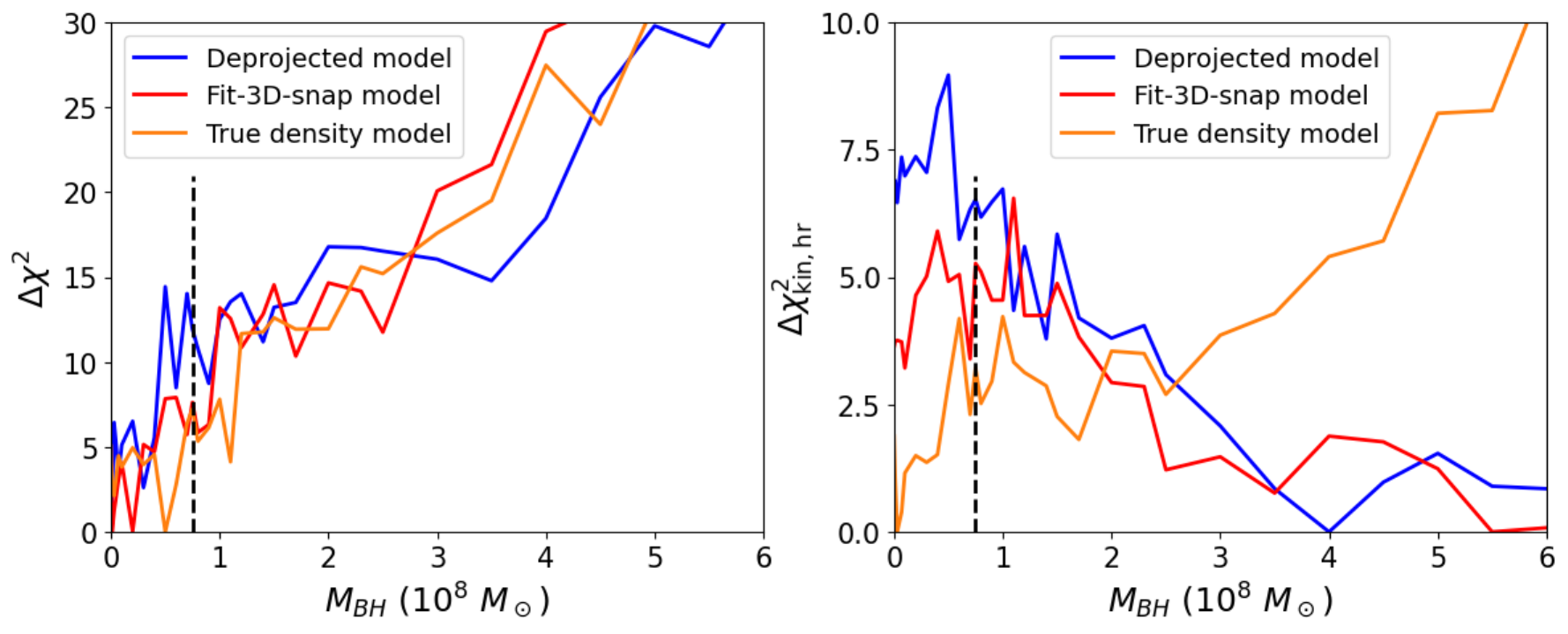}
    \caption{One-dimensional marginalized $\Delta \chi^2$ (left) and $\Delta \chi_{\rm kin,hr}^2$ (right) vs. $M_{\rm BH}$ curves for noisy input data. These curves are significantly more noisy than figure \ref{fig:mbh} (note that the y-axis scales are different). With the addition of noise, the weak constraints that were previously obtained in the error-free models are no longer present.}
    \label{fig:mbh_noisy}
\end{figure*}

However, the recovery of $M_{\rm BH}$ is quite different. Figure \ref{fig:mbh_noisy} shows the marginalized one dimensional curves of $\Delta \chi^2$ (left) and $\Delta \chi^2_{\rm kin,hr}$ (right) vs. $M_{\rm BH}$ for the noisy models (the analogue of figure \ref{fig:mbh}). It is clear that these curves are significantly more noisy than figure \ref{fig:mbh} and none of the models are able to recover the true value of $M_{\rm BH}$. Even the weak constraints that we were able to obtain in the error-free true density model (figure \ref{fig:mbh}) are no longer present. \\

In real observations, we can expect that the noise amplitude for the high-resolution kinematic dataset will be different than the low-resolution dataset, which may improve the results. A detailed investigation into the required signal-to-noise ratio required in order to accurately recover $M_{\rm BH}$ should be undertaken in the future. 

\bsp	
\label{lastpage}
\end{document}